\documentclass[10pt]{iopart}

\usepackage{iopams}
\usepackage{epstopdf}
\usepackage{graphicx}
\usepackage{cite}
\usepackage{textcomp}
\usepackage{color}

\expandafter\let\csname equation*\endcsname\relax
\expandafter\let\csname endequation*\endcsname\relax
\usepackage{amsmath}

\begin{document}
\newcommand{\unitm}[1]{\ensuremath{\,\mathrm{#1}}}
\newcommand{\unit}[1]{\,#1}

\topical%[WGM nonlinear and quantum optics]
{Nonlinear and Quantum Optics with Whispering Gallery Resonators}

\author{Dmitry V Strekalov\textsuperscript{1,2}, Christoph Marquardt\textsuperscript{2,3}, Andrey B Matsko\textsuperscript{4}, Harald G L Schwefel\textsuperscript{2,3,5} and Gerd Leuchs\textsuperscript{2,3}}

\address{\textsuperscript{1}Jet Propulsion Laboratory, California Institute of Technology, Pasadena, CA 91108, USA}
\address{\textsuperscript{2}Max Planck Institute for the Science of Light, G\"{u}nther-Scharowsky-Stra\ss e 1/Building 24, 90158 Erlangen, Germany}
\address{\textsuperscript{3}Institute for Optics, Information and Photonics, University Erlangen-N\"{u}rnberg, Staudtstr.7/B2, 90158 Erlangen, Germany}
\address{\textsuperscript{4}OEwaves Inc., 465 N. Halstead Str., Suite 140, Pasadena, California 91107, USA}
\address{\textsuperscript{5}Department of Physics, University of Otago, Dunedin, New Zealand}
\ead{dmitry.v.strekalov@jpl.nasa.gov}
\vspace{10pt}
\begin{indented}
\item[]\today
\end{indented}

\begin{abstract}
Optical Whispering Gallery Modes (WGMs) derive their name from a famous acoustic phenomenon of guiding a wave by a curved boundary observed nearly a century ago. This phenomenon has a rather general nature, equally applicable to sound and all other waves. It enables resonators of unique properties attractive both in science and engineering. Very high quality factors of optical WGM resonators persisting in a wide wavelength range spanning from radio frequencies to ultraviolet light, their small mode volume, and tunable in- and out- coupling  make them exceptionally efficient for nonlinear optical applications. Nonlinear optics facilitates interaction of photons  with each other and with other physical systems, and is of prime importance in quantum optics. In this paper we review numerous applications of WGM resonators in nonlinear and quantum optics. We outline the current areas of interest, summarize progress, highlight difficulties, and discuss possible future development trends in these areas.
\end{abstract}


\vspace{2pc}
\noindent{\it Keywords}: Microresonators, optical wave mixing, non-classical light
%
% Uncomment for Submitted to journal title message
%\submitto{\JPA}
%
% For two-column output uncomment the next line and choose [10pt] rather than [12pt] in the \documentclass declaration
\maketitle

\ioptwocol

\section{Introduction}\label{sec:Intro}

\subsection{The role of nonlinearity in optical science}\label{sec:Intro_nonlin}

Light does not interact with light in everyday life.  This interaction can only be induced by means of nonlinear physical systems and is only observable with strong enough electromagnetic fields. It is hardly possible to achieve such strong fields using incoherent light sources, so  the research in nonlinear optics commenced with the advent of high power light sources, namely the invention of lasers. Powerful pulsed lasers are often needed to observe nonlinear optical phenomena in bulk materials. Methods involving optical fibers \cite{agrawalbook}, plasmons \cite{maier2007plasmonics}, hollow-core photonic crystal fibers \cite{soukoulis2012photonic}, as well as metameterials \cite{engheta2006metamaterials,cai2010optical,joannopoulos2011photonic} allow reducing the light power requirement. This reduction is still insufficient for many applications where it is desirable to achieve a strong nonlinear response with faint light, ultimately at a few- or single-photon level.

Large nonlinear optical susceptibility is needed to observe nonlinear interactions at low light levels. There are two major ways of producing such a susceptibility: i) usage of a resonant, either natural or artificial, nonlinear media, and ii) usage of nonresonant but highly transparent nonlinear media embedded into optical cavities. Each of these approaches has its own advantages and disadvantages, and each should be evaluated in the context of the particular problem to be solved. In this contribution we focus on the second approach and discuss applications of monolithic optical microresonators in quantum and nonlinear optics.

Since their inception, monolithic microresonators became tools of nonlinear optics. Their major use is in the enhancement of the efficiency of nonlinear interactions occurring in transparent optical media. Unlike other types of optical cavities, monolithic ones can be integrated on a chip and multiplexed, which makes them indispensable in creation of chip-scale nonlinear optical devices able to generate optical harmonics, produce nonclassical states of light, process quantum information and so on. These resonators allow not only reducing the footprint of nonlinear optics experiments and moving them from the lab to industrial applications, but also facilitate nonlinear interaction at the single-photon level, representing one of the major goals of optical science nowadays.

The merit of a nonlinear optical system is often judged with respect to its optical attenuation introducing unwanted optical loss and decoherence. Resonant nonlinear media, such as atoms or plasmons, may have huge optical nonlinearity in a relatively narrow frequency band enabling interaction among single photons. However, the attenuation is also resonantly enhanced. Nonresonant nonlinear media, on the contrary, typically have relatively small optical nonlinearity, but also very small attenuation. As the result, photons confined in such a media for a long time have a better chance to interact without being absorbed. To support such a long interaction time optical cavities are utilized.

Cavity enhancement of the nonlinear interaction depends on the quality factor $Q$ and mode volume $V$ of the cavity. The quality factor defines the interaction time, while the mode volume characterizes the magnitude of the electric field of the confined photons. The smaller the cavity and the larger the quality factor, the stronger is the interaction. It is hard to reach the desirable values of both parameters at the same time. Usually, reduction of mode volume results in a decrease of the quality factor. For instance, the mode volume as small as $(\lambda/n_0)^3$ (where $\lambda$ is wavelength and $n_0$ is the refractive index of the cavity host material) was achieved in photonic bandgap nanocavities \cite{Akahane05oe,noda2007spontaneous,Dharanipathy14apl}, but quality factor does not exceed $2 \times 10^6$. Numerical optimization of the cavity shape show that Q-factor of a photonic bandgap cavity can reach $2 \times 10^7$ \cite{song2005ultra} if the fabricated cavity has ideal quality, but further increase is unlikely. This limitation arises from fundamental as well as from technical reasons. The confinement of light within a dielectric structure is reduced as the structure becomes smaller, while the tolerances of fabrication of such a structure become more stringent. It is possible to imagine a 3D photonic crystal nanocavity with a subwavelength mode volume, but fabrication of the cavity is out of reach of the existing technology.

Whispering gallery mode resonators (WGMRs) \cite{chang1996optical,vahala2003optical,vahala2004optical,matsko2006optical}, on the other hand, allow achieving three to four orders of magnitude higher quality factors at the expense of increased mode volumes. Since enhancement of the nonlinear conversion efficiency is usually proportional to $Q^2/V$ or $Q^3/V$, it may be reasonable to trade the small mode volume for larger $Q$-factor. On the other hand, higher $Q$ results in narrower linewidth and consequently in slower devices, so an optimization is usually needed. Another salient advantage of WGMRs is related to efficient coupling techniques developed for the resonators, which are extremely important in all the applications.

Here we review recent progress in both nonlinear and quantum optics applications of  WGMRs characterized with relatively small ($<10^6 \times (\lambda/n_0)^3$) mode volumes and ultra-high $Q$-factors ($Q>10^7$). These resonators have been widely utilized for efficiency enhancement of nonlinear optical processes such as three- and four-wave mixing, and also for observation of quantum effects associated with these processes. Long interaction time together with strong spatial confinement of light in nonlinear resonators result in a significant decrease of the thresholds of both lasing and stimulated scattering processes. Resonators made from optical crystals are excellent candidates for observation of broadband nonlinear optical phenomena, since these materials are highly transparent in the wavelength range of 150~nm to 10~$\mu$m. Finesse of these resonators can be very large. The highest demonstrated finesse in optics (${\cal F}>10^7$) has been achieved with a fluorite (CaF$_2$) resonator \cite{Savchenkov07bigF}.

The spectral properties of WGM resonators are well understood nowadays, and many means of design and control of WGM spectra are known (see sections \ref{sec:spectrum} and \ref{sec:tuning}). There are also many efficient techniques for coupling light in and out of these structures (see section \ref{sec:spectrum}). As all spatially extended systems, WGM resonators require phase matching for nonlinear interactions to be efficient. Phase matching in rotation-symmetric geometry, however, is very different from that in Cartesian geometry, and allows for greater flexibility. This topic is discussed in sections \ref{sec:chi2} and \ref{sec:Third-order} for the second- and third-order nonlinear-optical processes, respectively. Interaction of optical photons with solid-state excitations such as phonons is discussed in section \ref{sec:other}. Finally, in section \ref{sec:quant} we review quantum-optical phenomena accessible with monolithic microresonators. In the rest of this section, we briefly review the history and distinctive properties of WGM resonators.

\subsection{Discovery of WGM phenomena}

The first observation of a whispering gallery has faded from memory. In the realm of acoustics a smooth and circular wall features the characteristic that a soft spoken whisper can propagate along the wall to be heard by a listener far away from the source but close to the wall. Examples include the Gol Gumbaz in India, the Temple of Heaven in Beijing, and the St.\ Paul Cathedral in London. It was in London that Lord Rayleigh first described the phenomenon scientifically \cite{rayleigh_theory_1896,rayleigh_cxii._1910,rayleigh_ix._1914}, noting that the spherical wall reflects and continuously refocuses the sound wave and thus a whispering gallery is formed. An experimental observation of this effect was reported by Raman \cite{Raman1921WGM}. This concept of confining waves in the vicinity of a curved boundary was soon demonstrated in the scattering of electromagnetic radiation from gold spheres, which led to the establishment of Mie scattering theory \cite{mie_beitrage_1908}. It was furthermore found to be applicable to dielectric interfaces, which was important for understanding radar scattering from rain and hail. Debye simplified the underlying mathematical description \cite{debye_lichtdruck_1909} and applied it to the waves propagating along dielectric wires. For the next decades whispering gallery modes were only occasionally studied, such as by Richtmyer who considered dielectric wires bent into loops, i.e. ring resonators \cite{richtmyer_dielectric_1939}. One of the first solid state optical lasers was implemented in in a samarium doped fluorite sphere \cite{garrett_stimulated_1961}. On a much different length scale, the Earth troposphere and ionosphere can act as a closed waveguide supporting modes similar to WGMs \cite{wait_electromagnetic_1962}.

The WGM concept was picked up again in the 1980s and developed in two different directions comprising the study of properties of liquid droplets \cite{benner_observation_1980} in the optical domain, and properties of properly shaped pieces of high permittivity dielectrics \cite{vedrenne_whispering-gallery_1982} in the microwave domain. Let us discuss these directions in some more detail.

\subsection{Historic investigations of microwave WGM resonators }

Microwave WGM resonators were studied in parallel with the optical ones. Spherical \cite{gastine67tmtt,affolter73tmtt}, cylindrical \cite{wait67rs,vedrenne_whispering-gallery_1982,tobar91tmtt,krupka05tmtt}, and more complex resonator morphologies \cite{eremenko02tmtt} were considered. While there is no fundamental difference between the optical and microwave structures,
they are practically dissimilar because of i) the nature of attenuation in the resonator host material, and ii) the WGM excitation techniques.

The maximum achievable $Q$ factor of a microwave structure is determined by the dielectric loss tangent defined as  ${\rm tan}\, \delta=\epsilon''/\epsilon'$, where $\epsilon = \epsilon'+i\epsilon''$ is the complex permittivity of the material. Typically the loss tangent increases with microwave frequency \cite{fiedziuszko02tmtt,hartnett06uffc}. As a result, the product $Q\times f=f/{\rm tan}\, \delta$, where $f$ is the frequency of the microwaves, is considered as a constant. It means that available $Q$-factors are not very large at high microwave frequencies. For instance, the highest $Q$-factor attainable at $f=9$ GHz in a sapphire resonator at room temperature is $2\times 10^5$ \cite{hartnett13uffc}. This number can be improved to $10^9$ and possibly further at cryogenic temperatures \cite{Braginsky87cryo}.

Since the WGM quality factors are rather large compared to the quality factors of other types of dielectric structures, the microwave WGM resonators found use as filters \cite{jiao87tmtt}. They also are utilized as energy storage elements in ultra-stable microwave oscillators producing spectrally pure signals \cite{mcneilage04fcs,boudot06el,ivanov06mtt,Locke08mw,lefloch14rsi}. As a benchmark, a 9~GHz oscillator with phase noise of $-160$ dBc$/$Hz at 1~kHz offset frequency was demonstrated using microwave WGMs \cite{ivanov06mtt}. These oscillators are widely used. For instance, they were proposed for the tests of local Lorentz invariance by searching the difference in the speed of light in the directions parallel and perpendicular to the direction of the Earth motion around the Sun \cite{stanwix05prl,hartnett07fcs,tobar09prd}. Another application of the high-$Q$ structures is related to the study of weak attenuation of microwaves in nominally transparent materials \cite{krupka99tmtt,krupka99mst,hartnett06uffc,lefloch14rsi} and the material permitivity \cite{Tobar98rutile,Luiten98rutile}. A variety of nonlinear microwave phenomena can also be observed in WGM resonators \cite{Nand14rutile}.

The evanescent field of a microwave dielectric resonator may extend by as much as a millimeter. It means that a metallic antenna can be used to excite the WGMs \cite{lefloch14rsi}. For high radio frequency (RF) and THz radiation \cite{THzBook}, the excitation can be realized using dielectric \cite{DRW} or metal waveguides. Low order modes in the dielectric cavities are coupled to the free space and have significant radiative losses which allows applications of the structure as dielectric RF antennas \cite{long83tap}.

\subsection{Historic investigations of optical WGM resonators and their close relatives}

The requirement for a smooth resonator surface is much more stringent in optical than in the microwave domain. One approach to form a nearly perfect interface and reduce the scattering from the boundary roughness is offered by  surface tension. That is why WGMs have been initially studied in liquid droplets either caught in optical or ion-traps \cite{arnold_photoemission_1985} or planted onto hydrophobic surfaces \cite{Sennaroglu07R_las}. Their $Q$-factors  were of the order of $10^4-10^6$, thus stimulated phenomena such as Raman lasing was readily observed  \cite{Snow85Raman,Qian1986dropletRaman,Pinnick88SRS,Lin1994droplets,Qian1995droplets,Sennaroglu07R_las,Kiraz09R_las}. Conventional lasing was achieved as well by including dyes such as Rhodamine 6G \cite{Tzeng84las,Biswas89las,campillo91prl} or quantum dots \cite{schafer_quantum_2008} into the solution.

Surface tension  also defines shape and surface quality of microsphere resonators created by thermal reflow of glass. In 1989, Braginsky and co-workers \cite{Braginsky89nonlinWGM} were the first to realize that nearly perfect spheres can be formed by melting high-grade silica glass fibers. The potential of WGM resonators in quantum optics was emphasized already in this pioneering research: \textit{``With possible reduction of controlling energy of optical switching down to a single quantum and employment of the monophotonic states of light, the whispering-gallery microresonators can open the way to realize Feynman's quantum-mechanical computer."}

These microspheres are by now quarter-century old, and although they could not so far reach the parameters required to operate as quantum computer nodes, they allow observing essentially quantum effects. For instance, cavity QED experiments were performed with microspheres doped with nanoparticles \cite{fan00ol,park06nl}. The resonators also find such applications as single virus detection \cite{Vollmer08biosensing,Vollmer12review} and, by including rare earth dopants into the glass, lasing \cite{miura_laser_1996,Sandoghdar96laser,ilchenko2000microsphere,Cai2000las}.

Vahala and co-workers managed to combine the benefits of surface tension induced smoothness and lithographic production by creating micro-toroids out of silica on silicon \cite{Armani03chip}. These resonators showed an interaction of their optical WGMs and mechanical resonances via the radiation pressure \cite{kippenberg_analysis_2005}. Such opto-mechanical coupling allowed for cooling of a single \textit{mechanical} mode of an optical WGM resonator close to the quantum-mechanical ground state \cite{Kippenberg08optomech}. Toroidal fused silica cavities were used to achieve frequency comb generation \cite{Del'Haye07comb}, and very recently, to demonstrate the octave-wide comb operation \cite{Kippenberg11comb_rev,Okawachi11octave_comb,Del'Haye11octave}.

Along the same line is the fabrication of a different kind of resonator, the bottle resonator \cite{sumetsky04,Murugan10bottle}. These resonators are shaped by compressing a locally molten silica fiber in the axial direction, similarly to the first microtoroids \cite{ilchenko01ol}. The large tunability and potentially very large mode volume of bottle resonators make them interesting for coupling to atomic transitions \cite{louyer05}. Efficient nonlinear switching, envisioned for microspheres \cite{Braginsky89nonlinWGM}, was successfully demonstrated in bottle-shaped fused silica microresonators \cite{Pollinger10bottle,OShea11bottle}.

If a capillary is used instead of a fiber, hollow bottles or microbubble resonators \cite{sumetsky_optical_2010,Sumetsky10bubble,Han14bubble} can be created. They show promise for chemical sensing applications. Note that the whispering gallery \textit{waveguides} realized as straight (unstructured) capillaries also have been used for this purpose \cite{White06wgm_sensor,Zamora11wgm_sensor}.

Heating fibers by a CO$_2$ laser can be utilized for fabrication of Surface Nanoscale Axial Photonics (SNAP) resonators \cite{Sumetsky12snap} featuring sub-atomic precision of the surface profile control. The SNAP technology makes use of the frozen stress in the fiber, which, when locally heated, results into nanoscale deformation. Laser fabrication can be also applied to larger silica rods, yielding the resonators used for high quality frequency combs generation \cite{DelHaye13las_fab}. Precise shaping of optical rods can furthermore allow for experimental verification of interesting and counter-intuitive behaviors of WGM light which can form quasi-modes even in open systems  \cite{Sumetsky11cone,Strekalov15antiresonator}.

Crystalline materials with very low absorption can only be formed into high quality WGM resonators by mechnical polishing. One of the first solid state lasers reported in 1961 was a doped CaF$_2$:Sm$^{++}$ crystal sphere \cite{garrett_stimulated_1961}. Record quality factors of over $10^{10}$ have been achieved in undoped fluorite resonators \cite{Savchenkov04kHz,grudinin06highQ,grudinin06highQlin,Savchenkov07bigF}. These resonators also feature frequency combs \cite{Chembo10comb,chembo_modal_2010,Matsko12comb,matsko_generation_2011,Grudinin12comb,wang_mid-infrared_2013} and opto-mechanical interaction \cite{hofer_cavity_2010}. Another important aspect of crystalline materials is that they can be anisotropic and therefore support second order nonlinearities, paving the way for applications in quantum optics \cite{Ilchenko04SH}\footnote{In Eq.(5) of \cite{Ilchenko04SH}, both instances of $S(1+S)$ should be read as  $S(2+S)$.} and photonics \cite{ilchenko2000microsphere}. Generation of optical squeezing \cite{fuerst11sqz} and entanglement, coherent frequency conversion \cite{Strekalov09THzRF,rueda_efficient_2016}, and realization of quantum memories \cite{liu_ultra-small_2010} are all within reach to be implemented in the cavities. WGM resonators also can be useful for creation of efficient narrowband single photon sources \cite{Fortsch13NC} for quantum information protocols.

Lithographic fabrication of crystalline resonators of nearly any geometry and any material \cite{McCall92laser} has the benefit of mass-production, but still fails at achieving the surface qualities achieved with mechanical polishing. However with isotropic materials such as silica it was possible to achieve surface qualities en par with polishing and melting by chemical etching of wedge resonators \cite{Lee12etch}. The flexibility in the geometry opened the whole field of asymmetric resonant cavities, demonstrating a variety of interesting dynamic behaviors such as chaos and scar-type instabilities \cite{lacey01ol,rex_fresnel_2002,lacey03prl}. Asymmetric resonators have been used in high power quantum cascade lasers \cite{gmachl_high-power_1998} and sub-wavelength lasers \cite{song_near-ir_2009}. Theoretical descriptions of modal structures in deformed cavities cannot be done in the framework of Mie analysis, due to the non-integrability of the geometric system, leading to wave-chaotic formulations \cite{tureci_modes_2005}. The full interaction of such non-trivial resonance structures with non-uniform gain regions was only recently solved in the framework of a self-consistent lasing theory \cite{tureci_self-consistent_2006}.

Besides the conventional WGM resonators that are typically small, there exist larger monolithic crystalline resonators that are also based on total internal reflection (TIR) \cite{Schiller92monolithic}. They have been used for second harmonic generation \cite{Fiedler93monolit_cavity} and parametric down conversion \cite{Schiller93monoliticOPO}. Finally, low-order  WGMs can be observed around an irregularity in photonic bandgap crystals \cite{Ryu02pbg,Lee07pbg}.

\subsection{What is special about WGM resonators}\label{sec:benefits}

In summary, the WGM resonators are attractive because of their
\begin{itemize}
\item{\bf high $Q$-factor.} WGM resonators that are large with respect to the wavelength of light and have high enough surface quality suffer extremely low radiative losses. Depending on the resonator material they can also have low intrinsic losses.
\item {\bf wide spectral range.} WGM resonators provide high-$Q$ resonances throughout the entire transparency range of the dielectric they are made of.
\item {\bf low mode volume.} WGMs are localized close to the rim of the resonator and therefore have small volumes. A mode volume is conventionally introduced as the spatial integral over the field intensity normalized to the intensity maximum \cite{collot1993sphere}.
\item {\bf mechanical stability.} WGM resonators are monolithic and small, and therefore suffer only minimal mechanical instabilities.
\item {\bf tunable wavelength.} WGM spectrum can be engineered as well as dynamically tuned by a variety of techniques.
\item {\bf variable coupling.} Coupling to WGM resonators is usually achieved via frustrated TIR whereby the coupling rate depends on the distance between the resonator and the evanescent coupler. This allows for a simple control over the coupling rate.
\item {\bf strong nonlinear interaction} WGM resonators made out of nonlinear material allow for achieving strong nonlinear interaction at low light levels.
\end{itemize}

In the next section we formulate the mathematical description of the WGM spectrum and provide a strong footing of the special properties described above.

\section{Spectrum of WGM and ring resonators}\label{sec:spectrum}

\subsection{Mode structure and dispersion equation}\label{sec:dielectricResonators}

\begin{figure}[htb]
\vspace{-0ex}
\begin{center}
\includegraphics[clip,angle=0,height=4.6cm]{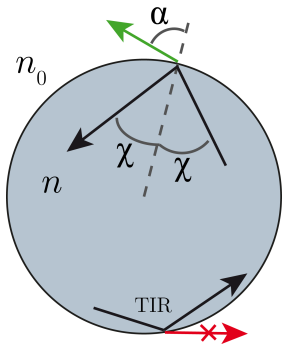}
\caption{Schematic of a dielectric whispering gallery mode resonator. In the ray-dynamical description, light incident at an angle $\chi$ can be either partially refracted out (if $\chi<\chi_{\mbox{\tiny crit}}$) or totally internally reflected (if $\chi>\chi_{\mbox{\tiny crit}}$). }
\label{fig:WGMschematic}
\end{center}
\end{figure}

Monolithic dielectric resonators have a refractive index larger than that of their surrounding $n>n_{\text{o}}$ (or, equivalently, the \textit{relative} index of refraction $\bar{n} \equiv n/n_{\text{o}}>1$) and confine the light by total internal reflection (TIR), see Fig.~\ref{fig:WGMschematic}. Here the complex angle of incidence and, therefore, the complex wave vector corresponds to the  evanescent field in the less optically dense medium. As TIR only depends on the angle of incidence and the refractive indices, it is a very broad-band process.

In a resonator with geometric path length $L$ and constant index of refraction $n$ the resonant frequency follows from the argument that an integer number of wavelengths $m$ needs to fit into the optical path length $nL$: $\lambda_m=nL/m$, or in terms of the wave number, $k_m=2\pi m/(nL)$. The optical path of large enough WGM resonators ($L\gg\lambda_m$) can be approximated by expression $L=2\pi R$, were $R$ is the resonator radius. This expression does not take the geometric dispersion of the WGM spectrum into account.

In order to understand the nature of this dispersion and to find a more accurate approximation for WGM spectrum  the Helmholtz equation needs to be solved for the boundary conditions set by the resonator geometry. The physical solutions are selected by requiring the continuity and the smoothness of the tangential derivatives of the electric field at the boundary. Furthermore, the so-called Sommerfeld boundary conditions should be selected at infinity, limiting the solutions to outgoing fields only. Under these conditions the solutions become complex, calling for some probing/external energy flow in order to excite and probe the resonances.

A WGM resonator belongs to a class of the so-called open resonators. Exact mathematical solution shows that modes of such a resonator include unbound spherical waves running away from the resonator. The modes are fundamentally unconfined, and defining volume of the mode is not straightforward. The problem is usually circumvented by neglecting the radiated wave part and considering the associated complex eigenvalue of the mode as radiative loss. The value of the loss is usually much smaller than the other types of attenuation. A calculation of the radiative Q-factor for a $100\ \mu$m water droplet results in $10^{73}$ at $\lambda = 600$~nm \cite{datsyuk01ufn}. As a rule, the radiation loss can be safely neglected in any resonator with the circumference exceeding a couple dozens of wavelength and a high enough refractive index contrast with the environment.

The most straightforward assumption is to consider a spherical resonator and thus introduce a spherical coordinate system \raisebox{1ex}{\footnotemark}.
\footnotetext{This is not the only possible choice of a coordinate system. Other coordinate systems, such as cylindrical, spheroidal, toroidal and ellipsoidal may also be used  \cite{Gorodetsky06geom,Breunig13eigenfunc}.} The transformation of the Laplacian into spherical coordinates introduces a centrifugal potential which provides a condition for bound states \cite{Oraevsky02WGM,little_analytic_1999}. The interaction of plane waves with such bound states was studied by Mie \cite{mie_beitrage_1908}. 
The analysis of six independent field components arising in this study can be simplified by realizing that only two orthogonal potentials are independent \cite{debye_lichtdruck_1909}. This yields two sets of solutions commonly known as the transverse (to the equatorial plane\raisebox{1ex}{\footnotemark}\footnotetext{TE and TM are sometimes defined in the opposite way, as transverse to the resonator \textit{surface}.}
) electric field (TE) and transverse magnetic filed (TM) polarization mode families. The main field component for each family is given by expression
\begin{eqnarray}
E_{mlq}(r,\theta,\varphi)&\sim & j_m(nk_0r)\times P_l^m(\cos\theta)e^{im\varphi}\nonumber\\
&=&j_m(nk_qr)\times Y_l^m(\theta,\varphi),\label{FieldSphere}
\end{eqnarray}
where $j_m$ is the spherical Bessel function of order $m$, $P_l^m$ are the associated Legendre polynomials, or respectively, $Y_l^m$ are the spherical harmonics \cite{Oraevsky02WGM}.

\begin{figure}[h]
\vspace{-0ex}
\begin{center}
\includegraphics[clip,angle=0,width=\linewidth]{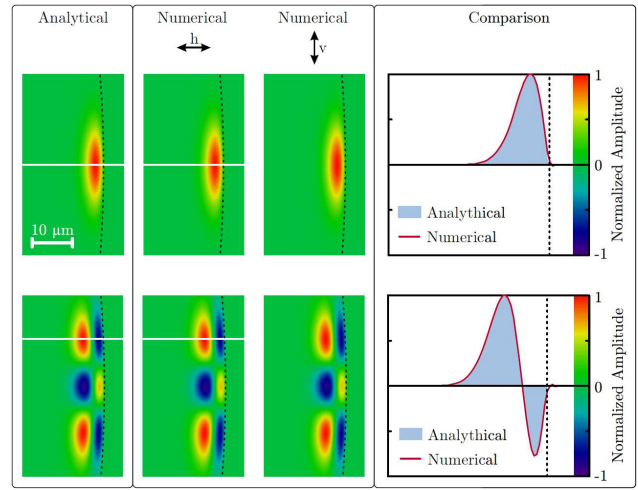}
\caption{Field distribution of whispering gallery modes found analytically and numerically. The resonator rotation axis is vertical, and the dashed line indicates the resonator boundary. The top row represents the fundamental equatorial WGM with $q=1,\,p=0$. In the bottom row, $q=2,\,p=2$. The right-hand panel shows the field cross section along the lines shown on the left-hand panel. Note again that the analytical solution does not take the evanescent field into account. The numerical however does.  Reprinted from \cite{Breunig13eigenfunc}.}
\label{fig:Transversalmoden}
\end{center}
\end{figure}

Here a nomenclature for the different integer separation constants is at order. The spherical Bessel function is an oscillating function of the radius $r$, whose zeros can be numbered by the integer $q=1,2,3,\ldots$, where $q=1$ designates the fundamental WGM. The angular momentum $m$ describes the number of wavelengths that fit around the equator. It can also be negative, pointing to a counter clockwise rotating mode. A positive integer $p=l-|m|$ corresponds to the number of nodes in the polar direction (see Fig.~\ref{fig:Transversalmoden}), indicating a transverse mode structure similar to that of a Fabry-Perot resonator.
Finding the eigenvalues and eigenfunctions for modes in large resonators, where the $m\approx l>10,000$ is not trivial but can be made feasible by suitable approximations to the high-index Bessel functions \cite{Lam92wgm_disp,schiller_asymptotic_1993} and Legendre Polynomials \cite{Breunig13eigenfunc}.

These approximations are useful for describing spherical resonators. However many WGM resonators are better described as spheroids with the major and minor semi-axes $a=R$ and $b$, respectively. Eigenfrequencies of such resonators can be found in the so-called semi-classical limit by the eikonal method \cite{Gorodetsky06geom,schwefel_polarization_2005}. In \cite{Gorodetsky06geom} we find the following dispersion equation:
\begin{eqnarray}
&2\pi R\frac{\nu n}{c}= nk_{lpq}R\simeq l-\alpha_q\left(l/2\right)^{1/3}\label{disp}\\
+&\frac{2p(R-b)+R}{2b}-\frac{\zeta n}{\sqrt{n^2-1}}+\frac{3\alpha_q^2}{20}\left(l/2\right)^{-1/3}\nonumber\\
-&\frac{\alpha_q}{12}\left(\frac{2p(R^3-b^3)+R^3}{b^3}+\frac{2n\zeta(2\zeta^2-3n^2)}{(n^2-1)^{3/2}}\right)\left(l/2\right)^{-2/3}\nonumber,
\end{eqnarray}
where $\alpha_q$ are the negative zeros of the Airy function, and $\zeta$ equals $1$ for TE and $n^{-2}$ for TM modes.
The leading term $(\sim l^{1})$ of this equation's right hand side clearly corresponds to the Fabry-Perot type resonance condition derived from the simple ray model. The second term $(\sim l^{1/3})$ is a correction taking into account change of the WGM diameter depending on its wavelength as well as $q$ number. It may be said that higher-$q$ modes effectively see a smaller resonator, hence the positive  frequency shift. In the third term $(\sim l^{0})$ a correction for different curvatures in polar and azimuthal direction is taken into account. The fourth term  $(\sim l^{0})$ arises in the eikonal method from the polarization-dependent Fresnel phases and implicitly accounts for the evanescent field of the resonator (hence the factor $\zeta$ distinguishing TE and TM modes). Note that this term explodes as the resonator material index of refraction $n$ approaches that of the surrounding media ($n_{\text{o}} =1$ is assumed in Eq.~(\ref{disp})) and WGMs become poorly confined.

\begin{figure}[t]
\includegraphics[width=8.5cm]{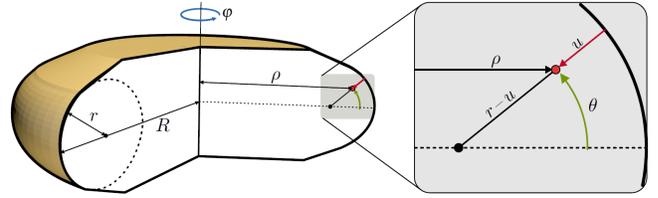}
\caption[]{\label{fig:coord}Surface shape of realistic WGM resonators may significantly differ from a sphere. In such cases introducing a local coordinate system such as shown here is more suitable. Reprinted from \cite{Breunig13eigenfunc}.}
\end{figure}

In large resonators the evanescent field can be neglected for the sake of simplicity. This is done by setting the \textit{metallic} boundary conditions at the rim of the resonator: $E(r=R)=0$. Then introducing a local coordinate system such as shown in Fig.~\ref{fig:coord} it is rather straightforward to find approximate expressions for the eigenfunctions \cite{Breunig13eigenfunc}:
\begin{eqnarray}\label{Eigenfunktion}
E &\sim  {\rm Ai}(u/u_m - \alpha_q)  \times {\rm e}^{ -\theta^2/2\theta_m^2}\;
H_p(\theta/\theta_m) \times  {\rm e}^{ {\rm i} m\varphi}, \nonumber\\
\theta_m&=\left(\frac{R}{\rho}\right)^{3/4}\frac{1}{\sqrt{m}},\quad\text{and}\quad u_m=\frac{R}{2^{1/3}m^{2/3}}.
\end{eqnarray}
Here $H_p$ is the Hermitian polynomial of the order $p$, e.g. $H_0 = 1$, $H_1 = 2\theta/\theta_m$, $H_2 = 4(\theta/\theta_m)^2 - 2$, and so on.  $\rm Ai$ is the Airy function with its negative zeros $\alpha_q$.

Note the relation between the local radius of curvature $r$ in Fig.~\ref{fig:coord} and the minor semi-axis of the approximating ellipsoid: $b=\sqrt{Rr}$. The results (\ref{Eigenfunktion}) agree well with numerical simulations that take the evanescent tail of the mode properly into account, see Fig.~\ref{fig:Transversalmoden}.

There are other ways to approximatly solve Laplase equation for large WGM resonators of an arbitrary shape. WGMs whose wavelengths are much smaller than the resonator size are localized in the vicinity of the equator of the resonator. Here cylindrical coordinates can be used for resonators of any shape, including spherical. The shape of the resonator can be presented in form $R=R_0+L(z)$, where $R_0 \gg L(z)$. The corresponding wave equation can be solved using separation of variables \cite{ferdous14pra}.

Dispersion relation (\ref{disp}) has a nontrivial dependence on the mode numbers $(l,p,q)$ and resonator parameters. This leads to complicated spectra that are difficult to interpret for larger resonators. However, identifying a WGM numbers, especially $q$, is critical for finding the phase matching conditions of various nonlinear-optical processes, as shown in section \ref{sec:Natural}. Various techniques of optical WGM identification have been developed, including the free spectral range (FSR) measurement by sideband spectroscopy \cite{Li12sb,Savchenkov12overmoded}, far-field emission pattern analysis \cite{Gorodetsky94cpl,Dong08patterns,Sedlmeir13angle}, or a combination of these techniques \cite{Schunk14modes}. The detailed spectrum of a WGM resonator can be measured using a reference optical frequency comb \cite{Herr14soliton}.

Up to now we only considered WGM resonators that had isotropic material properties. These properties are inherent to amorphous WGM resonators such as droplets \cite{benner_observation_1980,Snow85Raman,Qian1986dropletRaman,Lin1994droplets,Qian1995droplets,Sennaroglu07R_las,Kiraz09R_las,
Acker89,leach90ol,hill93josab,leach93josab,kasparian97prl}, molten silica spheres \cite{Braginsky89nonlinWGM}, toroids \cite{Armani03chip}, or wedge shaped resonators \cite{Lee12etch}. Moreover, all crystals with cubic symmetry like\ CaF$_2$ are isotropic. Most of the crystals fall, however, in one of the four other symmetry classes, where the refractive index varies depending on polarization and propagation direction of the light.
Among crystals of these four symmetry classes, the group of uniaxial crystals shows only one direction of light propagation, i.e.\ one {\em optic axis}, where the refractive index is independent of the polarization. The propagation of light in the plane perpendicular to the optic axis is then governed by the ordinary index of refraction $n_o$ for perpendicular polarization, while the extraordinary index of refraction $n_e$ is valid for the polarization parallel to the optic axis.

WGM resonators made out of uniaxial materials are very important in our following discussion. Usually they are made such that the optic axis is parallel to the rotational symmetry axis, i.e.\ in the so-called $z$-cut geometry. In this special case the TM (TE) polarized mode is mainly influenced by the ordinary (extraordinary) refractive index. Therefore the two mode families can be tuned independently, as the thermo-refractive, electro-optical, etc.\ effects all scale with the respective refractive indices, see sections \ref{sec:ThermalTuning} and \ref{sec:ElectricalTuning}.

A less common is the $x$-cut geometry, when the optic axis lies within the equatorial plane of the WGM resonator,  i.e. is perpendicular to the axis of rotation. In this case the TE polarized mode is governed by the constant ordinary refractive index, however the TM polarized mode experiences the refractive index oscillating \textit{approximately} harmonically between the ordinary and extraordinary values. We discuss applications of such resonators in section~\ref{sec:cycPM}.

The general case when the optic axis makes an arbitrary angle with the symmetry axis is highly non-trivial. It is not clear if there still exist two mode families in such resonators. The refractive index experienced by the light varies along the path of the WGM, and thus its polarization becomes position dependent. Position-dependent walk-off further complicates the analysis. In spite of several theoretical and experimental investigations of WGM properties in the resonators with arbitrary orientation of the optical axis \cite{Prokopenko04anisotr,Ornigotti11anisotropic,Lin12_BBO,Sedlmeir13spie,Sedlmeir13angle,Ornigotti14wgm}, their behavior is not yet fully modeled and understood.

\subsection{Geometrical dispersion and effective index approximation}\label{sec:neff}

Comparing the WGM dispersion equation (\ref{disp}) to the dispersion equation describing an ideal one-dimensional resonator of the optical length $L=2\pi Rn$:
\begin{equation}
2\pi R\frac{\nu n}{c}=m,
\label{neff}
\end{equation}
we see that Eq.~(\ref{disp}) includes higher-order terms that can be attributed to the geometrical, or waveguide, dispersion. This dispersion arises from the boundary conditions and the resonator curvature.
It is often convenient to treat large WGM resonators as one-dimensional resonators described by dispersion equation (\ref{neff}) with the {\em effective} index of refraction $n\rightarrow \tilde{n}$ which incorporates all the geometrical and material dispersion effects \cite{hocker1977mode,snyder1983optical}. Note that $\tilde{n}$ depends on a variety of parameters, including the wavelength, the resonator semi-axes $a=R$ and $b$, and the WGM family specified by the polarization and mode numbers $q$ and $p$. To find $\tilde{n}$, we first numerically solve (\ref{disp}) for $l$, then substitute $m=l-p$ into the right-hand side of (\ref{neff}), and solve it for $\tilde{n}$. In doing so we may give up the requirement for $l$ and $m$ to be integer, which means that we treat the resonator spectrum as continuous.

The effective index approximation is convenient for finding the phase matching conditions for nonlinear frequency conversion, e.g.\ spontaneous parametric down conversion (SPDC) $\nu_p\rightarrow\nu_s+\nu_i$, supported in resonators with nonzero $\chi^{(2)}$ nonlinearity. In this case the phase matching condition takes a simple form
\begin{equation}
\nu_p\tilde{n}_p=\nu_s\tilde{n}_s+\nu_i\tilde{n}_i,
\label{neffpm}
\end{equation}
which allows us to find the suitable temperature and wavelengths numerically. The benefits and limitations of this approach are further discussed in section \ref{sec:exoticPM}.

This approach has been successfully used for evaluating the phase matching in WGM resonators for second harmonic generation (SHG) \cite{Xiong11GaN_SHG}, SPDC \cite{Beckmann11,Werner12BlueOPO}, and direct third harmonic generation from a fused silica microsphere \cite{Carmon07tripling}. In \cite{Strekalov15jmo}\raisebox{1ex}{\footnotemark}\footnotetext{In the left-hand sides of Eqs. (1) and (2) in \cite{Strekalov15jmo}, the factor $2\pi$ should be in numerator rather than in denominator.} it was used to infer the parameters for the double phase matching, such that the signal mode excited in the SPDC process can generate its own second harmonic.

Geometrical dispersion contribution $\Delta n\equiv \tilde{n}-n$ is always negative in a spheroidal resonator. This is evident from the mode structure shown in Fig.~\ref{fig:Transversalmoden}. Here the optical field is concentrated inside the resonator, some distance away from its rim. The resonator optical path is shorter than $2\pi R n$ because of that. As a function of wavelength, the geometrical contribution to dispersion is \textit{normal}: $d\Delta n(\lambda)/d\lambda<0$. In part, this is because at longer wavelengths more of the optical power is carried by the evanescent field outside the resonator, and consequently the effective index of refraction is lower. However, the major reason is that the power is localized further from the resonator surface at longer wavelength, so the effective radius of the mode is smaller there.

In small resonators geometrical dispersion can be significant. This opens up an interesting opportunity for the WGM dispersion engineering by varying parameter $b$, or more profoundly, by machining the resonator rim into various non-spheroidal shapes. Such shapes usually do not allow for analytic dispersion equations such as  (\ref{disp}), and finite-element numerical techniques are required to find their spectra, see e.g. \cite{Grudinin12FEM,kaplan_finite_2013}.

The effect of the resonator rim shape on the Kerr optical comb properties has been experimentally studied \cite{Del'Haye11octave,Grudinin12comb,Grudinin13comb,Grudinin14comb}. In particular, the ``belt" resonators of rectangular crossection have been shown to compensate the chromatic dispersion of the the group velocity in magnesium fluoride, resulting in generation of extended-range optical frequency combs \cite{Grudinin2015}. The group velocity dispersion can also be optimized by engineering the waveguide cross section dimensions \cite{Moss13cmos}, adding a proper material cladding \cite{Riemensberger12coated}, or using slotted waveguide structures \cite{zhang10oe,zhang13ol,bao15josab}.

\subsection{Coupling, loss and quality factors}

Coupling WGM resonators to external light is achieved by frustrated TIR. This coupling can be realized via optical waveguides, prisms, gratings, other resonators, etc. Coupled in this way resonators have been used to build efficient narrow-band add/drop filters, including those enabling narrow linewidth light sources \cite{Sprenger09filter,Sprenger10las_stab,Collodo14lasing}, division multiplexers, and other optical devices \cite{Little97microring,Monifi13filter}.

The most efficient coupling to date was realized with tapered fiber
couplers \cite{knight97ol,chin98jlt,little_analytic_1999}. Fiber taper is a single mode bare waveguide with diameter optimized for phase matching with selected WGM. The coupling efficiency can approach 99.96\% with an adiabatically tapered fiber \cite{Cai2000cpl}.

Prism coupling to WGMRs was investigated both theoretically and experimentally
\cite{schiller91ol,rowland93jo,Gorodetsky94cpl,gorodetsky99coupling}, reaching some $80 $\% efficiency with microspheres \cite{gorodetsky99coupling}. The efficiency can be optimized by adjusting the shape of the optical beam as well as resonator morphology. Coupling efficiency exceeding $97 $\% was achieved in elliptical LiNbO$_3$ resonators \cite{mohageg07ellipt}, and nearly perfect coupling (more than 99\% criticality, \cite{Strekalov09THzOL}) was achieved in a LiNbO$_3$ WGM resonator with optimized shape of the rim.

Planar integration of WGMRs with waveguides is deemed to be the most practical approach
\cite{blom97apl,rafizadeh97ol,little99ptl,rabiei02jlt,tishinin99ptl,poulsen03oe,choi04ptl,tee06jstqe,le06ptl}.
For example, strip-line pedestal antiresonant reflecting waveguides have been utilized for robust coupling to microsphere resonators as well as to photonic circuits \cite{little00ol,laine00ptl,white06apl,white07oe}.  Integration of a lithium niobate resonator with a planar waveguide made with proton exchange in X-cut lithium niobate substrate was realized \cite{conti11oe}. Strong coupling between a resonator and a waveguide integrated on the same chip but lying in different planes was realized as well \cite{ghulinyan13prl}.

\begin{figure}[ht!]
\centering\includegraphics[height=6cm]{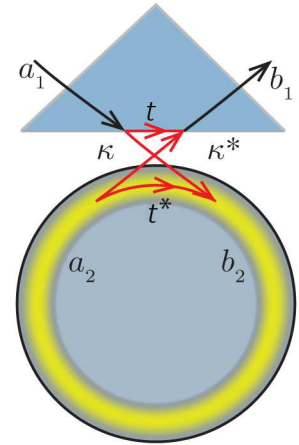}
\caption{Schematic of the WGM resonator coupling. The incoming field $a_1$ is related to the reflected field $b_1$ and the internal cavity field $b_2$. The coupling coefficients are $r$ and $\kappa$. }
\label{fig:couplingSchematic}
\end{figure}

Theoretically coupling of WGM resonators can be described via a simple transfer matrix scheme \cite{yariv_universal_2000} illustrated in Fig.~\ref{fig:couplingSchematic}. In this approach the normalized single mode outside fields $a_1, b_1$ are related to the internal fields $a_2,b_2$ by the complex coupling parameters $\kappa$ and $t=|t|\exp(-i\phi)$:
\begin{equation}
\begin{pmatrix}
b_1\\b_2
\end{pmatrix}
=
\begin{pmatrix}
t & \kappa\\
-\kappa^* & t^*
\end{pmatrix}
\begin{pmatrix}
a_1\\a_2
\end{pmatrix},
\end{equation}
where $|\kappa|^2+|t|^2=1$.

The internal field amplitudes are additionally related by $a_2=\alpha \exp(i\theta)b_2$, where $\alpha\ll 1$ is the loss per round trip and $\theta$ the phase change due to the round trip. The system is said to be in resonance and a stationary mode can form if the round-trip phase increment
is an integer multiple of $2\pi$.  The system is {\em critically coupled} and no field exits the cavity ($|b_1|^2=0$) if the internal losses equal the coupling losses: $\alpha=|\kappa|$.
At the critical coupling, the full width at half maximum of WGM resonance is given as
\begin{equation}
\delta\omega=\frac{2|\kappa|^2c}{nL},
\end{equation}
where $L=2\pi R$ is the length of the resonator circumference.

The \textit{finesse} of a critically coupled resonator is the loss per round trip: ${\cal F}=\pi/(1-|t|^2)=\pi/|\kappa|^2$. It is also equal the ratio of the FSR $\Omega =2\pi c/nL=c/nR$ to the linewidth $\delta\omega$ (full width at the half maximum) of the mode. The quality factor is related to the linewidth and finesse as
\begin{equation}
Q=\frac{\omega}{\delta\omega}
={\cal F}\cdot  m.
\end{equation}

There are many various loss channels in a WGM resonator that account for the loss factor $\alpha$:
\begin{equation}
\alpha=\alpha_{\text{ss}}+\alpha_{\text{material}}+\alpha_{\text{rad}}.\label{alphas}
\end{equation}
The terms of (\ref{alphas}) represent the surface scattering ($\alpha_{\text{ss}}$), material absorption ($\alpha_{\text{material}}$), and radiative loss ($\alpha_{\text{rad}}$).
They can all be lumped into the resonator-limited (i.e. independent of coupling) quality factor $Q_0$:
\begin{equation}
1/Q_0=1/Q_{\text{ss}}+1/Q_{\text{material}}+1/Q_{\text{rad}}.
\end{equation}

Strictly speaking, a curved boundary does not permit TIR \cite{Oraevsky02WGM}, so, as we already mentioned, a WGM resonator always radiates into space. The radiative loss is usually negligible and becomes observable only if the radius of curvature of the boundary becomes comparable to the wavelength of the light confined within the resonator \cite{garrett_stimulated_1961,nockel_resonances_1997,braginsky_optical_1993,Oraevsky02WGM,kippenberg_nonlinear_2004}.
The corresponding quality factor of a sphere is given by \cite{garrett_stimulated_1961}:
\begin{eqnarray}
Q_{\text{rad}}&\approx& \frac{2\pi R}{\lambda}\frac{\zeta \,e^{2T}}{\sqrt{\bar{n}^2-1}}, \label{Qrad}\\
T&=& \frac{2\pi R}{\lambda}\left(\cosh^{-1}\bar{n}-\sqrt{\bar{n}-\bar{n}^{-1}}\right)\nonumber
\end{eqnarray}
where $\bar{n}$ is the \textit{relative} index of refraction ($\bar{n}=n/n_o$ in terms of Fig.~\ref{fig:WGMschematic}) and $\zeta$ equals 1 for TE and $n^{-2}$ for TM modes defined according to our convention.
This loss channel is irrelevant for crystalline WGM resonators which are usually much larger than the wavelength.

Surface roughness, on the contrary, can be an important loss channel. The corresponding quality factor scales with the root mean square (rms) size of the surface inhomogeneity $s$, and the correlation length of the roughness at the resonator surface $B$ as \cite{gorodetsky2000}
\begin{equation}
Q_{\text{ss}}\approx \frac{3\lambda^3R}{8\pi^2\bar{n}B^2s^2}\label{QualitySS},
\end{equation}
The fact that this term depends on $\lambda^3$ can be used to test if the quality factor is mainly limited by the surface quality simply by testing it at different wavelengths.

Material absorption is usually characterized by the loss coefficient $\alpha_{\text{material}}$ describing power attenuation per unit length, which leads to the following expression:
\begin{equation}
Q_{\text{material}}=\frac{2\pi n}{\lambda \alpha_{\text{material}}}.
\end{equation}
The transparency of an ideal dielectric in the optical domain can be defined by the tails of multi-phonon absorption on the long wavelength side, and by  electronic transitions absorption (the so called Urbach tail) on the short wavelength side \cite{grudinin06highQ,grudinin_fundamental_2007}
\begin{equation}
\alpha_{\text{material}}(\lambda_0)=\alpha_{\text{uv}}e^{\lambda_{\text{uv}}/\lambda_0 } + \alpha_{\text{ir}}e^{-\lambda_{\text{ir}}/\lambda_0}.
\end{equation}
Here the coefficients $\alpha_{\text{uv}},\alpha_{\text{ir}}$ as well as $\lambda_{\text{uv}}, \lambda_{\text{ir}}$ are experimentally found values. The importance of the multiphonon absorption in the mid-IR was realized recently \cite{grudinin16ol,lecaplain16arxiv}. 

\section{Tuning the resonator spectrum}\label{sec:tuning}

In section \ref{sec:neff} we noticed that WGM spectra can be engineered by modifying the resonator rim profile. We also mentioned that the engineering and reversible dynamical tuning of the WGM spectra are important for achieving the phase matching in various nonlinear conversion processes. In this section we review some other available techniques to achieve both permanent and dynamical WGM frequency tuning.

\subsection{Mechanical stress and deformation}

Mechanical deformation of resonators can change their spectral properties rapidly and, with a due caution, reversibly. The leading-order effect here is the physical change of the resonator length, although  the pressure dependence of the refraction index should also be taken into account. Higher-order effects may include stress-induced birefringence, heating, and piezoelectric effect in some materials.

In an early experiment by Ilchenko \textit{et al.} \cite{Ilchenko98strain}, a fused silica microsphere ($R=80\,\mu$m) was squeezed between two copper pads actuated by a piezo. The spectrum was shifted by 150 GHz (approximately 3/8 of the FSR) while maintaining the quality factor over $10^8$. A small variation of the FSR was also observed.
The same technique was later applied \cite{grudinin06highQ} to a crystalline magnesium fluoride resonator with $R=250\,\mu$m. In this case multiple-FSR tuning was easily achieved. Note that a continuous tuning of the WGM spectrum over one FSR allows to couple the resonator to any optical wavelength. Such a resonator is called \textit{fully tunable}. Pressure-tuning study of highly non-equatorial WGMs in polystyrene
microspheres  is reported in \cite{Wagner13pressure}.

Bubble resonators offer a rather unique way of dynamic tuning: they can be not only stretched or compressed  \cite{Sumetsky10bubble}, but also inflated  \cite{Han14bubble}. All these techniques allow for achieving the fully tunable operation.

\subsection{Thermal}\label{sec:ThermalTuning}

Tuning a resonator frequency by changing its temperature is perhaps the most common tuning technique. It broadly applies not only to WGM, as in e.g. \cite{tapalian02ptl,rabiei03ptl}, but also to practically all other kinds of resonators. This type of tuning is based on the combination of two major effects: the thermal expansion and the temperature dependence of the resonator index of refraction. While the thermal expansion usually leads to a negative frequency shift with increasing temperature (as the resonator becomes larger), the thermorefractive contribution can be either positive or negative depending on the resonator material and temperature. Remarkably, the ordinary and extraordinary thermorefractive coefficients in birefringent materials can significantly differ. This allows for the \textit{differential} tuning of the TE and TM WGM spectra, a capability very important for some sensor applications \cite{Strekalov11T,Baumgartel12T}\raisebox{1ex}{\footnotemark}\footnotetext{In the left-hand part of Eq.~(1) in \cite{Strekalov11T}, $2\pi$ should be in the numerator rather than in denominator.} and particularly for  achieving the phase matching in various nonlinear optics applications \cite{Sturman12nl}. Practically all nonlinear WGM optics studies that we cite below relied on the temperature tuning.

High-$Q$ WGM resonator spectra can be very sensitive to variations of the temperature. It is easy to see that the  frequency tuning rate given in the units of the WGM linewidth $\delta\omega$ is related to the material thermal expansion and thermorefraction coefficients $\mu_L$ and $\mu_n$ as
\begin{equation}
\frac{1}{\delta\omega}\frac{d\omega}{dT}=-Q(\mu_L+\mu_n).\label{thermalshift}
\end{equation}
For the typical parameters of $Q=10^8$ and $\mu_L+\mu_n=10^{-5}\,{\rm K}^{-1}$, a milidegree temperature change can easily shift the spectrum by a full linewidth. This makes the consideration of the optical power dissipation inside the resonator relevant even for very weak light. For a critically coupled resonator all in-coupled optical power is dissipated and eventually converted to heat. Therefore a rapid thermal tuning can be achieved by controlling the input optical power. This control parameter has been put to use in a temperature stabilization application \cite{Strekalov11T,Baumgartel12T,Weng14nK}, allowing to reach a few nK temperature stability at above room temperature set point. In very high-$Q$ resonators, the spectral response to the injected optical power via the mode volume heating may lead to thermal noise and instability which we will discuss in more detail in section \ref{sec:ThermalNonlin}.

\subsection{Electro-optical}\label{sec:ElectricalTuning}

Because of high temperature sensitivity of the WGM spectra, the temperature tuning range can be very large, reaching a few to several tens of nanometers. However, this method is not very convenient because it is slow. Much faster tuning can be achieved leveraging the electro-optical effect. The
electro-optical tuning is only possible in materials where such effect is present. All WGM modulators described in section \ref{sec:EOM} below were electro-optically tuned. Like the thermal, the electro-optical tuning leads to the differential TE-TM frequency shift, due to different values of the electro-optical tensor components. Differential tuning among different $q$-families of the same polarization can be enabled by creating circuar patterns of inverted domains near the resonator rim \cite{mohageg05ring}, which may be conveniently achieved in lithium niobate using the \textit{calligraphic}
poling technique based on ``drawing" inverted domains with a sharp electrode \cite{mohageg05poling,Meisenheimer15poled}.

Compared to the thermal, the electro-optical tuning has the advantage of higher speed. On the down side, it is difficult to control the quasi-static electric field in the region of the optical  mode, because the electrodes cannot be applied to this region without compromising the optical quality factor. As a result, both the magnitude and the direction of the tuning field is poorly controlled in the very region where it matters, and the tuning rates observed in experiment are rarely consistent with the theoretical estimates.

Ferroelectric materials, such as lithium niobate, have additional problems associated with electro-optical tuning. Visible light induces photocurrents in such materials \cite{Gerson86photoconduct,Peithmann99photorefr}, which means mobilizing the charges that may screen the optical field area from the control electric field. As a result, the WGM spectrum follows a rapid bias field variation for a short period of time but then returns to its original state, with the time constant ranging from sub-seconds to hundreds of seconds, depending on the injected optical power and wavelength \cite{Schunk16CsRb}.

External electric fields can affect the WGM spectrum not only via the electro-optical, but also via the  electrostriction effect. This effect does not require nonlinear response and is present in all materials. It has been observed with hollow solid polydimethylsiloxane microspheres \cite{iopollo09oe}. Let us point out that both electrosctriction and electro-optical effects can be used not only for configuring the WGM spectra, but also for sensitive measurements of quasi-static electric fields by monitoring the spectral changes \cite{Savchenkov14Esens}.

\subsection{Other methods of WGM spectrum engineering}

Dispersion engineering can also be realized by coating a resonator with another dielectric that has a different index of refraction \cite{ilchenko03dispersion,Riemensberger12coated}. Different WGMs penetrate into the coating layer differently, which gives rise to the differential TE-TM frequency tuning, as well as differential tuning of different $q$-familes of the modes that have the same polarization. Making the coating optically active opens up the possibilities for quick spectral tuning by means of an optical control. For example, coating a microsphere with three bacteriorhodopsin protein monolayers allowed for a reversible high-speed control of a target WGM frequency, thereby achieving all-optical switching of a 1310 nm with 532 and 405 nm control beams in 50 $\mu$s \cite{Roy10switch}. Optical tuning can be also implemented in complex WGM structures comprising a solid-core
microstructured optical fiber with magnetic fluids \cite{Lin15magn}.

A dynamically controlled, polarization discriminating tuning can be achieved with a dielectric probe moving in the resonator's evanescent field \cite{Teraoka06diel_tuning}. In some configurations they may impart the anomalous blue shift to one polarization family, while the other experiences the normal red shift \cite{Foreman16tuning}. This technique has been used for fine frequency-tuning of WGM-based SPDC source \cite{Schunk15CsRb}.

Closely related to dielectric coating is immersing a resonator in a lower-index liquid \cite{Sedlmeir14swim}. This technique allows for index engineering, and also forms a basis for various sensing techniques, including those used for detection of organic molecules and viruses \cite{Vollmer08biosensing,Vollmer12review}.

Permanent or quasi-permanent spectral changes can be induced in WGM resonators by various photochemical processes in the host material that affect the index of refraction. For example, spectra of germanium-doped silica micropheres have been tailored by a UV-exposure for an optical filtering application \cite{Savchenkov03Ge}. The advantage of this approach is that the spectral changes can be monitored in real time, and the process can be stopped when the desired frequencies are reached.

Photorefractivity in lithium niobate and tantalate offers another method for controlled modification of a WGM spectrum \cite{savchenkov06photorefraction,savchenkov06MgOLiNbO3,savchenkov07damage}. Unlike with the UV curing of Ge-doped silica, here one can selectively pump a chosen mode at a relatively high optical power (even in the infrared wavelength range \cite{savchenkov06photorefraction}) shifting its frequency with respect to the  other modes. The process is surprisingly mode-selective, considering a significant overlap between the low-order WGMs. Once the first mode is sufficently tuned, the process can be repeated with a different mode, and so on. One can in principle arrange a number of WGMs into a spectral pattern that could be used e.g. for ``fingerprinting" of complex atomic or molecular spectra. The underlying photorefractive patterns ``engraved" in the resonator material are only quasi-permanent and can be erased by e.g. UV exposure \cite{savchenkov06MgOLiNbO3}.

\subsection{Mode crossing}

Large WGM resonators have very dense spectra. Occasionally, modes from different families become nearly degenerate \cite{carmon_static_2008,savchenkov07ornaments}. Ideally, these modes would not interact. However even a minute scattering can facilitate their energy exchange comparable with the intrinsic loss rate. In such a case the two modes can interact, which leads to the so-called avoided-crossings. As we have seen earlier in this section, modes from different families respond differently to external parameters such as temperature, pressure, electric field, or presence of scatterers. This opens one path to reaching a degeneracy, which is observed in e.g. photorefraction experiments \cite{savchenkov06photorefraction,savchenkov06MgOLiNbO3,savchenkov07damage}.

Usually in such experiments one of the crossing modes is coupled to a probing light source more efficiently than the other, which may be due to the different polarizations or $q$'s. Therefore these modes can be named a {\em bright} mode  (designated photon annihilation operator $\hat{b}$) and a {\em dark} mode (designated photon annihilation operator $\hat{d}$), respectively, with their unperturbed resonance frequencies $\omega_b$ ($\omega_d$).

There is a number of mode-coupling mechanisms, such as e.g. scattering by sub-wavelength features \cite{gorodetsky2000,Mazzei07,Zhu10onchip,wiersig_enhancing_2014}, that may be realized as impurities or refractive index inhomogeneities. Scattering may also occur between modes with different polarizations. This type of scattering has been studied in silicon nitride micro resonators \cite{Liu14Kerr,ramelow_strong_2014}. It also was observed \cite{Strekalov11T} and studied in magnesium fluoride \cite{weng_mode-interactions_2015}.

Regardless of how the mode coupling is physically realized, a straightforward approach to describing it is given by the following Hamiltonian:
\begin{equation}
\hat{H}=\hbar\omega_b\hat{b}^\dagger\hat{b}+\hbar\omega_b\hat{d}^\dagger\hat{d}+
\hbar\kappa\left(\hat{b}^\dagger\hat{d}+\hat{b}\hat{d}^\dagger\right),
\end{equation}
where the modes interaction strength is given by $\kappa$. This Hamiltonian has two eigenfrequencies:
\begin{equation}
\omega_{1,2}=\frac{\omega_b+\omega_d}{2}\pm\frac{1}{2}\sqrt{(\omega_b-\omega_d)^2+4\kappa^2}.
\end{equation}

\begin{figure}[b]
\vspace{-0ex}
\begin{center}
\includegraphics[clip,angle=0,height=7.5cm]{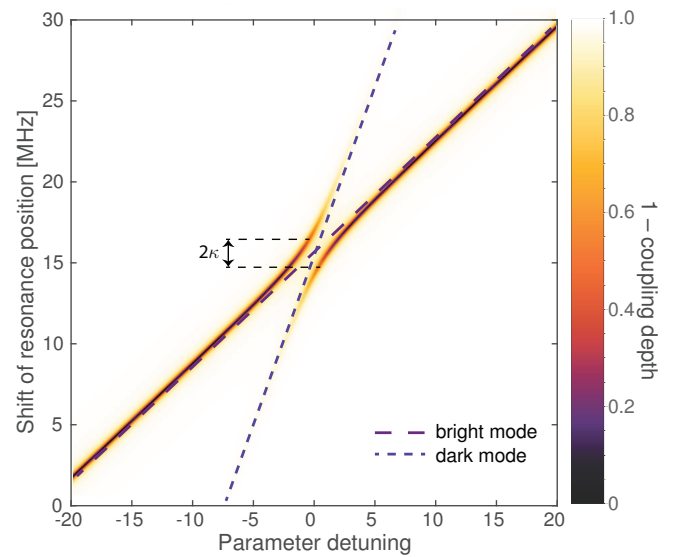}
\caption{Theoretical description of interaction between a {\em bright} and {\em dark} modes. The bright mode is critically coupled, and the external parameter controlling the modes' relative dispersion is tuned. Near the degeneracy the coupling rate $\kappa$ detunes the resonance frequencies, and a characteristic \textit{avoided-crossing} pattern is seen. Asymptotes correspond to the unperturbed modes. }
\label{fig:avoided_crossing}
\end{center}
\end{figure}

An example of avoided-crossing is shown in Fig.~\ref{fig:avoided_crossing}. Here the asymptotes  correspond to the standard change of the unperturbed resonances with the external parameter, which may be the temperature, exposure time in a photorefractive experiment, bias voltage, and so on. At the degeneracy point, the modes are separated by twice the coupling constant $2\kappa$.

WGM spectrum tuning by avoided mode crossing does not usually offer the versatile control or large tuning range available from other techniques we discussed earlier. However it may be efficient in the situations when the quasi-equidistant character of the WGM spectrum needs to be altered, e.g. for the design of single sideband (SSB) electro-optical modulators \cite{rueda_efficient_2016}. Furthermore, in non-Hermitian systems where there is a loss for each mode, the avoided crossing near the degeneracy allows for realizing the so-called {\em parity time symmetric} ($\cal PT-$Symmetric) systems \cite{peng_loss-induced_2014,chong_pt-symmetry_2011,jing_pt-symmetric_2014}, and even more generally, to study the \textit{exceptional points} \cite{Lee2009,liertzer_pump-induced_2012,wiersig_enhancing_2014,wiersig_sensors_2016}. This research leads to such counter intuitive observations as the onset of lasing by inducing loss \cite{peng_loss-induced_2014}.

\section{Second-order nonlinear processes}\label{sec:chi2}

Second-order nonlinear optical processes arise from the quadratic response of the media polarization $\vec{P}$ to the  external electric field $\vec{E}$ \cite{BoydBook}:
\begin{equation}
P^{(2)}_i=\sum_{j,k}\chi^{(2)}_{ijk}E_jE_k\label{P2},
\end{equation}
where $\chi^{(2)}$ is the nonlinear susceptibility tensor, and $i,\,j,\,k$ index the crystallographic axes $x,\,y,\,z$. %Generally, a sum or integral over the optical fields frequencies $\omega_1$ and $\omega_2$ is required for a complete description, but in the context of this review all optical fields are typically assumed quasi-monochromatic.

Second-order polarization (\ref{P2}) gives rise to a nonlinear term of the optical energy
\begin{equation}
H^{(2)}%=\int\,dV\vec{E}\cdot\vec{P}^{(2)}
=\frac{1}{3}\int\,dV\sum_{i,j,k}\chi^{(2)}_{ijk}E_iE_jE_k. \label{H2}
\end{equation}
Here the factor $1/3$ is necessary to account for the index permutations that preserve the frequency sum relation. The $\chi^{(2)}$ tensor elements involved in such permutations are postulated to be equal by the Kleinman symmetry convention \cite{BoydBook}. In many cases of interest polarization of each field is fixed with respect to the $\chi^{(2)}$ tensor axes, and then the sum in (\ref{H2}) can be omitted along with the factor $1/3$.

Following the canonical quantization of scalar single-mode electromagnetic field \cite{LoudonBook} inside a dielectric \cite{Crenshaw02quantization1,Crenshaw02quantization}, we write
\begin{equation}
\hat{E}(\vec{r},t)=in^{-1}\sqrt{2\pi\hbar\omega}\left(\hat{a}\psi(\vec{r})e^{-i\omega t}-\hat{a}^\dagger\psi^*(\vec{r})e^{i\omega t}\right), \label{quantization}
\end{equation}
where $n$ is the index of refraction for given optical frequency $\omega$ and polarization, $\hat{a}^\dagger$ and $\hat{a}$ are creation and annihilation operators for a photon in the given mode, $[\hat{a}^\dagger,\hat{a}]=1$, and $\psi(\vec{r})$ is the mode's eigenfunction normalized to unity: 
$\int\,dV|\psi(\vec{r})|^2=1$. Note that in the strict sense such normalization would be impossible because the integral diverges as $r\rightarrow\infty$. Indeed, the radial part of $\psi(\vec{r})$ for $r>R$ is given by Hankel functions that asymptotically correspond to diverging spherical waves representing radiative loss, as discussed in section \ref{sec:dielectricResonators} and in \cite{Foreman14sens}. The unity normalization therefore can be introduced only approximately, enforcing the metallic boundary condition (see section \ref{sec:dielectricResonators}) and limiting the integral to $r<R$.  

Substituting (\ref{quantization}) into (\ref{H2}) we derive a quantum-mechanical interaction Hamiltonian which governs all three-wave mixing processes in resonators:
\begin{equation}
\hat{H}_{int}=\hbar g(\hat{a}_1\hat{a}_2\hat{a}_3^\dagger+\hat{a}_1^\dagger\hat{a}_2^\dagger\hat{a}_3), \label{Hint}
\end{equation}
where
\begin{equation}
g= \tilde{\chi}^{(2)}\sigma_{123}\frac{\sqrt{(2\pi)^3\hbar\omega_1\omega_2\omega_3}}{n_1n_2n_3} \label{Omega}
\end{equation}
is the nonlinear coupling rate for interacting photons,
 $\omega_3=\omega_1+\omega_2$, the effective nonlinear susceptibility $\tilde{\chi}^{(2)}$ is determined by the fields polarizations, and
\begin{equation}
\sigma_{123}=\int dV\psi_1(\vec{r})\psi_2(\vec{r})\psi_3^*(\vec{r}) \label{ovlp}
\end{equation}
is the WGM overlap integral. In (\ref{Omega}) we used cgs units, with the following conversion of the standard second-order nonlinearity:
\begin{equation}
\chi^{(2)}{\rm [cgs\;units]}=\frac{3\times 10^{-8}}{4\pi}\,d{\rm [pm/V]}. \label{conv}
\end{equation}

Besides the overlap integrals (\ref{ovlp}), a different notation is often used, see e.g. \cite{Matsko02OPO,Ilchenko03parametric,Ilchenko04SH,Savchenkov07THz,Sturman11,Fuerst15SHG}, apparently ascending to the tradition of field quantization in the plane waves. In this notation the eigenfuctions $\psi(\vec{r})$ are normalized to the quantization volume (mode volume) $V$ rather than to unity. The electric field expression (\ref{quantization}) acquires in this case an extra factor $V^{-1/2}$ and the mode overlap (\ref{ovlp}) is measured in the units of volume, e.g. cm$^3$, instead of cm$^{-3/2}$. Just like the unity normalization, the volume normalization can only be done approximately by assuming the metallic boundary condition.

The disadvantage of normalizing WGM eigenfunctions to the mode volume is that, except for the case of plane waves quantization in a box, this volume lacks a rigorous first-principles definition. Usually it is defined as the volume integral of the optical intensity distribution normalized to the maximum intensity value \cite{collot1993sphere,VahalaBook}. With this definition equivalence of the volume- and unity-normalized approaches can be easily proven, however it may not be a good definition for the eigenfunctions that have multiple nearly-equal maxima. 
On the other hand, the advantage of the volume-normalized approach is the ability to directly compare the overlap integral (\ref{ovlp}) to the mode volume and therefore to easily quantify the coupled modes overlap in space as the ratio of these volumes. Other possible approaches to  eigenfunctions normalization and defining the mode volume are discussed Kristensen \textit{et al.} \cite{Kristensen15modVol}, who also prove them to be equivalent.

Interaction Hamiltonian (\ref{Hint}) leads to a set of ordinary differential equations (ODEs) describing the nonlinear processes and the input-output relations in a resonator. For any mode labeled $j$ we have
\begin{equation} \label{setz}
\dot{\hat a}_j=-(\gamma_j+i\omega_j)\hat a_j+ \frac{i}{\hbar} [\hat H_{int},\hat a_j]+ F_{0} e^{-i\omega_{0 j} t}
\delta_{j0,j},
\end{equation}
where $\delta_{j0,j}$ is the Kronecker's delta, $j_0$ labels the externally pumped mode, $\gamma_j=\gamma_{c j}+\gamma_{i j}$ is the half width at the half maximum for the optical modes, and $\gamma_{cj}$ and $\gamma_{i j}$ stand for coupling and intrinsic losses, respectively. $F_{0}$ represents the external pumping at a frequency $\omega_{0 j}$:
\begin{equation} \label{f0}
F_0=\sqrt{\frac{2\gamma_{c}P_{in}}{\hbar \omega_{0}}} e^{i\phi_{in}}
\end{equation}
where $P_{in}$ is the input power and $\phi_{in}$ is the phase of the pump.

Below we discuss various three-wave mixing processes described by Hamiltonian (\ref{Hint}). Let us also mention a very recent review \cite{Breunig16rev} dedicated to this subject.

\subsection{Electro-optical phenomena and applications of WGMRs}\label{sec:EOM}

We start our discussion of the second-order nonlinear optical phenomena from reviewing the interaction of optical and static or quasi-static (on the optical cycle time scale) electric fields. Such interaction enables a variety of important optical modulation, sensing and frequency conversion applications.
Perhaps the most important of these applications is the electro-optical modulator (EOM). Operation of an EOM can be considered as based on a strongly nondegenerate parametric process \cite{Matsko02OPO}. It upshifts or downshifts the frequency of a pump photon by the frequency of a microwave photon. The upshift (anti-Stokes) process corresponds to absorption of the pump photon and microwave photon and emission of the higher frequency photon. The downshift (Stokes) process corresponds to absorption of the pump photon and emission of both the microwave and lower frequency photon.

While the efficient modulation usually requires an external microwave field, a \textit{spontaneous} Stokes parametric frequency conversion has been predicted \cite{Matsko02OPO} and demonstrated for the frequency shifts ranging from sub-THz to approximately 20 THz \cite{Savchenkov07THz}. When such frequencies are efficiently out-coupled from the modulator, this process may be used as a narrow-band all-optical THz source. A similar type of THz source has been realized in a conventional cavity-assisted single-resonant optical parametric oscillator (OPO) \cite{Kiessling09THzSrc}. Let us also point out that the spontaneous Stokes process results in unavoidable quantum noise background present in EOMs \cite{Matsko07mod_noise}.

Utilizing microwave and optical resonances can enhance the parametric process efficiency at the cost of limiting the bandwidth %\cite{gordon63bstj,alferness82mtt,ho93ptl,kawanishi01el,gheorma02ptl,kato04el,benter05ao,kato05ao,gan06apl}.
\cite{gordon63bstj,ho93ptl,kawanishi01el,gheorma02ptl,kato04el,benter05ao,kato05ao,gan06apl}. Efficiency of the existing commercial EOMs still remains very low compared to the theoretical limit of combining each microwave photon with an optical photon. The reason is relatively short interaction length as well as insufficient spacial overlap of the interacting fields. Recent developments of highly efficient resonant EOMs based on WGM resonators allow to circumvent these problems.

Electro-optically active WGM resonators are attractive for the EOM application because they can provide a good overlap between the microwave and optical fields, which usually requires an additional microwave cavity. They also have low loss and high quality factors in a wide range of optical as well as microwave frequencies determined by the transparency window of the resonator material
\cite{ilchenko00spie,ilchenko01spie,cohen01el-a,cohen01el-b,cohen01sse-a,cohen01sse-b,ilchenko02ptl,ilchenko03josab,hosseinzadeh05sse,hosseinzadeh06tmtt,rabiei02jlt,
weldon04ptl,tazava05el,tazawa06ptl,tazawa06jlt,bortnik07jstqe,Ilchenko08qtz,
Gould11oe,Padmaraju12oe,Rabiei13oe,Kondratiev13bras,qiu14oe}.
These modulators operate either within the mode bandwidth \cite{tazawa06ptl} or involve different high-$Q$ optical modes of the same \cite{cohen01el-a,ilchenko02ptl,bortnik07jstqe} or different polarizations \cite{savchenkov09rc,savchenkov09ol,savchenkov10pmtt}.

The efficiency and directionality of the modulation process can be regulated by the phase matching conditions.
Realization of the phase matching between light and microwaves is complicated because the index of refraction of the electro-optical materials is very different at the optical and microwave frequencies. It can be achieved by optimizing the geometrical shape of both the microwave and optical parts of the modulator as well as using different materials for these parts.

The significant dissimilarity of the optical and microwave wavelengths gives a lot of flexibility for such optimization. For WGM resonators, this approach was initially proposed in \cite{ilchenko00spie,ilchenko01spie}. It was shown that it is possible to confine the microwave field in a metal resonator built on a top of an optical resonator to achieve the desirable phase matching \cite{ilchenko00spie,ilchenko01spie,cohen01el-a,cohen01el-b,cohen01sse-a,cohen01sse-b,ilchenko02ptl,ilchenko03josab,hosseinzadeh05sse,hosseinzadeh06tmtt}.
In a similar way it is possible to control the modulation process with high flexibility, e.g. to suppress the Stokes process nearly completely and create an SSB modulator \cite{savchenkov09ol} that is able to upconvert a microwave photon to the optical frequency domain with nearly 100\% efficiency \cite{Matsko08THz}. This type of modulator can be utilized for counting microwave or THz photons at room temperature \cite{Strekalov09THzOL,Strekalov09THz_LasPhysLett,Strekalov09THzRF,rueda_efficient_2016}.

A WGM EOM can be characterized by a modulation coefficient, defined as the ratio of the output power of the first optical harmonic and the optical pump power. The modulation coefficient is proportional to $P_{mw}Q^2Q_{mw} r_{ij}^2$ \cite{ilchenko03josab}, where $Q$ and $Q_{mw}$ are the loaded quality factors of the optical and microwave modes respectively, $r_{ij}^2$ is the relevant electro-optical coefficient of the material, and $P_{mw}$ is the applied microwave power. Therefore, the higher the quality factors and the nonlinearity, the lower is the microwave power required to achieve the same modulation efficiency. WGM resonators made out of crystalline LiNbO$_3$ and LiTaO$_3$, characterized by the optical bandwidth ranging from hundred kilohertz (weakly coupled) to gigahertz (fully loaded) as well as by large electro-optical coefficients \cite{ilchenko11chapter}, are particulary attractive for WGM EOMs. Tunable and multi-pole filters, resonant electro-optical modulators, photonic microwave receivers, opto-electronic microwave oscillators, and parametric frequency converters were realized using such EOMs.

Let us compare the efficiency of a conventional running wave electro-optical phase modulator and a WGM-based modulator. The optical field $E_{out}$ emerging from a phase modulator is related to the input field $E_{in}$ as
\begin{equation}
\frac{E_{out}}{E_{in}}=\exp \left [ i \pi \frac{\rm V}{\rm V_\pi} \cos \omega_{mw}t \right ],
\end{equation}
where ${\rm V}$ is the voltage of the RF signal at the modulator electrode, ${\rm V_\pi}$ is the characteristic voltage of the modulator imparting a $\pi$ phase shift to the optical carrier, and $\omega_{mw}$ is the modulation frequency. The relative power of the first modulation sidebands is
\begin{equation}
\frac{P_\pm}{P_{in}} = \frac{P_{mw}}{P_{sat}},
\end{equation}
where the characteristic (saturation) power can be expressed via the ${\rm V_\pi}$ and resistance of the microwave circuitry ${\cal R}$ as
\begin{equation}
P_{sat}=\frac{4}{\pi^2} \frac{\rm V_\pi^2}{2 {\cal R}}.
\end{equation}

For a WGM EOM based on z-cut lithium niobate resonator we have \cite{ilchenko11chapter}
\begin{equation}
P_{sat}=\frac{n_{mw}^2V_{mw}\omega_{mw}}{32\pi Q_{mw} Q^2n_e^4r_{33}^2 \sigma^2},
\end{equation}
where $n_{mw}$ and $n_e$ are the extraordinary refractive indices at the microwave and optical frequencies, respectively,  $V_{mw}$ is the microwave  mode volume, $r_{33}$ is the electro-optical coefficient, and $\sigma$ is the overlap integral for the process. It is easy to verify that for a typical running wave modulator with ${\rm V_\pi}=3$ V and ${\cal R}=50$~Ohm the saturation power exceeds 100~mW, while for a lithium niobate WGM EOM with 10~MHz bandwidth it is only about 2.5~$\mu$W \cite{ilchenko2010spienr,rueda_efficient_2016}. It means that the equivalent ${\rm V_\pi}$ of such WGM EOM is less than 20~mV.

In addition to the light modulation leading to generation of optical harmonics, resonant WGM EOMs can be used as quadratic receivers of the microwave field \cite{hosseinzadeh05sse,hosseinzadeh06tmtt}. The operation principle of such devices is based on a nonlinear absorption of the input light, which increases proportionally to the microwave signal power. The increase of the optical loss changes the coupling conditions towards under-coupled and reduces the coupling contrast. Such a behavior resembles operation of an opto-electronic transistor-like device, where a low-power (a few tens of microwatts) microwave signal  changes transmission of a higher-power (a few milliwatts) optical signal. Similarly a modulated microwave input results in modulation of the light passing through the resonator.

Optimally modified WGM EOMs can be furthermore used as efficient electric field sensors \cite{passaro06jstqe,wang06oe,sun07sj,iopollo09oe,passaro12s,zhang14jlt}.
A sensor based on a microwave dielectric cylindrical antenna concentrating the microwave field within a lithium niobate WGM resonator was originally proposed in \cite{hsu07np,ayazi08oe}. This sensor operates on similar principles as the fiber-based running wave \cite{semertzidis00nmpra} and resonant \cite{runde07apb}  electric field sensors. A more efficient all-resonant WGM configuration of a dielectric E-field sensor was introduced theoretically \cite{matsko10jlt} and validated experimentally \cite{Savchenkov10VCO}. High sensitivity of these devices has inspired the application proposals in areas such as e.g. biomedical studies \cite{TAMEE13psycho}.

Concluding the section on electro-optical phenomena in WGM resonators, we should also mention the possibility of  \textit{magneto}-optical phenomena. Such a possibility has been discussed theoretically \cite{Smirnov02magnOpt,Deych11magnOpt}, and experimentally explored in ferromagnetic microspheres made from yttrium
iron garnet \cite{Zhang15magnon,Haigh15mo,Osada16magnon,Haigh16mo}. The conventional Faraday effect as well as non-reciprocal sidebands generation at the magnon frequency is observed in these experiments. Although most typical materials with significant Faraday effect are lossy, and their $Q$-factor is limited to approximately $10^6$ \cite{Zhang15magnon}, the emerging field of WGM magneto-optics holds a great promise for building ultra-sensitive, room-temperature compact magnetometers as well as for efficient microwave-to-optics conversion.  

\subsection{Natural phase matching and selection rules for second-order processes}\label{sec:Natural}

Now let us turn to discussing second-order nonlinear interaction of optical fields.
The overlap integral (\ref{ovlp}) is responsible for all resonator-specific features of the three-wave mixing, including the mode selection rules. These rules reflect the symmetries inherent to the resonator and its eigenfunctions. The fundamental WGM symmetry is associated with rotation, which is expressed in the $e^{im\varphi}$ term in the eigenfunction (\ref{FieldSphere}). Therefore the fundamental selection rule arising from (\ref{ovlp}) corresponds to the angular momenta conservation: $m_1+m_2=m_3$. Further selection rules are discussed in \cite{Kozyreff08sufaceSHG,Fuerst10SH,Fuerst10PDC,Strekalov14SFG}\footnote{Note that \cite{Strekalov14SFG} required a Corrigendum \cite{Strekalov14SFGerr}.}. In particular, it is shown \cite{Fuerst10SH,Fuerst10PDC,Strekalov14SFG} that in large resonators the overlap integral $\sigma_{123}$ factors into the radial and angular parts as a very good approximation. The angular part yields Clebsch-Gordan coefficients $\langle l_1,l_2;m_1,m_2|l_3,m_3\rangle$ that describe the photon's angular momenta conversion and invoke well-known selection rules:
\begin{eqnarray}
m_1+m_2=&m_3&,\nonumber\\
|l_1-l_2|\leq& l_3&\leq l_1+l_2,\label{orb_sel_rules}\\
l_1+l_2+l_3 &=& 2\mathcal{N},\nonumber
\end{eqnarray}
where $\mathcal{N}$ is a natural number. For equatorial modes ($l_i=m_i\;{\rm for} \;i=1,2,3$) in large resonators, one can use the following asymptotic relation \cite{Fuerst10PDC}\raisebox{1ex}{\footnotemark}\footnotetext{In Eq. (6) of  \cite{Fuerst10PDC}, the numerical factor should be 4, not 8.}:
\begin{equation}
\langle m_1,m_2;m_1,m_2|m_3,m_3\rangle\approx 0.5(m_3/\pi^3)^{1/4},\label{CGassympt}
\end{equation}
which indicates a slow increase of the angular overlap, scaling as approximately the fourth-power root of the radius-to-wavelength ratio.

\begin{figure}[b]
\includegraphics[width=9cm]{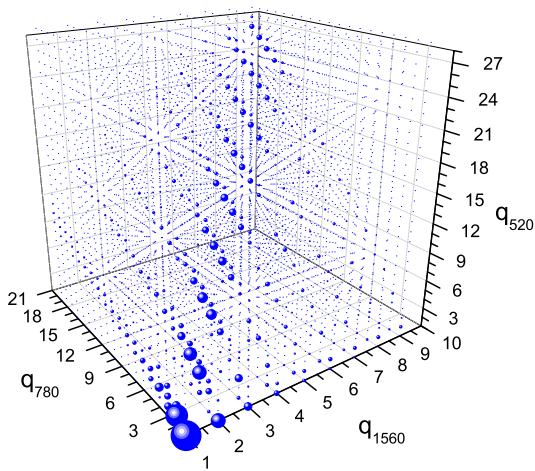}
\caption[]{\label{fig:ovlp}Absolute-square of the overlap integral (\ref{ovlp}) radial part is represented by the dots size for various mode numbers combinations $\{q_1,q_2,q_3\}$, for the $780\,{\rm nm}\,+1550\,{\rm nm}\,\rightarrow\,520\,{\rm nm}$ frequency-sum generation in a Lithium niobate WGM resonator of 0.65 mm radius. Reprinted from \cite{Strekalov14SFG}.}
\end{figure}

The radial part of the overlap integral given by the Airy functions overlap does not lead to any strict selection rules. However for large radial mode numbers $q_i$ and respective wavelengths $\lambda_i$ it favors such conversion channels that $q_1/\lambda_1+q_2/\lambda_2\approx q_3/\lambda_3$ \cite{Strekalov14SFG}, see Fig.~\ref{fig:ovlp}. This occurs because for large $q$ the radial part ${\rm Ai}(u/u_m-\alpha_q)$ of a WGM eigenfunction (\ref{Eigenfunktion})
takes on oscillatory form, and the selection rules begin to resemble those for coupled harmonic waves.

\begin{figure}[t]
\vspace*{-0.2in}
\includegraphics[width=9cm]{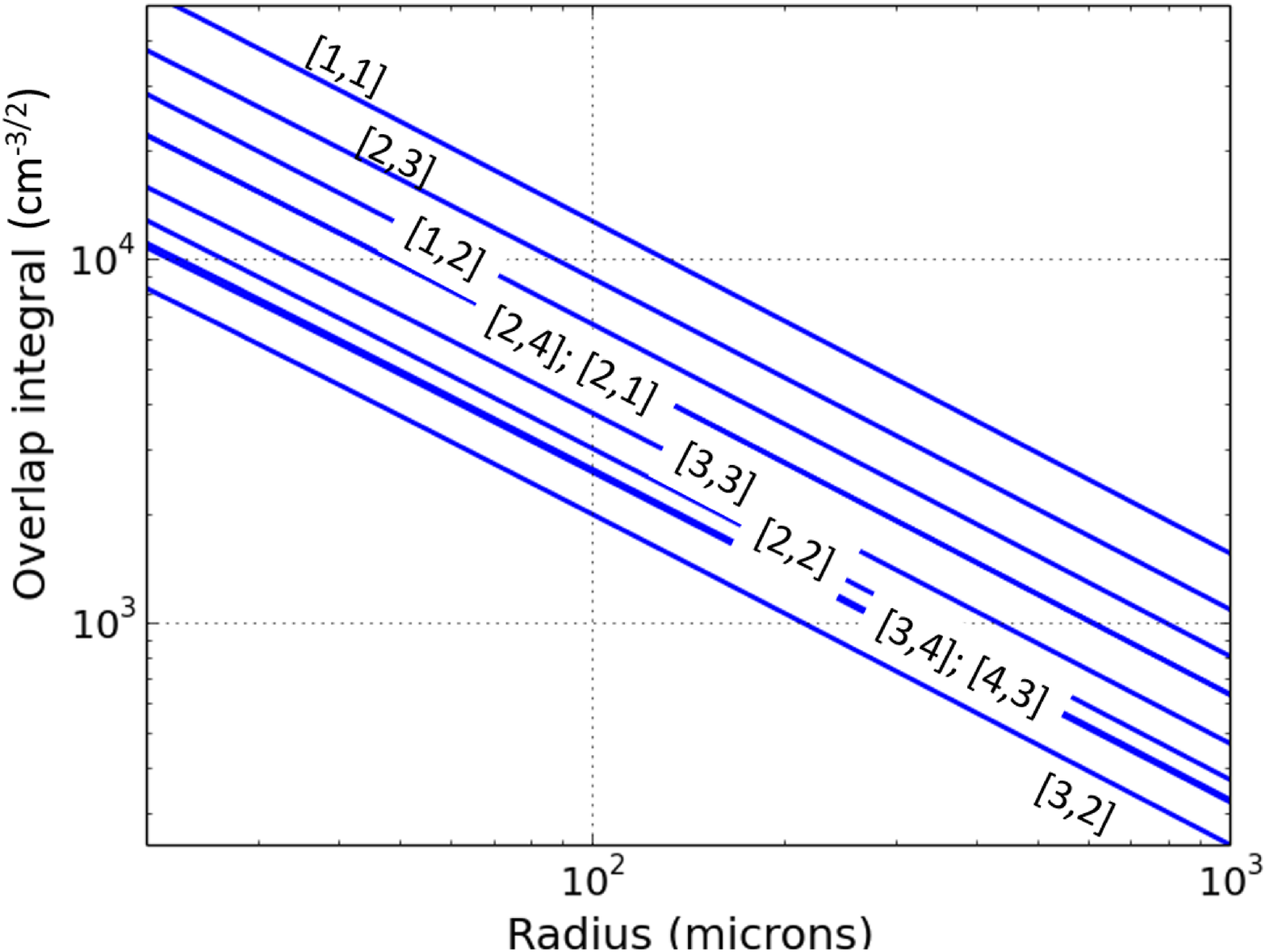}
\caption[]{\label{fig:Rscale}Overlap integrals $\sigma$ for $\lambda_p=1\, \mu$m to $\lambda_{SH}=$ 500 nm SHG are shown as functions of the resonator radius for various conversion channels $[q_p,q_{SH}]$.  Reprinted from \cite{Strekalov15jmo}.}
\end{figure}

Analyzing the overlap integrals (\ref{ovlp}) one can also learn how the nonlinear conversion efficiency depends on the resonator size. For example in \cite{Strekalov15jmo}, this dependence was theoretically studied for the frequency doubling of several pump wavelengths via various equatorial channels in a lithium niobate WGM resonator. Curiously, in all studied cases the same scaling law was found: $|\sigma|^2\propto R^{-1.8}$, see Fig.~\ref{fig:Rscale}. The same power scaling law was established, although not reported, for the  $780\,{\rm nm}\,+1550\,{\rm nm}\,\rightarrow\,520\,{\rm nm}$ frequency-sum generation \cite{Strekalov14SFG}.  Moreover, this is consistent with the $V_p^{-1}\propto R^{-1.83}$ WGM volume scaling in micro spheres \cite{Buck2003CQED,Kippenberg04Raman}\raisebox{1ex}{\footnotemark}
\footnotetext{Ref. \cite{Kippenberg04Raman} quotes  $V_p\propto R^{1.83}$ as calculated in \cite{Buck2003CQED}, but \cite{Buck2003CQED} only provides a plot for $V_p(R)$ in its Fig.3. We assume that in  \cite{Kippenberg04Raman} the power-law fitting of this plot for large $R$ was performed.}.

\subsection{Quasi-phase matching in periodically-poled WGM resonators}\label{sec:QPM}

\begin{figure}[htb]
\includegraphics[width=8cm]{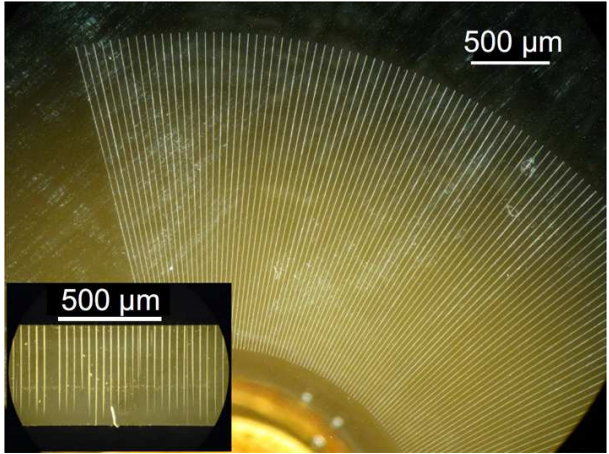}
\caption[]{\label{fig:poling}Radial poling pattern on a lithium niobate wafer, visualized by etching.  Reprinted from \cite{Breunig12spie}.}
\end{figure}

Similarly to bulk crystals and straight waveguides, phase matching in WGM resonators can be modified by periodical poling. Here the ``periodic"  means a radial pattern such as shown in  Fig.~\ref{fig:poling} rather than a series of parallel lines. Periodical poling in crystals such as lithium niobate changes the local direction of the crystallographic $z$-axis and the sign of those $\chi^{(2)}_{ijk}$ tensor components that have only one or all three indices $i$, $j$ and $k$ matching $z$. As a result, the $\tilde{\chi}^{(2)}$ factor in (\ref{Hint}) becomes coordinate-dependent and enters the overlap integral (\ref{ovlp}).

Although various radial poling patterns have been discussed \cite{Haertle10}, it is easy to see that what really matters is the Fourier-transform of this pattern with respect to the azimuthal angle $\varphi$. The most efficient quasi-phase matching (QPM) is therefore achieved with a radially-poled structure consisting of $K$ equidistant (in $\phi$) lines, see  Fig~\ref{fig:poling}. In this case the first phase matching condition (\ref{orb_sel_rules}) is modified as follows:
\begin{equation}
m_1+m_2=m_3\pm {\cal N}K,\quad{\cal N}=0,1,2,...,
\label{m4qpm}
\end{equation}
where in close analogy with periodically poled straight waveguides and bulk crystals ${\cal N}$ is the the phase matching order.
It should be mentioned that a fabrication error leading to eccentricity between the resonator and poling pattern may have a considerable phase matching broadening effect that may be desirable or undesirable depending on the specific application \cite{Beckmann11}.

Let us also point out that the early and some of the later demonstrations of QPM in WGM resonators were proposed \cite{Ilchenko03parametric} and  carried out \cite{Ilchenko04SH,Moore11_4th_harm} with the poling patterns consisting of parallel, rather than radial, lines. Phase matching in such resonators allows interesting interpretation in terms of the effective index approach discussed in section \ref{sec:neff}. In this approach a large WGM resonator can be ``unfolded" into a straight waveguide with a modified index of refraction. A linear equidistant poling pattern then becomes variable, with the period ranging from its nominal value (where the poling lines are radial) to practically infinity (where the poling lines are tangential). If the nominal period is sufficiently small, there will be four (in degenerate case, two) locations where the poling period is just right and the QPM is \textit{locally} achieved for the desired process at the desired wavelengths. Only these narrow segments of the resonator will contribute to the nonlinear conversion. Importantly, their contributions will add coherently and may lead to either constructive, or destructive interference. A wide range of wavelengths can be nonlinearly converted in such a resonator, each at its own four locations. The cost of the wavelength versatility is the reduced conversion efficiency due to a limited interaction length at each location.

Using the linearly-poled resonators was enabled by the commercial availability of periodically poled lithium niobate (PPLN)wafers. Since then, several lithography-based radial poling techniques have been developed \cite{Breunig12spie} specifically for WGM resonators, including the already mentioned calligraphic poling technique  \cite{mohageg05poling,Meisenheimer15poled}.

\subsection{Crystal symmetry based and ``cyclic" quasi-phase matching}\label{sec:cycPM}

QPM in WGM resonators may arise even without artificial domain inversion. As an optic field propagates around the resonator, its polarization orientation generally changes with respect to the crystallographic axes. This may lead to $\varphi$-dependent modulation of the effective nonlinearity  $\tilde{\chi}^{(2)}$ and/or of the index of refraction. The former effect can be observed e.g. in the SHG experiments when $\tilde{\chi}^{(2)}$ couples the $x$ and $y$ projections of the optic field. This process has been predicted \cite{Dumeige06SH,Yang07sh,Kuo11GaAs_prop} and observed \cite{Kuo14SHG_GaAs} in the ${\bar 4}$ symmetry crystals, such as GaAs, GaP, ZnSe and others. In this case harmonic variation of the field projections leads to the following modification of the azimuthal phase matching condition:
\begin{equation}
m_1+m_2=m_3\pm 2.
\label{m4sym_qpm}
\end{equation}
This modification may have an appreciable effect on the phase matching wavelength in small resonators where the FSR is large.

In a birefringent resonator whose optical axis does not match the axis of symmetry, the index of refraction for TM-polarized WGMs also becomes a periodic function of $\varphi$, as we discussed in section \ref{sec:dielectricResonators}. This leads to a more complicated situation than just the $\varphi$-dependent $\tilde{\chi}^{(2)}$. Because this situation is not analytically tractable as yet, only a special case of a uniaxial crystal with the optical axis lying in the resonator equatorial plane has been utilized in nonlinear optics. This type of resonators, usually called the x-cut or xy-cut resonators, manufactured from crystal quartz \cite{Ilchenko08qtz}, lithium niobate and tantalate \cite{Grudinin13}, and BBO \cite{Grudinin13,Lin13cyclic,Lin14tuning} have been shown to support high-$Q$ modes that can be identified as TE and TM at least at the coupling location.

An SHG \textit{cyclic} phase matching in an x-cut BBO resonator has been demonstrated in \cite{Lin13cyclic,Lin14tuning}. In such resonators, the  refraction index for TM modes $n_{\rm TM}$ oscillates between the ordinary and extraordinary values $n_o$ and $n_e$ according to
\begin{equation}
\left(\frac{1}{n_{\rm TM}(\varphi)}\right)^2=\left(\frac{\cos(\varphi)}{n_e}\right)^2+\left(\frac{\sin(\varphi)}{n_o}\right)^2,
\end{equation}
where we chose to measure the azimuthal angle $\varphi$ from the optical axis.
Since this resonator has no (infinite-order) rotation symmetry, the quantum numbers $q,\,L,\,m$ cannot be introduced in the strict sense, and selection rules  (\ref{orb_sel_rules}) no longer apply.  Phase matching in this resonator can be formulated following the effective index of refraction approach as
\begin{equation}
\beta_3=\beta_1+\beta_2,\label{betaPM}
\end{equation}
where
\begin{equation}
\beta(\lambda)=\frac{2\pi}{\lambda} \tilde{n}(\lambda)
\end{equation}
is the effective wave number for each mode. The local phase detuning for up-conversion of a TE polarized pump into a TM polarized second harmonic $\Delta\beta(\varphi)=2\beta_p-\beta_{SH}(\varphi)$ depends on the azimuthal angle $\varphi$ or on local coordinate $z = R\varphi$ measured along the ``unfolded" waveguide. This detuning governs the propagation equation for the second harmonic field  \cite{Lin13cyclic}:
\begin{equation}
\frac{dE_{SH}}{dz}=i\frac{\beta_{SH}(z)}{n^2_{SH}(z)}\tilde{\chi}^{(2)}(z)E_p^2\exp\left\{i\int_0^z\Delta\beta(z^\prime)dz^\prime\right\}.\label{SHGcyclic}
\end{equation}

As in the previously discussed example of linearly poled WGM resonators, only four short waveguide regions may contribute to the second harmonic field build-up. However now these regions are determined not by the local QPM, but by a stationary-phase condition $\Delta\beta(z_{pm})=0$. Other $z$-dependent parameters of equation (\ref{SHGcyclic}) can be evaluated at $z=z_{pm}$ as a very good approximation.  Again, contributions from different regions  $z_{pm}$ may interfere constructively or destructively, and a wide-range cyclic phase matching comes at the price of reduced conversion efficiency.

\subsection{Exotic phase matching in WGM resonators}\label{sec:exoticPM}

The phase matching conditions, or mode selection rules (\ref{orb_sel_rules}) in WGM resonators are significantly more relaxed compared to free space. Indeed, the usual phase matching requirement $\vec{k}_1+\vec{k}_2=\vec{k}_3$ is represented by a set of \textit{three} equations imposing constraints, whereas (\ref{orb_sel_rules}) has only \textit{one} such equation. This flexibility allows for
achieving in WGM resonators such types of phase matching that cannot be achieved in bulk crystals. In particular, Type 0 and Type II SPDC was predicted in lithium niobate, lithium tantalate and BBO z-cut crystals in a wide range of pump wavelengths \cite{Strekalov15jmo}, although the overlap penalty for using high-order modes with large $q$ may be significant.
It is even possible to find a double phase matching, e.g. for nondegenerate SPDC with simultaneous frequency-doubling of the signal, idler or both \cite{Strekalov15jmo}. Double phase matching in strongly nonlinear systems is interesting in the context of quantum-optics applications. It can lead to multipartite entanglement \cite{Tan11competing} and to control of the photon pair statistics in SPDC via quantum Zeno blockade. % \cite{??} \textbf{- Abijith will submit the paper}.

A search for exotic phase matching in WGM resonators can be conveniently accomplished using the effective index of refraction approach introduced in section \ref{sec:neff}. This approach leads to a simple but approximate analytical expressions for the phase matching conditions in the wavequide form (\ref{betaPM}). They are approximate because the WGM frequencies are treated as continuous, whereas in an actual resonator they are discrete. Therefore the accuracy of this method is limited by the resonator's free spectral range. The simplicity of this approach comes from the same approximation, which allows us to solve a continuous-value set of equations (\ref{betaPM}) rather than to look for the discrete-value solutions to the WGM dispersion equations set (\ref{disp}) constrained by selection rules (\ref{orb_sel_rules}).

An important observation regarding the effective index phase matching is that it strongly depends on the radial mode numbers $q$, especially when the resonator is small. In section \ref{sec:neff} we already mentioned that the geometric correction to the refractive index is negative: $\Delta n= \tilde{n}-n<0$. Moreover, $|\Delta n|$ becomes progressively larger for larger $q$, because the optical field of a higher-order  mode effectively has a shorter path to travel, see Fig.~\ref{fig:Transversalmoden}. In small resonators this effect can be quite strong even in comparison with natural birefrigence, which explains the unusual phase matching. We should point out that while the double phase matching can be relatively easily found by the effective index of refraction method, its realization in a fully-resonant system such as a WGM resonator is difficult to achieve in practice. It depends on such control parameters that cannot be easily tuned, e.g. the resonator radius and rim shape.

\subsection{Dynamics of the second-order processes in triply-resonant systems}\label{sec:dynam}

Analysis of nonlinear optical processes dynamics in phase-matched WGM resonators has been reported for the SHG \cite{Ilchenko04SH,Kozyreff08sufaceSHG,Kozyreff11review,Sturman11,Sturman12nl} and OPO \cite{Matsko02OPO,Ilchenko03parametric,Savchenkov07THz,Breunig11OPO_rev,Sturman11,Sturman12nl}, as well as for the sum-frequency generation (SFG) \cite{Huang12switch,Sun13switch,Strekalov14SFG,Strekalov14SFGerr} and difference-frequency generation (DFG) \cite{Huang10switch_prop}.  Usually this analysis
is carried out in terms of coupled ODEs such as given by Eq.~(\ref{setz}). It shows the dynamics similar to other triply-resonant systems with second-order nonlinearity, see e.g. \cite{Ashkin66,Graham68opo,Smith70HSG,Fabre89opo,Debuisschert93,Berger97SHG}. Its details strongly depend on the in/out coupling regime for the nonlinearly coupled modes, as well as on the relation between the resonator linear loss and nonlinear conversion rates. There are however some universal features. For example, the SHG conversion efficiency reaches a maximum  at a saturation pump power $P_{s}$, and then drops. The OPO has a pump power threshold $P_{th}$. For a degenerate OPO these two parameters are closely related: $P_{s}=4P_{th}$ \cite{Fuerst10PDC}, where
\begin{equation}
P_{th}=%\frac{c\,n^6\,V_{int}}{4\lambda_pQ_p\,(8\pi\chi^{(2)}Q_s)^2},
\frac{c}{\lambda_pQ_p}\left(\frac{n^3}{16\pi\sigma \chi^{(2)}Q_s}\right)^2,
\label{threshold}
\end{equation}
where $\lambda_p$ is the pump wavelength and $n = n_e(\lambda_p)=n_o(\lambda_s)$ is the phase-matched refractive index.

Degenerate Type I SPDC in WGM resonators cannot be described with just a pair of coupled-mode equations(\ref{setz}). Indeed, because the WGM spectral lines are locally equidistant, degenerate down conversion of the pump mode with an orbital momentum $2m_0$ into the signal (and idler) mode with $m_s=m_i=m_0$  also implies the possibility of the same pump conversion into non-degenerate modes $m_s =m_0\pm \Delta m,$  $m_i=m_0\mp \Delta m$. Here $\Delta m=1,2,...$, and up to some maximum number determined by the WGM linewidths and the group velocity dispersion of the resonator. Conversion channels with small $\Delta m$ will be nearly on-resonance and hence almost equally efficient. A SPDC optical comb is therefore expected \cite{Wu12comb-prop} to form around the degenerate wavelength.

As a higher-order effect in a multi-mode near-degenerate SPDC, the tresholdless sum-frequency generation process among the parametric comb lines may lead to building up another comb around the pump mode. For this secondary conversion to be efficient, the FSRs at the pump and degenerate SPDC wavelengths must be nearly equal, which is granted by the phase matching. This kind of a double-comb structure has been observed \cite{Ulvila13OPOcomb,Ricciardi15two_combs} and studied theoretically \cite{Leo16shg} in straight waveguide cavities.

The presence of multiple near-equidistant modes also makes the conventional WGM SHG description based on two coupled-modes equations insufficient beyond the low-power linear regime. Once the circulating SHG power exceeds the OPO threshold, the non-degenerate multiple-line down conversion commences. This phenomenon has been predicted to lead to self-pulsing in triply-resonant \cite{Martle94competing}, and observed in single-resonant SHG processes \cite{Bache02competing}. It may be responsible for the anomalous SHG signal behavior observed in \cite{Fuerst10SH}.

\begin{figure}[t]
\vspace*{-0.2in}
\includegraphics[width=9cm]{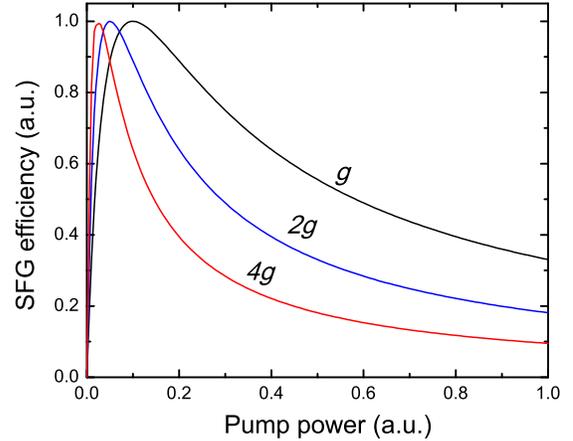}
\caption[]{\label{fig:SFGeffcy} A qualitative dependence of the SFG conversion efficiency on the pump power for various nonlinear coupling rates $g$. All three fields are assumed to be resonant.}
\end{figure}

Similarly to the SHG process, conversion efficiency saturation also occurs in SFG and DFG. Above the saturation power, efficiency of these processes decreases, asymptotically approaching zero as shown in Fig.~\ref{fig:SFGeffcy}. In this Figure we plot the SFG conversion efficiency (defined as the ratio of the SFG signal to the input \textit{probe} power) vs. the input \textit{pump} power calculated using the equations from \cite{Strekalov14SFG}. Both parameters are given in arbitrary units. The actual peak conversion efficiency depends on the relation between the linear and non-linear coupling rates of all involved modes. Three curves in Fig.~\ref{fig:SFGeffcy} illustrate the effect of increasing the nonlinear coupling rate $g$ introduced in Eq.~(\ref{Omega}).

The saturation behavior of SFG in triply-resonant structures is different from the non-resonant structures with undepleted pump \cite{Kumar90SFG}, where the oscillations of the probe and sum-frequency powers are expected. In high-$Q$, strongly nonlinear resonators the saturation opens up the possibility of realization an efficient low-power all-optical switch, which has been both predicted  \cite{Huang10switch_prop,Huang12switch,Sun13switch} and implemented in Fabri-Perot \cite{McCusker13switch} as well as in WGM resonators \cite{Strekalov14switch}.

\subsection{Experimental observations of the second-order processes}\label{sec:chi2expt}

Optical WGM and ring resonators with the second-order nonlinearity have been successfully used for optical  frequency doubling \cite{Yu99SHG,Ilchenko04SH,Fuerst10SH,Xiong11GaN_SHG,Levy11SiNmicroring,Pernice12shg,Lin13cyclic,
Lin14tuning,Kuo14SHG_GaAs,Wang14SGH,Lin15sh,Mariani14SH,Fuerst15SHG,Vukovic15SHG,Lake16SHG}, tripling \cite{Sasagawa09tripling,Levy11SiNmicroring}, and quadrupling \cite{Moore11_4th_harm}, as well as parametric down conversion above \cite{Savchenkov07THz,Fuerst10PDC,fuerst11sqz,Beckmann11,Beckmann12coupling,Werner12BlueOPO,Werner15OPO} or below\cite{Fortsch13NC,Fortsch15sm,Fortsch15jopt,Schunk15CsRb} the  OPO threshold. Generation of optical frequency sum \cite{Strekalov14SFG} and difference \cite{Andronico08diff_GaAs} also have been demonstrated. Some of these experiments \cite{Savchenkov07THz,Fuerst10SH,Fuerst10PDC,fuerst11sqz,Pernice12shg,Werner12BlueOPO,Fortsch13NC,Fuerst15SHG,Strekalov14SFG,Mariani14SH,Vukovic15SHG,Fortsch15sm,Fortsch15jopt,Schunk15CsRb,rueda_efficient_2016} have been relying on the natural phase matching whereas QPM with periodical domain inversion was harnessed in others \cite{Ilchenko04SH,Sasagawa09tripling,Moore11_4th_harm,Beckmann11,Beckmann12coupling,Werner15OPO}. Crystal symmetry based QPM \cite{Kuo14SHG_GaAs,Andronico08diff_GaAs} and the cyclic phase matching \cite{Lin13cyclic,Lin14tuning} have also been experimentally realized.

\begin{table}[t]
\begin{tabular}{c c c c c c}
		\hline
 Ref.					& $\lambda_p$ nm	& Crystal 	&$\eta$ W$^{-1}$	& $P_{s}$   & $\eta_m$\textperthousand  \\
		\hline
\cite{Fuerst10SH}		&1064				&LiNbO$_3$	&4500				&3.2		&90\\
\cite{Fuerst15SHG}		&490				&LB4		&37					&			&22\\
\cite{Lin13cyclic}		&634-1560			&BBO		&0.6-18				&			&0.7-16\\
\cite{Lin14tuning}		&634				&BBO		&7.4				&			&7.4\\
\cite{Ilchenko04SH}		&1550				&PPLN		&1.5 				&300		&500 \\
\cite{Xiong11GaN_SHG}	&1560				&GaN		&0.15				&			&0.03\\
\cite{Kuo14SHG_GaAs}	&1985				&GaAs		&0.5				&0.1		&0.005\\
\cite{Lake16SHG}		&1550				&GaP 		&0.38				&			&0.133\\
\cite{Lin15sh}			&800				&LiNbO$_3$	&0.135				&			&0.08\\
\cite{Wang14SGH}		&1546				&LiNbO$_3$	&0.135				&			&0.005\\
\cite{Pernice12shg}		&1550				&AlN		&0.001				&			&0.025\\
\cite{Mariani14SH}		&1580				&AlGaAs		&7e-4				&3			&0.002\\
\cite{Vukovic15SHG}		&1548				&ZnSe		&8e-4				&			&0.001\\
		\hline	
\end{tabular}
\caption{Summary of reported SHG observations in WGM resonators: pump wavelength $\lambda_p$ in nanometers, resonator material (PPLN is periodically poled lithium niobate, LB4 is lithium tetraborate), slope efficiency $\eta$, saturation power $P_{s}$ in mW, and maximum observed pump power conversion  $\eta_m$.}
\label{tbl:SHGeffcy}
\end{table}

WGM resonators have been shown to have a high nonlinear conversion efficiency, which is due to the exceptionally high $Q$-factor and small interaction volume. For SHG of the pump wavelength $\lambda_p$ in low-power regime the conversion efficiency is proportional to the pump power, and therefore can be characterized by the slope efficiency $\eta$. At a higher power regime saturation occurs and the conversion efficiency reaches its limit $\eta_m$. These parameters are listed in Table~\ref{tbl:SHGeffcy} for the SHG experiments discussed here. In the absence of the saturation power data, $\eta_m$ is evaluated at the maximum pump power used in a particular experiment.

WGM OPOs can be characterized by the threshold power $P_{th}$ and maximum observed conversion efficiency $\eta_m$.  These parameters are summarized in Table~\ref{tbl:PDCeffcy}. All these experiments have been carried out with lithium niobate resonators:  not poled for \cite{Fuerst10PDC}, and periodically poled in radial direction for the others. Strong variation in the listed threshold power is a consequence of the trade-off inherent to WGM OPOs: for strongly out-coupled resonators the maximum \textit{observed} conversion efficiency increase, but so does the threshold, due to reduction of the $Q$-factors. In different experiments different choices have been made regarding this trade-off.
\begin{table}[b]
\begin{tabular}{c c c c c}
		\hline
 Ref.						&$\lambda_p\rightarrow\lambda_s+\lambda_i$, nm		& $P_{th},\, \mu$W   	& $\eta_m$\%  \\
		\hline
\cite{Fuerst10PDC}			&$532\rightarrow1010+1120$						&7-28					&18 \\
\cite{Beckmann12coupling}	&$1037\rightarrow2011+2140$						&200					&30\\
\cite{Werner12BlueOPO}		&$488\rightarrow707+865	$						&66-330					&7\\
\cite{Werner15OPO}			&$1040\rightarrow2080+2080$						&86-2200				&55\\
\cite{Meisenheimer15poled}	&$1040\rightarrow1800+2500$						&21						&45\\
		\hline	
\end{tabular}
\caption{Summary of reported OPO observations in WGM resonators made from lithium niobate: pump, signal and idler wavelengths, threshold power $P_{th}$ and maximum observed pump power conversion  $\eta_m$.}
\label{tbl:PDCeffcy}
\end{table}

It should be mentioned that in many cases, especially among the most efficient SPDC and SHG processes observed in WGM resonators, the measured conversion efficiency falls considerably short of the theoretical predictions. The same applies to the SFG observations \cite{Strekalov14SFG,Strekalov14SFGerr}. Usually this is attributed to a sub-optimal selection of the conversion channel, e.g. coupling non-equatorial modes, or to photorefractive damage. A rigorous investigation of this discrepancy has not been carried out.

\section{Third-order nonlinear processes}\label{sec:Third-order}

Third-order nonlinear optical processes involve four interacting optical fields and generally described by the term \textit{four-wave mixing} (FWM). In this section we focus on two major FWM processes observed in microresonators -- hyper-parametric oscillation and third harmonic generation. Weak cubic nonlinearity of transparent optical solids requires high peak intensity of the pump to observe these processes in bulk. Such an intensity level can be reached with powerful pulsed lasers. Interaction length can be increased by using optical fibers, which allows to reduce the laser power requirements \cite{agrawalbook,stolen82jqe}. Usage of resonators allows to reduce it even further due to the intracavity field build-up, making the FWM effects observable with CW low power lasers.

\subsection{Interaction Hamiltonian and phase matching}\label{sec:pm3}

Third-order nonlinear optical processes result from the cubic response of the media polarization $\vec{P}$ to the  external electric field $\vec{E}$ \cite{BoydBook}, similarly to quadratic response (\ref{P2}):
\begin{equation}
P^{(3)}_i=4\sum_{j,k,l}\chi^{(3)}_{ijkl}E_j E_k E_l\label{P3},
\end{equation}
where $\chi^{(3)}$ is the third-order nonlinear susceptibility tensor.  The third-order polarization (\ref{P3}) corresponds to the interaction energy
\begin{equation}
H^{(3)}=\int\,dV\sum_{i,j,k,l}\chi^{(3)}_{ijkl}E_iE_jE_kE_l. \label{H3}
\end{equation}

This Hamiltonian can be simplified for the case of a hyper-parametric or third harmonic generation process in isotropic homogeneous medium. The hyper-parametric process involves transformation of a pair of photons at some frequencies $\omega_1$ and $\omega_2$  to another pair of photons at frequencies $\omega_3$ and $\omega_4$, so that $\omega_1+\omega_2=\omega_3+\omega_4$. Often  $\omega_1=\omega_2$ pertain to the same optical pump field, which suggests an analogy with the previously discussed parametric down conversion and justifies the term ``hyperparametric". In the case of a narrowband process ($|\omega_{i}-\omega_j| \ll |\omega_{i}+\omega_j|$, $i,j=1,\dots,4$) the Hamiltonian (\ref{H3}) in the rotating wave approximation for this process leads to
\begin{equation} \label{vsimp1}
\hat H_{HP}= -\frac{\hbar g}{2}  \sum_{i,j,k,l} \hat a_i \hat a_j \hat a_k^\dag \hat a_l^\dag ,
\end{equation}
where $g$ is the coupling parameter. Under the assumption of complete spacial overlap of the resonator modes
\begin{equation}\label{g}
g= \frac{\hbar \omega_0^2 c}{{\cal V} n^2}\, n_2.
\end{equation}
Here ${\cal V}$ is the effective mode volume \cite{Chembo10comb}, $n$ is the linear index of refraction and  $n_2$ is the cubic nonlinearity of the material at the carrier frequency $\omega_0$. We assumed that the medium is isotropic and homogeneous, so that the nonlinearity is polarization-independent and (see \cite{BoydBook})
\begin{equation}
\chi^{(3)}=\frac{n^2 c}{12 \pi^2} n_2.
\end{equation}

Resonant third harmonic generation (THG) can be described using a similar interaction Hamiltonian presented in the rotation wave approximation
\begin{equation} \label{vsimpth2}
\hat H_{THG}= -\frac{\hbar g}{3}  \sum_{i,j} ( \hat a_i^3 \hat a_j^\dag +\hat a_i^{\dag3} \hat a_j),
\end{equation}
where $g$ is defined by Eq.~(\ref{g}). Interaction Hamiltonians (\ref{vsimp1}) and (\ref{vsimpth2}) lead to the same general form ODEs (\ref{setz}) that are already familiar to us from the second-order interaction discussion.

\subsection{Self-Phase and Cross-Phase Modulation}

The simplest effect related to the FWM is the effect of self-phase modulation (SPM). It involves only one mode evolving in time in accordance with the Hamiltonian
\begin{equation} \label{vsimpspm}
\hat H_{SPM} =-\hbar \frac{g}{2} \hat a^\dag \hat a^\dag \hat a \hat a.
\end{equation}
SPM results in power-dependent frequency shift and corresponding distortion of the WGM resonance curve. At higher pump power it leads to the bistability with respect to the optical pump. It is easy to show using Eq.~(\ref{vsimpspm}) that for a lossless mode populated with $\hat N=\hat a^\dag \hat a$ photons at frequency $\omega_0$ the state evolution is described by the operator
\begin{equation}
\hat a(t)= e^{-i(\omega_0-g \hat N)t} \hat a(0).
\end{equation}
Therefore, $g$ represents an SPM frequency shift of the mode per photon.

\begin{figure}[t]
%\vspace*{-0.2in}
\begin{center}
\includegraphics[width=8cm]{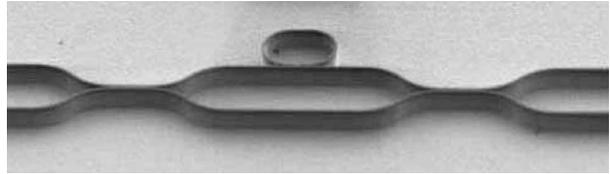}
\end{center}
\caption[]{\label{fig:MZint}A Mach-Zehnder interferometer coupled to an AlGaAs microring.  Reprinted with adaptations from \cite{Heebner04phase}.}
\end{figure}

Among the WGM resonators, the effect of bistability was initially studied in microspheres
\cite{Braginsky89nonlinWGM,Treussart98crioKerr}. A significant number of bistability-related research papers
was published very recently. For instance, optical bistability in Er-Yb co-doped phosphate glass
microspheres was investigated in \cite{ward07jap}. A regenerative pulsation arising from the competition between Kerr nonlinearity and thermal nonlinearity, when the two nonlinearities with very different timescales are comparable in magnitude but opposite in sign, was studied in cryogenically cooled microspheres \cite{park07ol}. Signal processing and all-optical switching by means of Kerr-based bistability in silica bottle microresonators with $Q>10^8$ was studied in \cite{Pollinger10bottle,OShea11bottle}. It was predicted that coupled microresonators can demonstrate stronger SPM effect than a single resonator in the chain \cite{heebner04josab}. This was later demonstrated in a setup combining an on-chip Mach-Zehnder interferometer with a microring \cite{Heebner04phase}, as shown in Fig.~\ref{fig:MZint}. Here, a nonlinear phase shift induced by the microring has lead to self-switching of an optical pulse from one Mach-Zehnder interferometer output to the other depending on the pulse power.

SPM can also lead to the quantum phenomena of photonic blockade and quadrature squeezing, discussed in sections  \ref{sec:quant}.

Cross-Phase Modulation (XPM) is another simple FWM effect that involves interaction of two modes. The interaction is represented by the Hamiltonian
\begin{equation} \label{vsimpxpm}
\hat H_{XPM} =-2\hbar g \hat a^\dag \hat b^\dag \hat a \hat b.
\end{equation}
XPM results in frequency shift of each mode that depends on the photon number in \textit{the other} mode, as described by the evolution equations
\begin{eqnarray}
\hat a(t)&=& e^{-i(\omega_0-2g \hat N_b)t} \hat a(0), \nonumber\\
\hat b(t)&=& e^{-i(\omega_0-2g \hat N_a)t} \hat b(0).
\end{eqnarray}

XPM was directly observed in an amorphous silicon carbide microdisk with $R=6\,\mu$m using a pump-probe technique \cite{Lu14Kerr}. This effect can be useful in quantum nondemolition measurements of photon number in WGM resonators \cite{xiao08oe}.

\subsection{Hyper-parametric oscillation}\label{sec:FWM}

Hyper-parametric optical oscillations \cite{KlyshkoPhotonsBook} involve, at the fundamental level, three optical modes populated with four photons: two pump, one signal, and one idler. This system is also prone to the SPM and XPM. Its complete Hamiltonian is

\begin{eqnarray} \label{ham}
\hat H&=& \hat H_0+\hat H_{HP}, \\ \nonumber \hat H_0 &=& \hbar \omega_0 \hat a^\dag \hat a+ \hbar \omega_+
\hat b_+^\dag \hat b_+ + \hbar \omega_- \hat b_-^\dag \hat b_-, \\
\nonumber \hat H_{HP} &=&\hat H_{SPM}+\hat H_{XPM}+\hat H_{P}, \\ \nonumber \hat H_{SPM} &=&
-\hbar \frac{g}{2} \bigl ( \hat a^\dag \hat a^\dag \hat a \hat a + \hat b_+^\dag \hat b_+^\dag
\hat b_+ \hat b_+ + \hat b_-^\dag \hat b_-^\dag \hat b_- \hat b_- \bigr ) , \\ \nonumber \hat H_{XPM} &=&-  2 \hbar g \bigl (
\hat b_-^\dag \hat b_+^\dag \hat b_+ \hat b_- + \hat a^\dag \hat b_+^\dag \hat b_+ \hat a + \hat a^\dag
\hat b_-^\dag \hat b_- \hat a \bigr ) , \\ \nonumber \hat H_{P} &=& -  \hbar g \bigl ( \hat b_-^\dag \hat b_+^\dag \hat a \hat a +
\hat a^\dag \hat a^\dag \hat b_+ \hat b_- \bigr ),
\end{eqnarray}
where $\omega_0$, $\omega_+$, and $\omega_-$ are the eigenfrequencies of the pump, signal, and idler optical cavity modes, respectively, $\hat a$, $\hat b_+$, and $\hat b_-$ are the annihilation operators for these modes, and $g$ is the coupling constant (\ref{g}). $\hat H_{SPM}$, $\hat H_{XPM}$, and $\hat H_{P}$ stand for the SPM, XPM and hyper-parametric processes, respectively. The SPM and XPM terms change spectral properties of the system, while the hyper-parametric term results in the oscillation process, in which the signal and idler optical sidebands grow at the expense of the pumping wave.

Using the same coupling constant $g$ in all terms of (\ref{ham}) is an approximation which is based on the mode volume concept and on the assumption of complete spacial overlap made while deriving (\ref{g}). A more accurate analysis requires evaluation of the overlap integrals similar to (\ref{ovlp}) but with four WGM eigenfunctions instead of three. The results may be different for different processes as well as for different modes.
These overlap integrals determine the selection rules arising for each of the third-order processes. The orbital selection rule only arises for the hyper-parametric process: $2m_a=m_{b_+}+m_{b_-}$. Other selection rules associated with the angular momentum conservation of four photons are not as broadly used as those arising in the three-wave mixing case from the Clebsch-Gordan coefficients, see Eq.~(\ref{orb_sel_rules}). The radial part of these selection rules is even less studied. We will not invoke this cumbersome analysis here and stay with the approximation (\ref{g}) throughout the rest of this paper.

A WGM resonator supporting hyper-parametric interactions may become unstable and begin to oscillate \cite{matsko05hyper}. The WGM hyper-parametric oscillations were first observed in liquid droplets \cite{Lin1994droplets} and then studied in solid state WGM resonators  \cite{kippenberg04Kerr,savchenkov04hyper}. Similarly to the case of $\chi^{(2)}$-based OPO discussed in section \ref{sec:chi2}, the onset of the hyper-parametric oscillations occurs when the stimulated conversion dominates the spontaneous conversion, which requires the pump power to exceed a certain threshold. The pump threshold of resonant hyper-parametric oscillation \cite{matsko05hyper} is
\begin{equation} \label{hp_th}
P_{HP} \simeq  0.4 \frac{\omega_0 n^2 {\cal V}}{c n_2 Q^2}.
\end{equation}
Equation (\ref{hp_th}) is derived under the assumptions that parameters of all interacting modes are identical, and that the pump mode is critically coupled. It is easy to see that the threshold decreases with the quality factor increase and mode volume decrease.

The hyper-parametric oscillations are different from the parametric ones. The parametric oscillations i) are based on $\chi^{(2)}$ nonlinearity coupling three photons, and ii) involve far separated optical frequencies. The hyper-parametric oscillations i) are based on $\chi^{(3)}$ nonlinearity coupling four photons, and ii) usually involve nearly-degenerate optical frequencies, although strongly non-degenerate hyper-parametric oscillations have been also observed \cite{Liang15OPO}.

Three-mode hyper-parametric process can be observed experimentally. Suitable mode structure can be created in a WGM resonator either by intentional engineering of its spectrum, as discussed earlier in sections \ref{sec:spectrum} and \ref{sec:tuning}, or by properly selecting the optical pump frequency so that only three modes become phase matched. Different mode families can be used to achieve this type of the hyper-parametric oscillation \cite{Liang15OPO}. Unless these special measures are taken, multiple modes of the same WGM family become phase-matched leading to a multi-mode hyper-parametric oscillation and formation of an optical comb structure.

\subsection{Cascaded hyper-parametric oscillation as frequency comb generation}\label{sec:cascadedFWM}

Frequency combs are important in contemporary physics. Two-point stabilization of a frequency comb produced in a mode-locked laser resulted in a revolution in metrology and many other fields \cite{yebook}, providing a precise link between oscillators operating in optical and microwave frequency domains. A stabilized frequency comb is an important part of optical clocks \cite{yebook} and of high spectral purity microwave photonic oscillators \cite{fortier11np}. Frequency combs are used in the search for extraterrestrial planets  \cite{chihhao08n}, in optical communications  \cite{Levy10comb} and sensors \cite{coddington09np}.

In WGM resonators, frequency combs can be generated via hyper-parametric processes. Similarly to the degenerate SPDC discussed in section \ref{sec:dynam}, hyper-parametric conversion of a monochromatic pump may occur simultaneously into multiple modes, forming a frequency comb structure. Microresonator-based optical frequency combs have rapidly become a subject of extensive research as a simple alternative to the conventional frequency combs produced with femtosecond  modelocked lasers \cite{Del'Haye07comb,DelHaye08comb,Savchenkov08comb,Grudinin09comb,Agha09comb,braje09prl,matsko09proc,Levy10comb,razzari10np,Chembo10comb}.  It was proven experimentally \cite{Del'Haye07comb} that the microresonator-based combs can have excellent uniformity as well as high repetition rate, which  makes them ideal for many practical applications.

Similarly to the simple three-mode hyper-parametric oscillation, generation of Kerr frequency combs occurs above a certain power threshold. While expression (\ref{hp_th}) for the hyper-parametric oscillation threshold power $P_{HP}$ can be used as an estimate, the actual threshold value may also depend on the material group velocity dispersion (GVD) and thermorefractive properties, as well as on the excitation regime \cite{Matsko12comb,hansson13pra}. Two different excitation regimes can be realized. In the ``soft" regime,  the oscillations may start from either the vacuum fluctuations of the field or nonzero initial conditions, while in the ``hard" regime, the oscillations can only start from nonzero initial conditions.

Under certain conditions the lines of the generated frequency combs are phase locked. Such frequency combs are considered coherent. The phase of the mutually locked  comb lines depends on the phase of the optical pump. While the \textit{relative} phase stability of the harmonics can be much higher than the pump phase stability, it is still prone to fundamental diffusion processes.

Generation of a Kerr frequency comb is a multi-mode nonlinear process that does not need an optical amplifying medium to sustain itself. This process can be conservative, in the sense that the same optical power exits the resonator as enters it, for the case of an overloaded WGMR with vanishing attenuation. This energy conservation law does not prevent power redistribution among the comb lines.

In some cases both hyper-parametric oscillations that involve only a few frequency harmonics  \cite{kippenberg04Kerr,savchenkov04hyper} and broad frequency combs \cite{Del'Haye07comb,Savchenkov08comb,Kippenberg11comb_rev} observed in WGM resonators can be considered as generalized hyper-parametric oscillations because the high order harmonics smoothly grow from the lower order ones, as described in \cite{herr12np}. This allows to define three stages of Kerr comb development. At the first stage, a hyper-parametric oscillation starts. At the second stage, each oscillation harmonic forms its own, secondary, independent oscillation. At the third stage, all the secondary harmonics phase lock due to the XPM effect. The locking occurs if the primary oscillation spacing is nearly an integer multiple of the secondary oscillation spacing. This simplified model has several deficiencies. It does not describe the general hard excitation regime \cite{Matsko12comb} of the oscillations when the harmonics cannot be generated from quantum fluctuations. Furthermore, the frequency combs corresponding to optical solitons confined in the resonators have a different growth mechanism, in which case all harmonics are formed simultaneously.

\subsection{Mode-locked Kerr-comb generation}\label{sec:comb}

\paragraph{Observations.}

The major difference between the cascaded hyper-parametric oscillation and the mode-locked Kerr frequency comb generation is the natural formation of high-contrast optical pulses within the microresonator pumped solely with CW light, in the latter case.
The pulse generation is usually associated with the hard excitation regime in which the pulses \cite{Matsko12comb} emerge as dynamic normal modes of the nonlinear structure. Duration of the pulses is much shorter than the round trip time of the cavity. In contrast, the hyper-parametric oscillation corresponds to low contrast pulses with duration comparable with the round trip time \cite{Coillet14ol}.

Anomalous GVD is considered optimal for the mode-locked Kerr frequency combs corresponding to formation of bright optical solitons. The first experimental demonstration of the mode-locked Kerr comb in a WGM resonator was reported in \cite{herr14np}. The pulses with 35.2~GHz FSR and 450~kHz loaded bandwidth were generated in a magnesium fluoride resonator and recorded using frequency-resolved optical gating technique. Existence of the mode-locked (soliton) regime was independently confirmed for magnesium fluoride \cite{Herr14soliton}, silicon nitride  \cite{Huang15comb} and fused silica \cite{yi15o} resonators. A broad, $2/3$ octave, coherent frequency comb was observed in a silicon nitride resonator \cite{Brasch15s}. The optical spectra resembling the mode-locked regime were observed by other groups as well  \cite{grudinin15spie,liang2015nc}.

Normal GVD resonators can support generation of dark solitons. This phenomenon was predicted theoretically \cite{Matsko12ol1,Liang14ol,godey14pra,hansson13pra} and observed experimentally \cite{xue15np}. Both narrow \cite{coillet13pj,Henriet15ol} and broad Kerr frequency combs were demonstrated for normal GVD \cite{Liang14ol,xue15np}. The stability range of the mode-locked combs in normal GVD resonators is rather narrow \cite{Matsko12ol1,soltani15lpr}. A different type of stable pulses can exist in this case, enabled by irregularities of the WGM spectrum due to the modes anti-crossing \cite{lobanov15arch}.

A resonator can support one or several solitons as independent solutions of the corresponding set of Hamilton equations. It is possible to excite a preferable number of solitons by selecting proper initial conditions or by dynamically manipulating the pumping light parameters \cite{papp13oe,taheri15pj,karpov16arch}. Each of these independent solutions corresponds to a separate frequency comb with no fixed phase relation to others. Multiple solitons are not always independent. They can interact via e.g.~Raman scattering \cite{Milian15soliton} which leads to their synchronization. In \cite{Brasch15s} two- and three-soliton combs were experimentally investigated and coherence properties of the two-soliton combs were demonstrated. In this case the solitons propagating around the resonator rim are separated by exactly 180$^\circ$.

\paragraph{Theoretical Models.}

Kerr frequency combs generation can be described  in terms of the ODEs such as (\ref{setz}), with the number of equations equal to the number of the comb lines. This approach was proposed in \cite{chembo_modal_2010} considering the theoretical model of  three-mode hyper-parametric oscillation \cite{matsko05hyper}. Numerical simulations of the multimode  Kerr frequency comb were advanced in \cite{Chembo10comb,chembo_modal_2010,chembo10ol} and then followed by other groups \cite{Matsko12comb,Savchenkov12overmoded,matsko12ol2,Savchenkov12comb,herr14np,hansson14oc}.

Another, equivalent to the ODE \cite{chembo13pra}, approach proposed for theoretical study of the Kerr frequency comb generation involves Lugiato-Lefever (LL) equation. The original LL equation \cite{lugiato87prl} has been adapted for the description of frequency combs. Similarly to the ODE set, the LL equation is valid at times longer than the ring-down time of the resonator. Solutions of this equation have been studied, and existence of the stationary mode-locked regimes of the Kerr frequency comb has been confirmed \cite{wabnitz93ol,blow84prl,ghidaglia88aihp,barashenkov96pre,afanasjev97pre}. The analyisis predicts that optical pulses can emerge from the nonlinear microresonator pumped with CW light, without any pulse seeding \cite{matsko09proc,matsko11ol,chembo13pra,herr14np,coen13ol1}.  The power loss of the pulses is compensated by their nonlinear interaction with the background within the resonator \cite{barashenkov96pre}.

The LL equation for the optical field inside a nonlinear resonator with constant second-order GVD reads   \cite{matsko11ol,chembo13pra}
\begin{eqnarray} \label{basic1}
 && \frac{\partial A}{\partial \tau} + \frac{i}{2\tau_0}\beta_{2\Sigma}\frac{\partial^2
A}{\partial t^2}  = \\ && ig|A|^2A  - \left [\gamma_{0c}+ \gamma_{0} +i (\omega_{j0}-\omega) \right ] A+
F_0, \nonumber
\end{eqnarray}
where $A(\tau,t)$ is the slowly varying envelope of the electric field, $\tau$ is the slow time, $t=t-z/v_g$ is the retarded time, $v_g$ is the group velocity, $\beta_{2\Sigma}=(2\omega_{j0}-\omega_{j0+1}-\omega_{j0-1})(\tau_0/\Omega^2)$ is the GVD parameter, $\Omega \simeq (\omega_{j0+1}-\omega_{j0-1})/2$ is the FSR of the resonator. By definition, time scale $\tau$ is much longer than the round trip time $\tau_0$. Eq.~(\ref{basic1}) can be used to find the time dependent amplitude of light exiting the resonator.% if $\hat E$ is replaced with $A(\tau,t)$.

The similarity between the LL and ODE approaches was shown in \cite{matsko09proc,chembo13pra}. An optimization of the numerical algorithm makes the computation time for the LL and ODE approaches also equivalent \cite{hansson14oc}. The possibility of the direct pulse generation in the resonator was demonstrated by the numerical solution of the ODE set \cite{Matsko12comb}. Analytical solutions of the LL equation describing the mode-locked Kerr comb regime were provided in \cite{coen13ol1,matsko13oe,herr14np} and technical as well as fundamental quantum noise associated with the repetition rate of the pulses was analyzed in \cite{matsko13oe,matsko15josab,matsko15pra}.

\paragraph{Impact of the modes anti-crossing.}

The first observed phase-locked Kerr frequency comb, obtained in an on-chip  fused silica cavity \cite{Del'Haye07comb}, was most likely impacted by the mode GVD modification due to the anti-crossings \cite{Savchenkov12overmoded}. A similar assumption can be made regarding other frequency comb observations \cite{ferdous11np,herr12np,Saha13oe}. The envelope shape of these frequency combs  is significantly different from that of a mode-locked (soliton) frequency comb. Modes anti-crossing results in an asymmetry of the frequency comb envelope. It can suppress or enhance generation of comb lines within some frequency bands. Introduction of the avoided mode crossings seems to be important to explain observation of the majority of Kerr frequency combs, and a further study is needed for a complete understanding of this process.

The first observation of the low repetition rate Kerr frequency combs in fluorite resonators was also based on modes crossings  \cite{Savchenkov08comb}. Initially a primary comb emerged, and then a broad frequency comb with irregular envelope was generated. The repetition rate of the primary comb was determined by the interaction between different families of the WGMs. Similar results were obtained with another fluorite resonator \cite{Savchenkov08oe}. A truly mode-locked frequency comb generated in a normal GVD resonator has a significantly different envelope \cite{Liang14ol}. A controlled anti-crossing of modes from two \textit{different} coupled resonators \cite{Miyazaki2000prb} can be used to adjust the repetition rate and achieve mode locking of a WGM comb \cite{Miller15comb,Xue15comb}.

\paragraph{Unsolved problems.}

There are many unsolved problems related to Kerr frequency combs. Low efficiency of the nonlinear process involving the mode-locked Kerr comb is one of them \cite{bao14ol}. The efficiency, defined as the ratio of the pump power and averaged pulse train power, degrades with growth of the comb spectral width, and is inversely proportional to the number of comb lines. A solution has to be developed to circumvent this restriction and to enable generation of the spectrally broad, high power Kerr frequency combs with a relatively low repetition rate. Other problems include generation of a \textit{coherent} octave-spanning Kerr frequency comb, self-stabilization of the comb, generation of frequency combs in visible and ultra violet parts of the optical spectrum (mid-IR frequency combs were recently demonstrated \cite{savchenkov15ol,lecaplain15arch}), as well as complete integration of the comb generator with the pump laser on a chip. The complete understanding of the coherence properties of a Kerr frequency comb is also yet to be achieved.

\subsection{Forced frequency combs}

It is possible to generate thresholdless frequency combs using not monochromatic, but bichromatic \cite{Strekalov09comb} or multi-chromatic \cite{Fulop15ol} pumping light. The repetition rate of such forced  combs is determined by the modulation frequency if the pump power is low enough. However, if the power exceeds a certain value, the comb becomes unstable. The major signature of this process is that the generated comb lines are no longer locked to the frequencies of the pumps \cite{hansson14pra}. The forced comb can also become chaotic.

The impact of the pump spectrum on the comb formation was studied under different conditions \cite{papp13oe,taheri15pj,Hu15ao,lobanov15arch,Lin15ao}. These include flat-topped dissipative solitonic pulse generation \cite{lobanov15arch} and parametric seeding for stabilization of a conventional Kerr frequency comb \cite{papp13oe,taheri15pj}. The parametric seeding can be used to improve the combs stability and achieve low frequency and phase noise.  It was also suggested that advantage may be taken of the injection locking properties of the Kerr frequency comb oscillator to create an opto-electronic oscillator that involves an active Kerr comb oscillator as a part of the optical loop \cite{ilchenko13ifcs}.

\subsection{Frequency dependent absorption in mode locking}

As we mentioned above, the mode-locked Kerr frequency combs can be generated in a WGM resonator with either anomalous or normal GVD. In the case of anomalous GVD the intracavity pulses corresponding to the combs are bright, and in the case of normal GVD -- dark. It was shown both theoretically and experimentally \cite{Huang15comb} that this situation can change if the resonator modes have wavelength dependent loss. Such loss can play a role of a built-in bandpass filter that mode locks the Kerr frequency comb.

In this experiment \cite{Huang15comb} a Si$_3$N$_4$ ring resonator with 115.6~GHz TE$_{11}$ and 111.2~GHz TE$_{21}$ FSRs was used. The group velocity dispersion of the fundamental mode family (TE$_{11}$) was normal, as was verified by coherent wavelength interferometry and numerical simulations. The $Q$-factors of the TE$_{21}$ mode family was an order of magnitude lower than of the fundamental TE$_{11}$ mode family ($1.2\times 10^5$ versus $1.1\times 10^6$).  The TE$_{21}$  mode family also had a larger mode volume. Other higher-order mode families had even lower $Q$-factors and larger mode volume. Therefore, these modes did not support efficient generation of Kerr frequency comb and also did not couple with the fundamental modes in a way that would sufficiently alter the GVD. Still, the resonator produced short (74~fs) bright pulses when pumped with CW light.

This result was explained by the presence of the wavelength-dependent attenuation in the resonator.  The intrinsic bandpass filtering imposed by the H-bond absorption of Si$_3$N$_4$ in the short wavelength range and the increased coupling loss in the long wavelength range served to achieve a clean and short bright pulses in spite of the globally normal GVD of the fundamental mode family. This is a different mechanism of mode locking of Kerr frequency comb compared to the conventional early experiments. A simple analytical model confirms this conclusion \cite{Huang15comb}.

\subsection{Third harmonic generation and up-conversion via four-wave mixing}

It is more difficult to achieve phase matching for the resonant THG than for the resonant hyper-parametric process because of chromatic dispersion of the resonator material. This dispersion usually is not critical in a hyper-parametric process when all the involved fields have nearly the same optical wavelengths. However the wavelengths involved in the THG process differ by a factor of three.
The consequent difference of the index of refraction can be compensated by the geometrical dispersion discussed in section \ref{sec:neff}, similarly to how it is done in optical microfibers \cite{akimov03apb,grubsky05oe,grubsky07oc,afshar09oe,wiedemann10oe,collet10josab,lee12oe}.

The pioneering WGM THG observations were carried out in various organic and inorganic liquid micro-droplets \cite{benner_observation_1980,Acker89,leach90ol,hill93josab,leach93josab,kasparian97prl}. Recently, efficient third harmonic generation was demonstrated in silica micro toroids \cite{Carmon07tripling}, Si$_3$N$_4$ micro rings \cite{Levy11SiNmicroring}, and in silica micro spheres \cite{Farnesi14THG}. All these resonators were pumped in a low-power CW regime at 1550-1560 nm  wavelength, producing visible (green) third harmonic. The phase matching in WGM resonators \cite{Carmon07tripling,Farnesi14THG} was achieved between the fundamental pump and higher-order third harmonic modes. Because of the small resonator size, the geometrical dispersion of these modes was sufficient to compensate for chromatic dispersion of the resonator material.

Observing the third harmonic emission from evanescently-coupled resonators is not trivial. Because of different evanescent field decay lengths, a significant output coupling of the third harmonic wavelength leads to strong overcoupling of the pump. This causes a loss of conversion efficiency, which scales as the cube of the loaded $Q$-factor for the pump \cite{Carmon07tripling}. For this reason surface scattering was observed instead of the waveguide output to confirm THG  in \cite{Carmon07tripling,Farnesi14THG}. Evidently, to make practical application of the third harmonic WGM sources one needs to solve a problem of the \textit{selective coupling}, such that allows the pump in-coupling rate to be adjusted independently from the signal out-coupling rate. This type of coupling based on polarization dispersion has been demonstrated in a monolithic cavity \cite{Fiedler93monolit_cavity}. It is also possible to use waveguides optimized for coupling at particular wavelengths. For example, a ``pulley" shaped waveguide bending around a short segment of a resonator may show strongly suppressed coupling at a desired wavelength \cite{Li16conv}.

Along with the third harmonic, series of discrete blue-shifted emission peaks arising from four-wave mixing of the input radiation and the stimulated Raman-scattered radiation were reported \cite{Acker89,Farnesi14THG}. A similar phenomenon was observed in a large, 7 mm in diameter, crystalline MgF$_2$ resonator \cite{Liang15OPO}, in addition to the hyper-parametric comb and stimulated Raman scattering. In a more controlled way, up-conversion via four wave mixing was studied by injecting two different pump wavelengths (1553 nm and 1674 nm) in a silica micro toroid \cite{Carmon07tripling}. Emission at 542 nm was observed, which corresponds to combining a 1553 nm photon with two 1647 nm photons. A similar experiment was perfromed in a doped high-index silica glass microring, combining a 1558.02 nm photon with two 1553.38 nm photons \cite{Ferrera08}.  We are not aware of any quantitative analysis of such processes in WGM resonators, although the ODEs describing this process in waveguides, see e.g. \cite{Roussev04}, can be relatively easily adapted for this purpose.

An added degree of freedom makes the four-wave mixing based up-conversion  more flexible compared to a direct third harmonic generation. One may expect this approach to allow for efficient conversion of quantum states from infrared to visible range, or from visible to ultraviolet range. As a motivation for this research, let us mention that a narrow-line, tunable and efficient ultraviolet source is highly desirable for compact and low-power spectroscopy applications. In section \ref{sec:qconvet} we will also see how this up-conversion technique may be useful in quantum optics applications.

\subsection{$\chi^{(2)}$-$\chi^{(3)}$ processes}

Generation of Kerr frequency combs in microresonators made of materials characterized with both quadratic and cubic nonlinearity is especially interesting since it allows observing both the second and third order nonlinear effects in the same microcavity.  In addition, the noncentrosymmetric materials characterized with nonzero $\chi^{(2)}$ usually demonstrate electro-optic effect, which enables simple electric manipulation of the Kerr frequency combs. Generation of the Kerr combs was demonstrated in aluminium nitride (AlN) microrings \cite{Jung13ol,Jung14o,Jung14ol}. However it was not achieved in crystalline quartz \cite{Ilchenko08qtz}, nor in lithium tantalate \cite{Savchenkov10VCO} resonators for reasons yet unknown.

AlN is a wide-band semiconductor that has significant second order nonlinearity and approximately two orders of magnitude larger thermal conductivity than Si$_3$N$_4$. It has both strong Kerr nonlinearity and electro-optic Pockels effect. The microring resonators with $Q$-factor approaching  $6\times10^5$ at $1.5\ \mu$m wavelength \cite{xiong12njp} allow for strong power enhancement, leading to Kerr frequency comb generation and cascaded frequency conversions in the visible range. Three frequency comb sets in IR, red and green were simultaneously generated in an AlN microring pumped by a single telecom IR pump laser \cite{Jung14o}. These combs arise from a combination of the three- and four-wave mixing in the same resonator. High-resolution spectroscopic study of the visible frequency lines indicates matched free spectrum range over all the bands. Therefore, the observed process simplifies self locking of the frequency combs. Electro-optic switching of the frequency comb also becomes feasible \cite{Jung14ol}.

\section{Other nonlinear processes}\label{sec:other}

\subsection{Raman and Brillouin scattering}\label{sec:RB}

Light scattering by various solid state excitations, such as optical and acoustic phonons and polaritons, leads to nonlinear-optical phenomena known as Raman and Brillouin scattering. Usually these excitations do not form modes with any specific spatial symmetry, and the corresponding processes do not require phase matching. However they still have to conserve energy, and furthermore can only occur for the optical frequencies supported by the resonator.

The threshold of a resonant Raman or Brillouin laser \cite{Spillane2002Raman,Grudinin09Brillouin} is
\begin{equation} \label{SRS_th}
P_{R,B} \simeq \frac{\pi^2 n_p n_s V_p}{G \lambda_p \lambda_s Q_p Q_s},
\end{equation}
where $\lambda_p=2 \pi c/\omega_p$ is the pump wavelength and $G$ is the Raman or Brillouin gain.
Here the oscillation threshold is proportional to the factor $V_p/Q_pQ_s$ just as for the
hyper-parametric oscillations (\ref{hp_th}). It means that these three
processes compete. While $P_{HP} \simeq P_{R}$ \cite{matsko05hyper}, the threshold of Brillouin
laser is usually much lower than the other two \cite{Grudinin09Brillouin}, and only phase mismatch can make
it less favorable in a microresonator.

Spontaneous and stimulated Raman scattering in liquid micro-droplets \cite{Snow85Raman,Qian1986dropletRaman,Lin1994droplets,Qian1995droplets,Sennaroglu07R_las,Kiraz09R_las} was among the earliest nonlinear-optical WGM experiments. Raman scattering was also observed in solid crystalline \cite{Grudinin07Raman,savchenkov07ringdown,Grudinin08Raman,Liang10RamanCmb} and amorphous \cite{Spillane2002Raman,Min03Raman,Kippenberg04Raman,Kippenberg04Raman1,Vanier13Raman} WGM resonators, as well as studied theoretically \cite{Jouravlev12theory}. The interest in this process is stimulated by a wide range of potential application for Raman lasers. In Table~\ref{tbl:Ramaneffcy} we summarize parameters and performance of such lasers implemented in solid WGM resonators.

\begin{table}[t]
\begin{tabular}{c c c c c c}
		\hline
 Ref.						& $\lambda_p\rightarrow\lambda_R$ nm	& Media 			& $P_{th}\,\mu$W   	& $\eta_m$ \%\\
		\hline
\cite{Grudinin07Raman}		&1064$\rightarrow$1102					&fluorite			&1				&24 \\
\cite{savchenkov07ringdown}	&1320$\rightarrow$1378					&fluorite			&1600				& \\
\cite{Grudinin08Raman}		&1064$\rightarrow$1102					&fluorite			&78				&65 \\
\cite{Liang10RamanCmb}		&1532$\rightarrow$1610					&fluorite			&300				&50 \\
\cite{Spillane2002Raman}	&1555$\rightarrow$1670					&silica				&62				&35 \\
\cite{Min03Raman}			&976$\rightarrow$1021					&silica				&56				&6.5 \\
\cite{Kippenberg04Raman}	&1550$\rightarrow$1660					&silica				&62				&45 \\
\cite{Kippenberg04Raman1}	&1550$\rightarrow$1660					&silica				&74				&45 \\
\cite{Vanier13Raman}		&1550$\rightarrow$1636					&As$_2$S$_3$		&13				&10.7 \\
\cite{Melnikau11j}			&485$\rightarrow$535					&J-aggr.		&				& \\
\cite{Li13coated}			&685$\rightarrow$850					&PDMS			&1300				& \\
\cite{Deka14Raman}			&765$\rightarrow$796					&silica:Ti		&52.6				& \\
\cite{Yang05laser}			&1561$\rightarrow$1679					&TEOS		&640				& \\
		\hline
\end{tabular}
\caption{Summary of reported Raman WGM lasers: pump and Stokes wavelengths in nanometers, Raman-gain media, threshold power $P_{th}$ in $\mu$W, and maximum observed pump power conversion $\eta_m$. In cases when multiple Raman lines are observed, the strongest conversion is shown.}
\label{tbl:Ramaneffcy}
\end{table}

An important feature of WGM Raman lasers is their high gain, which can be further enhanced by a special coating \cite{Yang05laser,Melnikau11j,Li13coated,Deka14Raman} or by making the resonator out of chalcogenide glass \cite{Vanier13Raman}. The high Raman gain allows for a cascaded scattering, when the lower-order Stokes signal serves as a pump for the higher-order process. In
\cite{Min03Raman} such cascaded Raman lasing up to the fifth order was demonstrated in a fused silica microsphere.  It was also demonstrated in a large fluorite resonator \cite{Grudinin07Raman}, where a record threshold as low as 1 $\mu$W was reported. The cascaded Raman scattering was furthermore shown to form a mode-locked optical comb \cite{Liang10RamanCmb}.
Cascaded \textit{hyper}-Raman conversion was demonstrated in a lithium niobate resonator \cite{Simons11Raman}. This is a higher-order process when instead of one, two pump photons are absorbed to generate a (higher-frequency) Stokes photon and to excite a phonon.

Distinctly from Raman scattering, generalized Brillouin scattering involves \textit{acoustic} phonons that typically have lower frequencies and narrower bandwidths, but larger wave numbers than the optical photons or polaritons.  Brillouin interactions are responsible for one of the strongest known optical nolinearities. However they are difficult to observe in microresonators whose large FSRs are incompatible with small frequency shifts arising from Brillouin scattering. In other words, it is difficult to find a pair of modes with close enough frequencies and yet vastly different local wave number (i.e., different effective index of refraction). In WGM resonators, two solutions to this problem are possible: i) backward scattering, or ii) forward scattering into a different mode family, e.g. accompanied by a large change of the radial mode number $q$.

Both these approaches have been explored.  In Table~\ref{tbl:Breffcy} we summarize the key parameters for backward \cite{Grudinin09Brillouin,Tomes09optomech,Li13Brillouin} and forward \cite{Lin14Brillouin,Bahl11optomech,Dong15Brillouin} scattering experiments in resonators ranging from approximately 100 $\mu$m \cite{Tomes09optomech,Bahl11optomech,Dong15Brillouin} to 5-6 mm \cite{Grudinin09Brillouin,Li13Brillouin} in diameter. Many of these experiments also show cascaded Brillouin scattering.

\begin{table}[htb]
\begin{tabular}{c c c c c c}
		\hline
 Ref.						& Material 	& $f_B$, GHz	& $\Delta f_B$, kHz	& $P_{th},\,\mu$W  \\
		\hline
\cite{Lee12etch}			&silica		& 10.8		&   				&50 \\	
\cite{Grudinin09Brillouin}	&fluorite	& 9-20		&					&3 \\
\cite{Tomes09optomech}		&silica		& 10-11		& 700 				&26 \\
\cite{Li13Brillouin}		&silica		& 21.7		& 0.01-0.2			& \\
\cite{Lin14Brillouin}		&BaF$_2$	& 8.2		&  27 				&7000 \\
\cite{Bahl11optomech}		&silica		& 0.06-1.4	&  0.58				&22.5 \\
\cite{Dong15Brillouin}		&silica		& 0.04		&  4 				& \\
\cite{Loh15Brillouin}		&silica		&   		&  0.1-3.3 			&100 \\
		\hline
\end{tabular}
\caption{Summary of reported Brillouin WGM lasers: resonator material, Brillouin frequency shift $f_B$, linewidth $\Delta f_B$, and pump threshold power $P_{th}$. The linewidth in \cite{Li13Brillouin} is reduced by a feedback loop; in \cite{Lin14Brillouin} it is measured at -20 dB level.  }
\label{tbl:Breffcy}
\end{table}

The backward Brillouin scattering reported in \cite{Grudinin09Brillouin,Tomes09optomech,Li13Brillouin} relies on bulk acoustic phonons. In contrast, the forward scattering relies on a surface acoustic wave which may form its own high-$Q$ WGM \cite{Matsko09saw,Savchenkov11saw,Zehnpfennig11optomech,Sturman15ac_wgm}. In this respect the forward Brillouin scattering is closely related to opto-mechanical processes. Its phase matching conditions are similar to (\ref{orb_sel_rules}).

The relatively low frequency of the surface acoustic waves allows to observe Brillouin scattering on the wing of the optical pump WGM line. Alternatively and more efficiently, an ``overmoded" resonator may be used \cite{Lin14Brillouin}. Such resonators have very high spectral density of modes and can potentially function as high-$Q$ white-light resonators \cite{Savchenkov06WhiteLight}, capable of supporting optical fields at practically any wavelength.

Brillouin scattering enables very narrow-line and low-noise, low-threshold lasers \cite{Loh15Brillouin}, which explains a rapidly growing interest in studying this process in high-$Q$ WGM resonators. These studies also lead to an exciting and very active topic of optomechanics. Perhaps the most remarkable aspect of this topic is the possibility to optically cool a mechanical oscillator to the quantum ground state \cite{Tomes11cooling,Bahl12cooling}. Finally, let us note that Brillouin scattering of light may occur not only on phonons but also on other solid-state excitations, such as magnons that can form their own WGMs \cite{Zhang15magnon,Osada16magnon,Haigh16mo}.

\subsection{Thermal nonlinearity}\label{sec:ThermalNonlin}

The thermo-optical frequency shift is a fundamental property of WGM resonators, important in many applications  \cite{ilchenko92lp,Savchenkov04kHz,Carmon04thermal_osc,fomin05josab}. As we mentioned  in section \ref{sec:ThermalTuning}, this property originates from the increase of the resonator temperature due to  absorption of the light confined in a resonator mode. The frequency shift results from thermorefraction and thermal expansion of the resonator as described by Eq.~(\ref{thermalshift}).
For a resonator with $Q=10^{10}$, one degree temperature change usually leads to the frequency shift exceeding the WGM resonance bandwidth by as much as five orders of magnitude \cite{Savchenkov04kHz}. Given extremely small modal volume, even a small absorbed optical energy may be sufficient to ``heat the WGM out of resonance". This effect is clearly nonlinear with respect to the optical power, and may be compared to SPM. An important distinction, however, is that unlike Kerr nonlinearity, thermal nonlinearity is very slow on the optical time scale.  Also note that this effect is solely due to heating of the mode volume by the optical power absorbed in the material; scattering does not lead to the thermo-optical effect.

In the simplest approximation the evolution of a single-mode thermo-optical system can be described by a pair of equations \cite{fomin05josab}
\begin{equation} \label{thermal1}
\dot E + E\left [ \gamma  +i \left ( \omega +
\delta  \right ) \right ] = F(t), \quad \dot \delta + \Gamma \delta = \Gamma \xi |E|^2,
\end{equation}
where $E$ is the complex amplitude of the field confined in the resonator mode, $\gamma$ is the mode bandwidth, $\omega = 2 \pi \nu$ is the unperturbed mode frequency, $\delta$ is the thermal frequency shift, $F(t)$ stands for the external pump, $\Gamma$ characterizes the thermal relaxation rate, and $\xi$ is the thermal nonlinearity coefficient.

\begin{figure}[b]
\vspace{-0.3in}
\includegraphics[angle=0,width=1.1\linewidth]{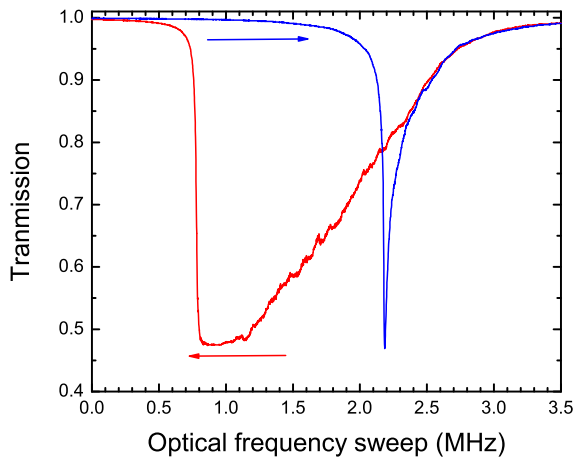}
\caption{A typical signature of the thermo-optical effect is the line shape dependence on the direction of the probe laser frequency sweep. In this case a 120 mW pump was swept across a WGM resonance of an $R=4$ mm, $Q=10^9$ magnesium fluoride resonator in 34 ms in directions indicated by the arrows. This measurement was performed as a part of experiment \cite{Strekalov11T} but not previously reported.}
\label{fig:bi}
\end{figure}

Equations (\ref{thermal1}) describe a variety of interesting behaviors. An immediate and most direct manifestation of the thermo-optical effect is the  non-Lorentzian asymmetric line shape of WGM resonances such as shown in Fig.~\ref{fig:bi}. Here the frequency of a strong probe laser is slowly scanned through a WGM resonance in two directions.  The observed line shape depends on the input optical power and the sweep speed, but most dramatically on the laser frequency sweep direction \cite{fomin05josab}. For a sufficiently high laser power, thermo-optical line distortion leads to bistability and hysteresis \cite{collot1993sphere,matsko05hyper}. Both these phenomena are clearly present  in Fig.~\ref{fig:bi}.

The speed of the probe laser sweep is important because a fast sweep does not lead to a significant heating of the resonator mode \cite{rokhsari05ol,Schmidt08oe}. If furthermore the scan period is much longer than the thermo-optical time constant $\Gamma^{-1}$ (which is typically of the order of a millisecond), the mode volume can efficiently exchange heat with the bulk of the resonator.  Therefore to ensure that the true shape of the resonance is observed, it is important to reduce the input power and sweep the laser frequency fast enough and with low duty cycle to minimize the pulling effect of the thermal nonlinearity.

Besides the bistability, numerical modeling of the systems described by equations (\ref{thermal1}) predicts the oscillatory instability \cite{fomin05josab}. This instability is commonly observed in high-$Q$ resonators, see e.g. \cite{fomin05josab,Savchenkov04kHz}, where it competes with soliton formation. Even in Fig.~\ref{fig:bi} we see a trace of this effect which appears as the noise ripples on the most gradual slope of the resonance curve. To minimize the thermo-optical effect one needs to reduce the absorption of light in the resonator or to adjust its temperature \cite{Weng15stab}. On the other hand, it is possible to find a regime when the interplay between the thermorefractive and Kerr effects leads to very strong thermo-optical relaxation oscillations \cite{Diallo15thermo}.

The thermo-optical effects enable the thermo-optical locking of a WGM frequency to the pump laser frequency \cite{Carmon04thermal_osc,Weng15stab}. This technique is used in cavity opto-mechanics, resonant nonlinear optics and microresonator lasers, thermo-optical cooling experiments, resonant optical sensors, and many other applications and systems. It relies on the frequency self-tuning of the resonator pumped with a CW light. Let us consider the normal thermo-optical effect corresponding to the decrease of the resonator mode frequency with the resonator temperature increase, and assume that the laser frequency is initially set at the blue slope of the WGM resonance, see Fig.~\ref{fig:bi}. The WGM frequency may slowly drift due to ambient temperature variations or other reasons. A blue drift would shift the mode towards the laser. Coupling, circulating power, and the consequent mode volume heating would increase, and the rising temperature will tune the mode frequency back to the red. Likewise, a red drift of the WGM frequency will cause the compensating blue tuning due to the temperature decrease. In this way the mode frequency can be locked to the frequency of the laser, and the temperature of the resonator can be stabilized. It is worth noting, that this locking strategy depends on the power stability of the laser. The locking point changes with the pump power.

There are numerous studies devoted to the thermo-optical effect. For some applications it would be desirable to reduce this effect. This can be achieved by dynamically varying the optical power during the sweep \cite{Grudinin11Tcompens}, or by using composite resonators. By precisely controlling the optical field overlap with the polymer film coating a silica resonator, an environmentally stable devices were demonstrated whose resonant frequencies were independent of the input power \cite{He08Tcomp,choi10apl}. Various polimer coatings reducing or enhancing the thermo-optical effect as well as the Kerr SPM effect were studied in \cite{Murzina12wgm}. Interplay between the negative thermo-optical effect, thermal expansion, and the Kerr effect may result in stable thermo-optomechanical oscillations, as was demonstrated with high-$Q$ ZBLAN WGM resonators \cite{Deng13ol}.

\subsection{Photorefraction}

The term ``photorefraction" describes the change of the refractive index resulting from light-mediated redistribution of charges within some optical materials.
There are multiple mechanisms of photorefractivity. In the most common one, absorption of a photon excites an electron from a donor site to the conduction band. The electron diffuses from the region of high light intensity to the regions of low light intensity and becomes trapped at an ionized deep donor site. Each photo-electron leaves behind a positive ionic charge.  Therefore a nonuniform space charge is created when the electron is trapped at a different site. This space charge generates a position-dependent electric field that changes the refractive index of the material via the electro-optic effect. The maximum photorefractive change of the extraordinary refractive index of the material (for instance, lithium niobate) arising from this mechanism is determined by the diffusion field $E_D$ \cite{Peithmann99photorefr}:
\begin{equation}
\Delta n_e = -\frac 12 n_e^3 r_{33} E_D.\label{n_photorefr}
\end{equation}
Here the electric field $E_D$ is linearly proportional to the light intensity at the beginning of the exposure. The saturated value of $E_D$ also depends on the intensity, but this dependence is not necessarily linear \cite{maleki14chap}. The maximum value of  $E_D$ is limited by the domain flipping.

Similarly to the SPM and thermal nonlinearity, photorefraction is a nonlinear optical phenomenon leading to the refractive index change in response to the optical field. However while the thermorefractive response is much slower than Kerr response, the photorefractive response is much slower than the thermorefractive response and also can be saturated.

The magnitude of photorefractivity depends on the energy of photons that induce the charge redistribution, one one hand, and on the dark current that reverces this process, on the other. The dark curent depends on the temperature, crystal composition, and other factors. Photorefractivity of lithium niobate, quite strong in the UV and visible parts of the spectrum, diminishes in the infrared and far infrared. High sensitivity of WGM spectra allows for measuring this effect with low-power infrared light. Photorefractivity was observed with a continuous wave 780 nm and 1550 nm light in WGM resonators made from lithium niobate (including the MgO-doped samples) as well as lithium tantalate   \cite{savchenkov06photorefraction,savchenkov06MgOLiNbO3,savchenkov07damage}. WGM resonators allowed to discover that not only light, but also low-power radio-frequency electromagnetic radiation can result in a significant modification of the refractive index of strontium barium niobate (SBN), one of widely used photorefractive materials \cite{savchenkov13spie}. The effects observed in SBN cannot be explained using existing theories of photorefractivity in bulk material.

The photorefractive changes observed in WGM resonators share the common features of other photorefractive experiments: i) the observed optical modification of the WGM spectrum does not disappear if the light is switched off; ii) the changes can be removed by illuminating the resonators with UV light; iii) the time scale of the observed effect is in the range of hours; and iv) the observed effects are more pronounced for shorter wavelengths of light but are present in the near infrared as well. Because of their high sensitivity, WGM experiments provide more insight into the long wavelength properties of the impurities of the photorefractive crystals than those with bulk samples. These observations support the point of view that the photorefractivity does not have a distinct red boundary in wavelength. This understanding is important for various  applications of lithium niobate resonators, and it points at a possible source of ``aging" of the telecom devices that use lithium niobate elements.

\section{From nonlinear to quantum optics}\label{sec:quant}

\subsection{General prerequisites for efficient quantum optics processes}

Nonlinear optics is a cornerstone in the generation and manipulation
of quantum states of light. The generation and processing of non-classical
states such as single-photon states, squeezed states or entangled
states requires efficient nonlinear interactions \cite{GerryKnight,bachor_book}. Both
second- and third-order nonlinear processes have been used to generate
non-classical light in WGM or ring resonators and will
be discussed below.
The requirements on these processes are more stringent than in the classical
applications because quantum states are very fragile.
%Any uncontrolled interaction with the environment
%such as loss, addition of noise or unknown coupling will lead to decoherence
%of the properties of the non-classical light.
There are three main concerns that make the generation of non-classical
light particularly demanding: loss, unwanted nonlinear processes,
and addition of noise.

Loss destroys the fragile quantum states. In the case of continuous-variable squeezed and entangled states it will lead
to a convolution of their Wigner functions with  that of a vacuum state. Hence the
quantum correlations will be reduced or lost. In discrete-variables measurements, loss will reduce counting and coincidence
rates. As we have mentioned in section \ref{sec:Intro_nonlin}, strong optical nonlinearity is often accompanied by a considerable
loss. When designing an efficient source for quantum states one has
to take special care of the balance between high nonlinearity and
loss involved.

Unwanted nonlinear processes are another source of degradation of
quantum states. These processes include stimulated Raman and Brillouin scattering,
photorefractive effects and nonlinear processes of a different order.
Second-order nonlinear interactions are much stronger than the third-order ones, and hence are more immune to such parasitic processes.

Even when all of the above concerns are resolved, the system can still
suffer from spontaneous scattering or from fluorescence. Spontaneous scattering (Raman, Brillouin,
and Rayleigh) can lead to additional noise that masks the features of the generated quantum states. Depending on the pump wavelength, fluorescence can also have an essential
impact. Therefore the resonator materials and the optical wavelengths have to be
chosen carefully. In very small resonators thermorefractive
noise can also have a considerable impact \cite{Gorodetsky2004lr}.

Let us first discuss the sources of non-classical light that use second-order nonlinear processes. The interaction part of the Hamiltonian
for three-wave mixing (\ref{Hint}) can lead to the variety of nonlinear processes introduced in section \ref{sec:chi2}. All these classical processes can be used to generate or process quantum states.

The most widely used process in quantum optics is SPDC. In terms of quantum optics this process can be interpreted
as a pump photon annihilation and two photons
in the signal and idler modes creation. A more detailed study shows
that these signal and idler beams are entangled
in their field variables (i.e. are in a continuous-variable entangled state), which leads
to correlations in photon number measurements and phase-sensitive measurements performed in the signal and idler beams. The dynamical behavior of SPDC depends on
the pump power. In resonators SPDC may lead to the OPO regime if the pump power exceeds the threshold (\ref{threshold}). Depending on the operation regime, different
quantum states can be generated. Very low pump power leads to the
regime of cavity assisted spontaneous parametric down conversion
\cite{Ou1999}. As a good approximation, such a system may be regarded as
emitting photon pairs. The efficiency and bandwidth of this process are
modified by the resonator. At higher pump power the generation
of multiple photon pairs has to be considered \cite{Christ2009lr}. Further
increasing the pump power to just below the threshold is widely used
to generate the \textit{squeezed vacuum} beams \cite{bachor_book,Wu86sqz}.
Above the threshold, the resulting OPO is known to emit bright beams which
are intensity-correlated (twin beams) \cite{Heidmann87sqz}.

In addition to varying the pump power, there is a possibility
to seed the system with light from an external source, leading to
an optical parametric amplification. This approach has been used to generate
squeezed states in conventional resonators as well as in bulk crystals.
Second harmonic generation is also known to generate squeezed states
of light, both in the up-converted \cite{Collett85sqz,Sizmann90SHGsqz,Kurz92SHsqz} and the pump frequency \cite{Drummond81SH,Pereira88SHGpumpSQZ}.
Not all of the above mentioned possibilites have
been tested with microresonators yet, although the trend of recent years shows that such applications may become common.

Third-order nonlinear processes can also generate non-classical states of the optical field. These processes are described by the
interaction part of the Hamiltonian (\ref{ham}). %Here the SPM and XPM processes do not require any special phase matching conditions and occur always, although with different efficiency among different sets of modes, as determined by their overlap. The hyper-parametric conversion term requires the orbital phase matching.
Note that with a classical pump field the hyper-parametric conversion term of this Hamiltonian is effectively reduced to the three-wave mixing
discussed above, leading to similar ODEs and similar quantum states that can be generated. For this reason in quantum optics papers the parametric and hyper-parametric processes are sometimes intermixed, disregarding the difference in the underlying optical nonlinearity.

Although the third-order processes are weaker than the second-order processes, they have some important advantages.
They can be realized in amorphous materials and materials that
are compatible with nano-manufacturing. They also do not usually lead to a very large wavelength difference between the pump and the non-classical light, which is typical in the second-order processes.  On the downside, in most of the quantum experiments involving third-order nonlinearities, scattering and parasitic nonlinear processes can have a severe impact on the purity of the states. Usually one has to carefully design the experimental parameters to minimize the detrimental contribution of these effects.

In the following we will investigate different experiments where microresonators
and micro-rings have been used to generate or process non-classical
states of light.

\subsection{Second-order processes above threshold: squeezing}\label{sec:2sqz}

One of the earliest experiments on producing quantum-correlated
beams of light used an optical parametric oscillator operating above the
threshold \cite{Heidmann87sqz}. The generated in this experiment beams of light were bright and quantum-correlated in intensity. The latter means that the intensity fluctuations of these two beams are correlated stronger than could be achieved by any classical modulation of coherent states originating from a laser.

As intensity correlations can be easily measured, this is a good test
for the generation of quantum states through a three-wave mixing process.
With this technique the first generation of squeezed light inside
a nonlinear crystalline whispering gallery resonator was demonstrated \cite{fuerst11sqz}.

\begin{figure}[b]
\includegraphics[width=8cm]{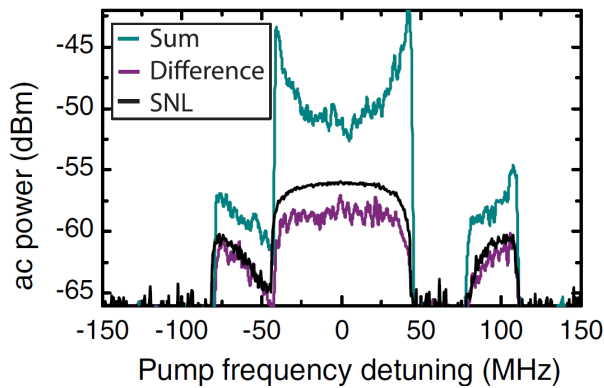}
\caption{\label{fig:twinbeams}Measurement of twin-beam intensity correlations. The sum and difference of the intensity noise of
the signal and idler beams of a WGM OPO are
plotted vs. the pump detuning from the WGM resonance. The difference noise falls below
the shot noise level (SNL), proving the quantum correlations. Reprinted from \cite{fuerst11sqz}.}
\end{figure}

The experiment \cite{fuerst11sqz} employed a lithium niobate WGMR that was pumped at
532 nm and operated as a non-degenerate OPO above the threshold. The Type
I phase matching was controlled via temperature and voltage tuning
as described in sections \ref{sec:ThermalTuning} and  \ref{sec:ElectricalTuning}. The signal, idler and pump wavelengths
were separated, and signal and idler beams were coupled into fast photodetectors.
Their photo currents were subtracted and added, and the resulting noise
power was measured at a suitable sideband frequency with an electronic
spectrum analyzer. A reduction of the difference photo current
below the shot noise level was observed, see Fig.~\ref{fig:twinbeams}.
%The shot noise level was calibrated by measuring the noise of a shot noise limited laser.

The two-mode photon-number squeezing, such as measured in the experiment \cite{fuerst11sqz}, does not depend on the pump power. The interaction Hamiltonian (\ref{Hint}) always induces correlation between
signal and idler photon numbers by generating them strictly in pairs. Other parameters however do significantly
alter the measured squeezing. Since the conversion only occurs to the light
inside the resonance bandwidth, the squeezing depends
on the ratio between measurement sideband frequency and this bandwidth.
Best squeezing is achieved at low frequencies that are well inside the total resonator linewidth. The observed squeezing also depends on the ratio between the parametric light out-coupling rate
and total resonator linewidth, that is, on the role of dissipative loss inside the resonator relative to the  out-coupling. Therefore one should try to employ high-$Q$ resonators
where the coupling rate can easily dominate the internal loss rate, i.e., the resonator can be strongly over-coupled.
As we will see later, this optimization is also beneficial for
generation of other quantum states.

A very low pump power threshold of the WGM OPOs helped to directly investigate the
phenomena that previously were outside the experimental capabilities.
It had been known that far above the pump threshold, in addition to the
quantum correlations in intensity of the twin beams, \textit{each} signal and idler
beam is also amplitude squeezed \cite{Collett85sqz,Reid88sqz,Fabre89opo}. However this phenomenon eluded a direct observation with conventional OPOs because the combination of the required high pump power and accessible linewidths caused relaxation oscillations \cite{Lee98opo,Porzio01opo,Zhang04sqz} hiding the effect. The OPO inside the WGMR made it possible to measure this behavior directly \cite{fuerst11sqz}.

\subsection{Second-order processes below threshold: photon pairs generation}

\begin{figure}[thb]
\vspace{-0ex}
\begin{center}
\includegraphics[clip,angle=0,width=\linewidth]{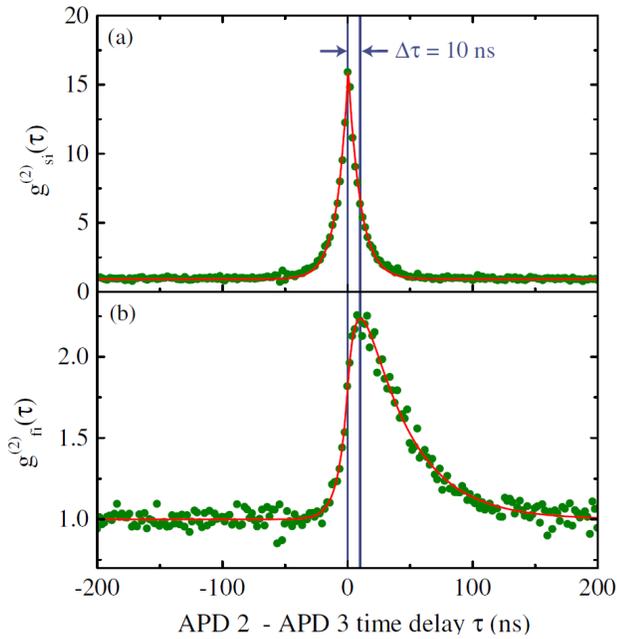}
\caption{A two-photon correlation function of the SPDC light emitted from a WGM resonator (a) is modified by re-emission from an atomic transition (b). The new correlation function is fit to a model combining both the resonator and the atomic transition ring-down times. Reprinted from \cite{Schunk15CsRb}.}
\label{fig:Cscorr}
\end{center}
\end{figure}

OPO pumped below the threshold can be used to generate photon pairs. More accurately,  the
generated quantum state can be described, to a good approximation, as
a superposition of vacuum states and a pair of single photons in the signal
and idler beams. This cavity-assisted \cite{Ou1999} SPDC is a
very efficient tool in optical quantum information processing.
The consequences of enhancing SPDC with a cavity are manifold. One is that the required pump power is dramatically
reduced, especially within the triply-resonant WGM systems. In the experiment \cite{Fortsch13NC}
the required pump power was in the range of 100~nW. Furthermore, the
bandwidths of the generated photon pairs are governed by the bandwidths
of their respective WGMs. In the
case of lithium niobate WGM resonators this may range from a few to a few hundred MHz in the visible and near-infrared wavelength range. Such bandwidths together with the continuous wavelength tuning capability allow for efficient coupling of SPDC photon pairs to various
atomic transitions. This coupling was used to perform single-photon time-resolved spectroscopy of cesium D1 transition, allowing to directly probe the selected transition lifetime \cite{Schunk15CsRb}. Fig.~\ref{fig:Cscorr} shows how the two-photon correlation function of the parametric light emitted from the resonator is transformed by the atomic re-emission process, which includes approximately a 10 ns peak delay and asymmetric broadening of the correlation function. The solid line is a theoretical model describing the resonator ring-down for panel (a) and its convolution with the atomic ring-down for panel (b).

Photon pair sources can also be used as heralded single photon sources
with e.g. detection of an idler photon heralding the arrival of a signal photon \cite{Fortsch13NC}. Photon heralding is a key to many quantum information protocols based on conditional measurements, in particular in linear quantum computing.
Flexibility in the signal and idler wavelengths available with the WGM approach can be leveraged to optimize the quantum efficiency of the heralding detection.

The resonator defines the electromagnetic modes in which
signal and idler are generated. It has been found that the phase
matching conditions discussed earlier in section \ref{sec:chi2} can be selective enough to constrain the photon pair source to strictly a single
pair of modes for the signal and idler \cite{Fortsch15sm}. Single-mode operation is important in quantum information
processing when photon pairs from different sources have to interfere, or when the mode-selective measurements such as
homodyne detection are required. Furthermore, as we have mentioned in section \ref{sec:dynam}, near degeneracy
the parametric down conversion in the WGM resonators leads to a comb structure of pair-wise quantum-correlated signal and idler modes. Such combs may be used for creating multipartite entangled states, similarly to Fabry-Perot or bow-tie resonators \cite{Hage10multipartite,Pysher11ent_comb}. Multipartite entanglement is a resource highly desired in many quantum information applications, e.g. in linear quantum computing.

Another important benefit of using a WGM resonator as a photon pair source is the
tunability of its central wavelength and bandwidth. Continuous tuning of these parameters was demonstrated and its significance was discussed in \cite{Fortsch13NC,Schunk15CsRb}.

\subsection{Third-order processes for generation of non-classical light }\label{sec:Kerrsqz}

The $\chi^{(3)}$ hyperparametric processes have been widely used for
generation of squeezed light and correlated photon pairs in nonlinear optical
fibers, see e.g. the recent review \cite{Andersen16sqz_rev} and references therein. With this process one can
generate photon pairs in the materials compatible
with on-chip integration techniques. This provides an opportunity to use microfabricated high-$Q$ resonators that can be easily  overcoupled to achieve a large heralding efficiency of single photons.

Recently there has been a number of experimental demonstrations of photon pairs generated via four wave
mixing in silicon on-chip fabricated microring and microdisk resonators  \cite{Clemmen:2009dn,Azzini:2012rc,Engin:2013qd,Guo:2014db,Grassani:2015zl,Wakabayashi:2015lq,Suo:2015fv,Rogers15pairs,Jiang15pairs}. All these devices operate in the telecom wavelength range and required very low (sub- to ten milliwatts) pump power, providing high pair-production rate within a narrow linewidth. In \cite{Grassani:2015zl,Wakabayashi:2015lq}  time-energy entanglement of the generated photon pairs was demonstrated by violating Bell's inequality, and in \cite{Suo:2015fv} time-energy and polarization hyper-entanglement was achieved, also verified by Bell's inequality violation.

Two-mode squeezing among multiple pairs of modes was demonstrated in a microfabricated  Si$_3$N$_4$ ring \cite{Dutt15on-chip_sqz}. In this experiment the resonator FSR was large enough to select a single pair of squeezed modes by spectral filtering. These modes were found to be squeezed at the level of 1.7 dB, or 5 dB when corrected for losses.
Broadband quadrature squeezing based on SPM in the same kind of resonator has been theoretically predicted \cite{Hoff15sqz}.

All on-chip devices described in this section are integrated with the coupling waveguides using on-chip processing technology. The efficiency of the overall on-chip device is tied to technological progress in
surface smoothness and overall loss in the photonic guidance.

\subsection{Quantum-coherent frequency converters}\label{sec:qconvet}

Converting a quantum state of an optical mode from one frequency to another is an important problem in quantum information processing and communications. This problem usually arises when the optical frequency of a quantum processing node (e.g., a trapped atom, ion, color center or an opto-mechanical system), defined by the system's physical nature, is poorly compatible with the communication or distribution channel, e.g. an optical fiber, quantum light source, or a high-efficiency and low-noise detector. To overcome these problems, SFG was proposed \cite{Kumar90SFG} and realized for frequency-conversion of squeezed states \cite{Huang92upconvert} and single photons \cite{Albota04,Langrock05converion} in PPLN waveguides. With strong enough pump, the conversion efficiency was shown to reach 46\%  in an open wave guide \cite{Langrock05converion}, and over 90\% in a cavity \cite{Albota04}. Non-degenerate four-wave mixing in a photonic cystal fiber used for a similar purpose has reached 29\% single-photon conversion efficiency \cite{McGuinness10}. This process involves two classical pump fields at two different frequencies. Annihilating a photon at one frequency while generating one at the other shifts a signal photon frequency by the difference of the pump frequencies. Such conversion was also demonstrated in an on-chip Si$_3$N$_4$ microring ($R=40\,\mu$m), in which case the conversion efficiency reached over 60\% \cite{Li16conv}.

A practical quantum-coherent frequency converter needs to have nearly unity efficiency, and also be free of spontaneous conversion processes and of other noise and decoherence mechanisms.
The unity efficiency requirement is very difficult to achieve in nonlinear optics with weak fields. On the other hand, using a high-power optical pump as in the experiments mentioned earlier may lead to excessive noise.  In this situation, the extremely strong enhancement of the nonlinear processes rate and prolonged interaction time offered by WGM resonators may provide a solution.
For example, the four-wave mixing up-conversion approach has been suggested \cite{Huang13} and demonstrated \cite{Li16conv}.  The second-order SFG \cite{Strekalov14SFG,Strekalov14switch} can also be used for this purpose.

A special case of quantum-coherent frequency conversion is up-conversion of microwave photons into optical domain. This case is important because many quantum systems proposed for quantum-logic implementation operate at such frequencies \cite{Schoelkopf08,Devoret13rev}. It is also important for sub-mm wave astronomy \cite{Leisawitz04}, where individual microwave photons detection may be desirable. Performing such detection efficiently and with low noise is very difficult due to low photon energy, while it presents a much lesser problem in the optical domain. A microwave-to-optics coherent converter can be realized as a unity-efficient, anti-Stokes singe sideband electro-optical modulator, such as we discussed in section \ref{sec:EOM}.

\subsection{Direct coupling of WGM light with quantum systems}

WGM resonators can facilitate strong interaction of their optical fields with quantum systems such as quantum dots, atoms, color centers, or condensed matter exitations such as phonons or plasmons.  The enhancement factor of spontaneous emission rate of a dipole interacting with a resonator mode is \cite{Faraon11diamond}
\begin{equation} \label{purcell}
F_p=\frac{3}{4\pi^2} \frac{Q\lambda^3}{V_pn^3},
\end{equation}
where the mode volume $V_p$ is defined slightly differently from the previously defined $V$:
\begin{equation} \label{vpurcell}
V_p=\frac{\int_V n^2({\vec r}) | E({\vec r})|^2 dV}{{\rm max} [n^2({\vec r}) | E({\vec r})|^2]}.
\end{equation}
In (\ref{vpurcell}) $n({\vec r})$ is the spatial distribution of the refractive index, and $|E({\vec r})|$ is the electric field amplitude.

Eq.~(\ref{purcell}) shows that the spontaneous emission rate can be
enhanced by factor proportional to $Q/V_p$. The enhancement scales linearly with $Q$ because only one mode of a {\em linear} optical resonator is used. It is worth noting, though, that
enhancement of stimulated Rayleigh scattering scales as $Q^2/V_p$, even though this scattering
is linear in the sense that it does not depend on the power of the scattered wave
\cite{gorodetsky2000}.

In order to reach the strong coupling regime, both high quality factor and small mode volume are required at the same time, see (\ref{purcell}), which means a high finesse. WGMs fulfill both requirements allowing for demonstration of strong coupling of the light field to various quantum systems. Optimization of resonator parameters for cavity QED applications is discussed in more detail in e.g.  \cite{Buck2003CQED,Spillane05CQED}. Theoretical analyis and experimental investigations have been performed with WGM coupled to atoms \cite{Mabuchi1994WGMatoms,Vernooy98QED,Aoki06atom,Dayan08turnstile,Aoki09switch,Junge13,Shomroni14switch,Rosenblum15single}, molecules \cite{Norris97molecule}, quantum dots \cite{Michler2000qdot1,fan00ol,Peter05qdot,Srinivasan07qdot} and  nitrogen vacancy (NV) centers in diamonds \cite{park06nl,Schietinger08nv,Larsson09nv,Li11nv,Faraon11diamond,Chen12Qgate}.

Alkali atoms were the first quantum systems coupled with the evanescent field of WGM resonators. In 1994, Mabuchi and Kimble proposed a scheme for trapping atoms in the evanescent field of a microsphere using dipole forces of two  WGMs blue- and red-detuned from an atomic transition \cite{Mabuchi1994WGMatoms}. Conveniently, the wavelength scaling of the evanescent fields in this case creates a repulsive potential near the resonator boundary and an attractive potential farther out. Best to our knowledge, this kind of atom trapping near WGM resonators has not been realized yet. But four years later the same group at Caltech realized a simpler experiment, demonstrating velocity-selective interactions between WGM field of a fused silica microsphere ($Q=5\times 10^7,\;V=10^{-8}\;{\rm cm}^3$) and cesium atomic vapor \cite{Vernooy98QED}. More recently the Caltech group repeated this experiment with a microtoroidal resonators and cold cesium atoms that were released from a magneto-optical trap to fall on the chip supporting the microtoroids \cite{Aoki06atom}. A strong coupling regime between an individual atom and the WGM field was reached in this experiment.

A similar technique was later used with cold rubidium atoms and a bottle WGM resonator by Junge \textit{et al.} \cite{Junge13}. In this experiment the focus was made on the study of polarization properties of the evanescent WGM field. Such fields can have a strong longitudinal polarization component, which fundamentally affects the interaction between light and atoms. This component was detected, and its effects studied by coupling the TE and TM WGM modes to different Zeeman levels of the $F=3\rightarrow F^\prime=4$ transition in rubidium.

Coupling a single atom to a microtoroidal resonator, Dayan \textit{et al.} \cite{Dayan08turnstile,Aoki09switch} have converted Poissonian laser light to single-photon state manifested by strongly sub-Poissonian statistics, thereby realizing a photon turnstile device. Later a similar device using an optical microsphere coupled with a single rubidium atom was shown to function as an all-optical switch, almost reaching the single-photon switching limit  \cite{Shomroni14switch}. Specifically, a transition from the highly reflective (R = 65\%) to the highly transmitting  (T = 90\%) state was achieved with an average of approximately 1.5 (3 with the linear losses included) control photons per switching event. A variation of this technique has allowed for deterministic single-photon subtraction from an optical pulse \cite{Rosenblum15single}. Remarkably, this photon is not absorbed but re-routed into a different optical channel and can be used e.g. for heralding of the modified pulse arrival.

Fluorescence of a single organic molecule attached to a WGM microsphere and excited by its near field was directly imaged in \cite{Norris97molecule}. This experiment required cryogenic temperatures to suppress inhomogeneous broadening of the molecular spectrum. Similar coupling of a single GaAs \cite{Peter05qdot} or InAs \cite{Michler2000qdot1,Srinivasan07qdot} quantum dots also required low temperatures, and has lead to the strong coupling regime. This approach has enabled a pulsed single-photon source with strongly suppressed high photon number amplitudes, characterized by the Glauber correlation function reaching as low as $g^{(2)}(0)=0.16$ \cite{Ates13qdot}.

In contrast with molecules and quantum dots, NV centers in diamond provide the most
stable quantum emitters at room temperature. Most of the experiments coupling NV centers to WGM and similar resonators aimed at the increase of the emission rate due to Purcell factor (\ref{purcell}) and achieving discrete emission spectra. Quantum properties of this emission have been demonstrated as well. For example in \cite{Schietinger08nv}  emission from a single NV center in a diamond nanocrystal was coupled to a tiny (4.84 $\mu$m in diameter) polystyrene microspherical WGM resonator. The non-classical character of the single quantum emitter light was verified by Glauber correlation function measurement $g^{(2)}(0)<0.5$, and the coupling to the WGMs by a discrete spectrum of the emission. Coupling NV centers to WGM resonators has been proposed for quantum information transfer \cite{Li11nv}, decoherence-free quantum gate implementation \cite{Chen12Qgate}, and  generating W-states \cite{Jin14W}.

\subsection{Quantum Zeno effect.}\label{sec:QZB}

The Quantum Zeno effect \cite{Misra77Zeno} arises when frequent state projections, usually realized as a series of measurements, inhibit continuous evolution of a quantum system. To understand this effect it is instructive to consider the following example from classical optics. A series of half-wave plates inserted in an optical beam can be aligned so as to rotate the beam polarization each by a small angle, adding up to a large angle. If however each plate is followed by a polarizer set to transmit the original polarization and to absorb the other, the final polarization obviously will not change.  Moreover, in the limit of very small rotation angle steps, the total optical loss in this system will asymptotically vanish.

\begin{figure}[b]
\includegraphics[width=8cm]{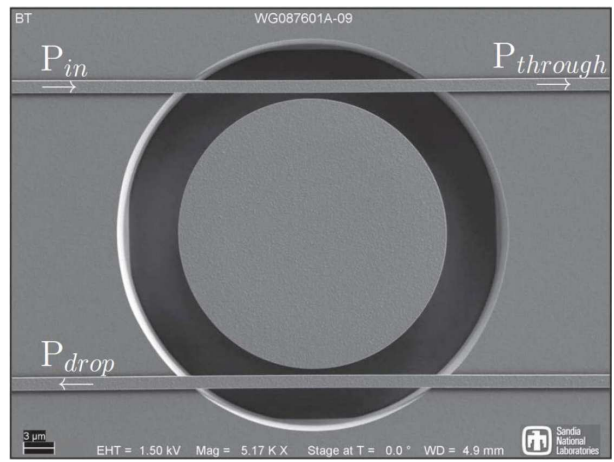}
\caption[]{\label{fig:SiNswitch}A Si$_3$N$_4$ microdisk used in the all-optical switch based on two-photon absorption in surrounding rubidium vapor. Reprinted from \cite{Hendrickson13switch}.}
\end{figure}

Quantum Zeno effect leads to a very interesting concept of interaction-free manipulation of a system's  evolution. This means that e.g. a photon can be affected not by a measurement (which would annihilate it), but by a \textit{possibility} of the measurement. From the practical perspective, this approach may be useful for manipulating fragile quantum states, e.g. in quantum logic implementations \cite{Franson04QZB,Franson07WGM_QZB,Huang10QZB}. Such applications have been proposed based on electromagnetically induced transparency of  WGM resonator's evanescent field in surrounding atomic vapor \cite{Clader13switch}, or its
two-photon absorption \cite{Hendrickson13switch}, see Fig.~\ref{fig:SiNswitch}. Either of these processes can be regarded as a ``measurement" sensitive to the number of photons in the optical mode. Even though there is no observer present to gain knowledge from such a ``measurement", it still enables quantum Zeno effect.

Not only a dissipative, but also a reversible Hermitian process can, under certain conditions, serve as a ``measurement" in quantum Zeno effect. Difference- and sum-frequency generation have been considered for this purpose \cite{Huang10switch_prop,Sun13switch,Strekalov14switch}. In this context the term ``Quantum Zeno Blockade" (QZB) is frequently used, implying that the presence of a control photon in a cavity prevents, via strong photon-photon interaction, the signal photon from entering the cavity. The exact mechanism of the QZB depends on the modal structure and decay time of the mediating (difference- or sum-frequency) field. It can range from  incoherent to fully coherent QZB \cite{Huang10QZB,Sun13switch}. A simplistic explanation of the incoherent QZB is the cross-phase modulation of the signal photon by the control photon. When this phase is shifted by $\pi$, interference of the amplitudes for the signal photon reflecting from the cavity and exiting it becomes constructive instead of destructive, and switching between the ``drop" and the ``through" ports (see Fig.~\ref{fig:SiNswitch}) occurs. The coherent QZB, on the other hand, can be explained in terms of Autler-Townes splitting \cite{Huang10switch_prop} which leads to equivalent switching operation.

A QZB single-photon switch is the quantum limit of the earlier mentioned all-optical switch. It requires, in the SFG-based version, that even a single photon should saturates the SFG process yielding near-zero SFG conversion efficiency, i.e. operates on the far-right side of Fig.~\ref{fig:SFGeffcy}. Furthermore, operation of the single-photon optical switch depends on deterministic coupling of a single photon into a resonator, which requires special pulse shaping \cite{Sun13switch}.

\begin{figure}[t]
\vspace*{-0.15in}
\includegraphics[width=8cm]{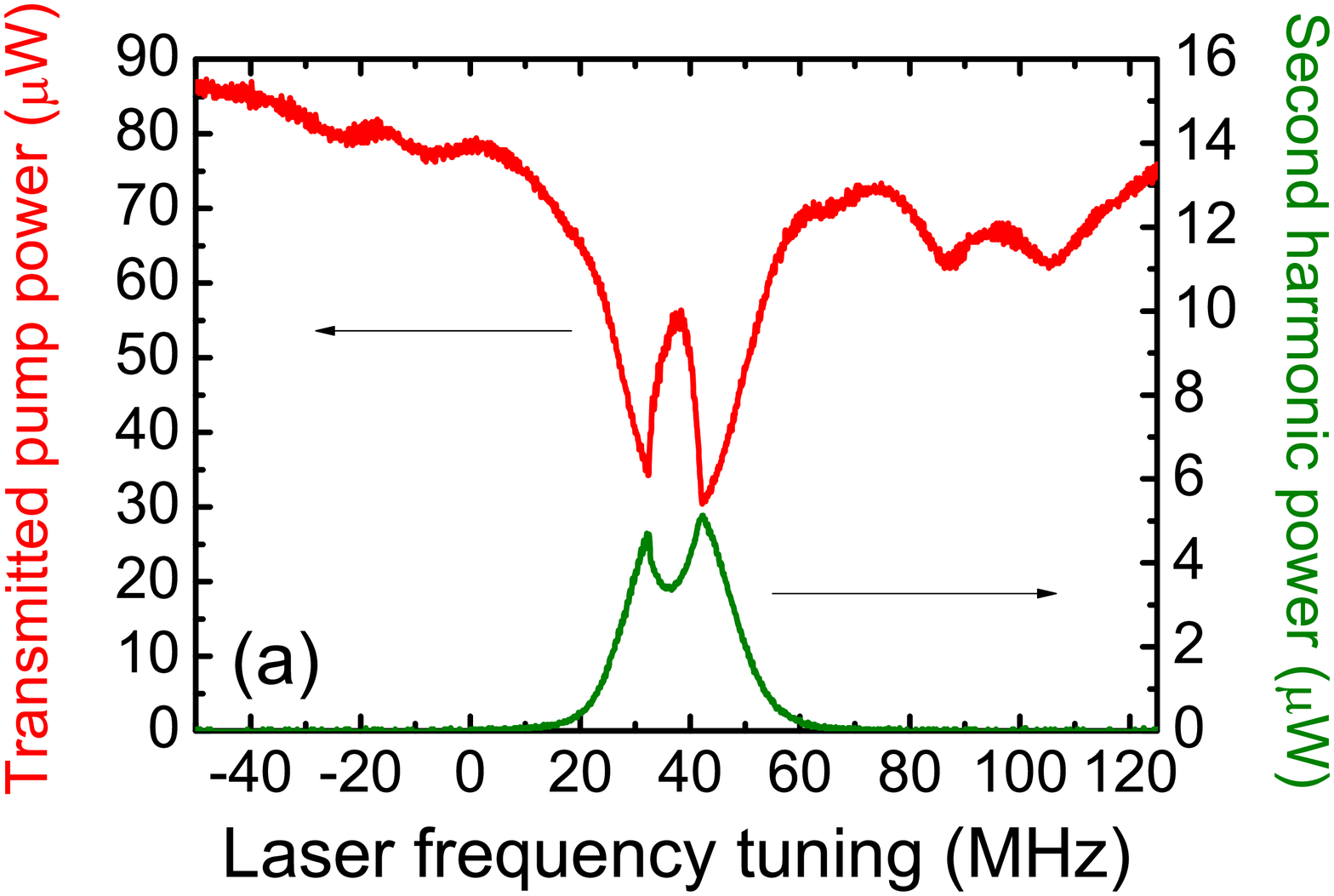}
\includegraphics[width=8cm]{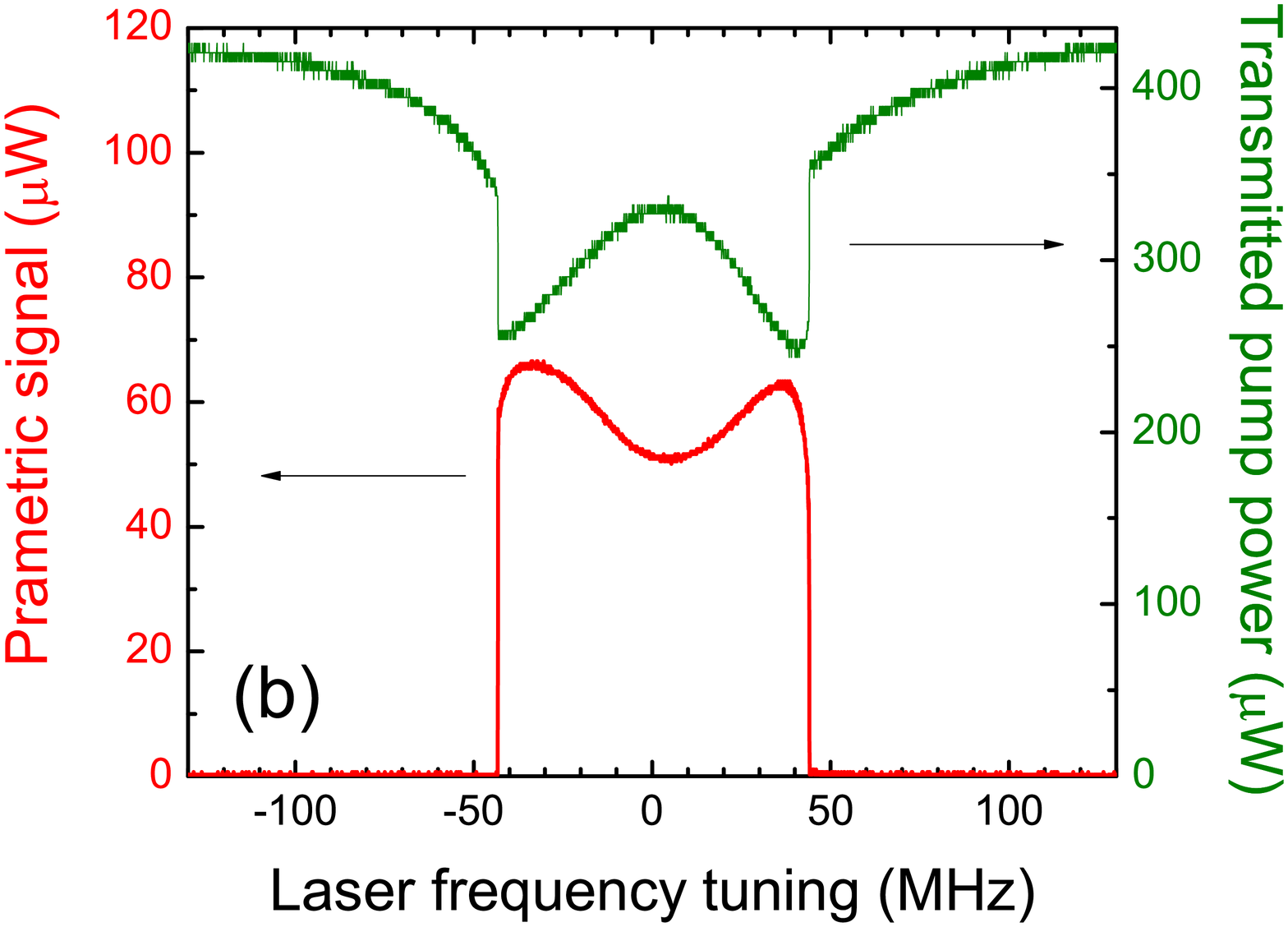}
\caption[]{\label{fig:nonlinear}Distortion of the transmitted pump and emitted SH (a) and SPDC (b) resonances indicates the presence a dynamical processes effectively decoupling the pump from the resonator and impeding the conversion efficiency. Plot (a) is reprinted from \cite{Fuerst10SH}, plot (b) shows previously unreported result of the experiment described in \cite{Fuerst10PDC}.}
\end{figure}

The single-photon optical switches have not yet been demonstrated. In the next section we will discuss their feasibility with the best available resonators. Let us point out here that the onset of Zeno blockade-like behavior can be observed in the all-optical switches operating in the low-power but still classical regime. In the WGM SFG-based switch \cite{Strekalov14switch}, counter intuitively, the SFG emission \textit{decreases} for higher control power, which means that the switching loss for the signal beam is reduced. Let us also recall the pump resonance distortion observed in efficient SHG \cite{Fuerst10SH}, SPDC \cite{Fuerst10PDC,Breunig13pump_res_struct} and SFG processes \cite{Strekalov14SFG} which may be due to a similar effect of self-decoupling the resonator from the pump (possibly aided by formation of a comb in the case of SHG), see Fig.~\ref{fig:nonlinear}.

Quantum Zeno blockade can not only facilitate the quantum gates functionality, but also modify the statistics of a mode occupancy by photons.  In \cite{Huang12antibunch} antibunched emission of photon pairs is predicted to occur via spontaneous hyper-parametric conversion in a microcavity coupled to rubidium vapor. Here the emission of multiple pairs will be suppressed due to the strong two-photon absorption in the vapor. Similarly, antibunched pairs can be expected to emerge from an SPDC process when the signal or idler is simultaneously phase matched for the SHG \cite{Strekalov15jmo}.

\subsection{Feasibility of strong coupling in the single-photon limit and of quantum logic with photons}\label{sec:Feasibility}

In the idealized case of just two WGMs coupled by the SHG process in a lossless resonator, the first nontrivial solution of the Schr\"odinger equation generated by Hamiltonian (\ref{Hint}) is
\begin{equation}
|\Psi\rangle=A(t)|2\rangle_p|0\rangle_{s}+B(t)|0\rangle_p|1\rangle_{s},
\label{psis}
\end{equation}
where the indices $p$ and $s$ stand for the pump and second harmonic, respectively. The quantum-mechanical amplitudes $A$ and $B$ oscillate as
\begin{equation}
A(t)=\sin(\sqrt{2}gt+\phi),\quad B(t)=\cos(\sqrt{2}gt+\phi),
\label{ab}
\end{equation}
and the phase $\phi$ is determined by the initial conditions.

The meaning of this solution is that starting at some point of time with a photon pair at the
fundamental pump frequency ($|A|^2=1,\, B=0$), we expect this pair to up-convert to a single photon at double frequency with certainty ($A=0,\, |B|^2=1$)
after $\Delta t = \pi/(2\sqrt{2}g)$, then return to the initial state after $2\Delta t$, and so on \textit{ad infinitum}.

In a resonator with a finite decay rate $\gamma$
this oscillation eventually decays to the ground state ($A=B=0$). In a realistic nonlinear resonator the oscillation period $2\Delta t$ is longer than the resonator ring-down time $1/\gamma$. Then the probability
of up-conversion of a photon pair, or of down-conversion of a double-frequency photon,  is
\begin{equation}
p_{s\leftrightarrow p}=\sin^2(\sqrt{2}g/\gamma)=\sin^2(\sqrt{2}gQ/\omega).
\label{p}
\end{equation}
It is easy to find that for a millimeter diameter lithium niobate resonator with $Q=4\cdot 10^8$  and
$\lambda_p = 1.5\ \mu$m, $p_{1\leftrightarrow 2}\approx 0.1$. This is a very impressive probability for interaction of individual photons, and it raises certain optimism regarding quantum-optical applications of nonlinear WGM resonators.

As we have seen in section \ref{sec:dynam}, a two-mode model for SHG is an oversimplification, and a more elaborate analysis is required. However the main message of the above example holds: the feasibility of single-photon all-optical switches and photonic quantum logic gates hinges on reaching the strongly coupled regime with single photons, when the nonlinear coupling rate defined in (\ref{Omega}) exceeds the resonator decay rate,  $g>\gamma$.

Let us turn to the discussion of an SFG-based QZB switch in \cite{Sun13switch}. Here equations of motion are derived for the signal, pump, and sum-frequency field operators inside and outside of the resonator using a Hamiltonian whose interaction part is equivalent to (\ref{Hint}). The resonator is assumed to be strongly over-coupled, which means that the cavity decay is dominated by the outcoupling while the absorption and scattering losses are negligible. The initial conditions for these equations of motion are set by the external signal and pump pulse shape, which may be Gaussian or the ringdown-matching exponential \cite{Cirac97}.

Interaction inside the resonator entangles the signal and pump states. The output joint signal-pump wave function can be decomposed into Schmidt modes, which allows to define the gate fidelity as the overlap of the input signal pulse with its first Schmidt mode at the output. It is shown that in the absence of the pump, the resonator-matching signal pulse has a very high fidelity, if one takes into account the time reversal and sign change of the output pulse\cite{Cirac97}. Moreover, in the strongly coupled regime in presence of the pump pulse, signal fidelity is also high. In this case the sign change and time reversal do not occur.

The analysis \cite{Sun13switch} predicts that fidelity of 0.99 can be reached in a lithium niobate resonator with $Q=10^8$ for $g>400$ MHz, which requires the resonator radius $R<25\,\mu$m.
The required combination of $Q$ and $R$ does not appear entirely unrealistic, at least in theory. But we have to note that the expression for the nonlinear coupling rate $\Upsilon\equiv g$ provided in the Supplementary material to \cite{Sun13switch} is equivalent to the incorrect expression for the coupling rate $g$ in \cite{Strekalov14SFG}, and may be also incorrect. The correct expression has an extra factor $(n_sn_pn_f)^{-1}$ \cite{Strekalov14SFGerr}. Therefore the estimates for $g$ provided in \cite{Sun13switch} may be unrealistically high, and the benchmark fidelity would in fact be much harder to reach.

The dependence of the coupling rate $\Upsilon$ on the resonator radius $R$ computed in  \cite{Sun13switch} is approximately captured by the expression $\log_{10}\Upsilon\approx 0.073(\log_{10}R)^2-0.77\log_{10}R+1.74$. If we limit the fitting function by the linear term of $\log_{10}R$, the approximation would be $\log_{10}\Upsilon\approx -0.86\log_{10}R+1.73$, which is again consistent with the ``magic" scaling of the overlap integral $\Upsilon\propto|\sigma|\propto R^{-0.9}$ discussed in section \ref{sec:Natural}.

Concluding this section, we need to mention a specific concern regarding using single-photon XPM for realization of quantum logic operations. Such operation would require very strong Kerr interaction that is not presently achieved. However, even if it is achieved, it has been shown \cite{Shapiro06KerrSucks,Dove14XPM} that delayed $\chi^{(3)}$ response of realistic Kerr media (which is a direct consequence of causality) will induce enough phase noise to make high-fidelity operation of a quantum logic gate impossible.

\section{Summary and conclusion}\label{sec:Summary}

WGM and ring resonators found applications in nearly every branch of nonlinear and quantum optics. Their advantages come from the exceptionally high quality factor maintained within a very large wavelength range, small mode volume, continuously tunable coupling, and inherent mechanical stability. On the other hand, complexity of the optical spectrum and its strong temperature dependence make some of nonlinear optics applications challenging, while the exponential behavior of the evanescent field makes simultaneous optimal coupling of different wavelength difficult. We have discussed some of the approaches allowing to solve or circumvent these problems. Perhaps the most important limitation of WGM resonators is their fabrication process. We will discuss this issue specifically in the end of this section.

\subsection{Applications and phenomena not covered in this review}\label{sec:NotCovered}

Some of nonlinear-optical phenomena observed in WGM resonators and associated with them applications have been left out from this paper because they were covered in recent reviews in a great detail. Here we would like to mention them briefly and to direct the readers to these reviews.

One such prominent application, dating back to 1960s, is WGM lasers \cite{garrett_stimulated_1961,Walsh63laser}. Their history and state of art is discussed in a recent review \cite{He13WGM_las_review}. It is worth mentioning that in such lasers the resonator can be either entirely made from the gain media \cite{garrett_stimulated_1961,Walsh63laser,Campillo1991droplets,McCall92laser,miura_laser_1996,Sandoghdar96laser,Chang99las,Cai2000las,Polman04Er,Kippenberg06Er,Murugan11laser}, or coated with it \cite{Yang03solgel,Yang03Er_laser,Yang05laser}.

The coating technique reminds us of an important role  surface physics plays in WGM optics. Besides introducing laser gain media, various molecular coatings can be used to facilitate SHG  \cite{Dominguez-J11coatedSHG,Levy11SiNmicroring} as well as SPDC \cite{Xu2008parametric} in resonators made from inversion-symmetric materials. Theoretically, even a perfectly clean surface of a WGM resonator is expected to enable such processes \cite{Kozyreff08sufaceSHG}, as it too breaks the inversion symmetry.
Optical response of molecules and micro-objects inadvertently attached to WGM resonators immersed in liquids or gases gives rise to efficient bio-sensing techniques recently reviewed in \cite{Vollmer08biosensing,Vollmer12review,Foreman14sens,Foreman15sens}.

Discussing the interaction of WGM photons with acoustic phonons, we alluded to the new and rapidly developing  field of optomechnics. One of the main goals in this field is to reach the quantum regime with mechanical oscillators, which is normally hard to achieve because of low phonon energy and strong coupling to a thermostat. Many spectacular results have been demonstrated in this field lately, see reviews \cite{Kippenberg08optomech,schliesser10aamop,Aspelmeyer14rev}. 
Opto-mechanically induced transparency \cite{dong13pra} and light storage \cite{fiore13pra}, as well as storing optical information as a mechanical excitation \cite{fiore11prl}, were demonstrated in silica microresonators. Formation of opto-mechanical dark modes is reported in \cite{dong12s}. Resolved-sideband and cryogenic cooling of an opto-mechanical resonator is reported in \cite{park09np}. Generation of high quality radio frequency signals by optical means is another significant achievement in the field \cite{Rokhsari05oe}. Of the most direct relevance to the scope of our review is the proposal \cite{Peano15optomech} to couple mechanical oscillations of a WGM resonator with the squeezed light internally generated in it as discussed in section \ref{sec:2sqz}.

\subsection{Anticipated development of the field}\label{sec:Anticipated}

One of the most attractive goals that may be achieved using WGM resonators is strong interaction between individual photons. Achieving this goal would constitute a breakthrough in quantum logic and quantum computing with photons. With the state-of-art nonlinear WGM resonators this goal appears to be close enough but not yet within reach. Its feasibility hinges on the relation between the linear loss rate $\gamma=\omega/Q$ and non-linear conversion rate $g$. To make this relation favorable ($g>\gamma$) one can increase the overlap integral $\sigma$, the $Q$-factor, or the nonlinear susceptibility. Unfortunately, only limited resources are available for the progress in either of these directions.

The overlap integral has almost a universal power scaling with the resonator size, which is captured in Fig.~\ref{fig:Rscale}. Therefore making smaller resonators is the most direct way to increase their nonlinear response.
But very small resonators suffer radiative loss. From Eq.~(\ref{Qrad}) we find that in the near infrared $Q_{\text{rad}}$ of lithium niobate spherical resonators drops to the absorption-limited value of $10^8$ when $R$ ranges from 15 to 20 $\mu$m (the exact radius value depends on the wavelength and polarization), which appears to set a limit for the minimal useful resonator size. For smaller resonators $Q$ is limited by  $Q_{\text{rad}}$ which drops exponentially with the radius. However, microring resonators have been reported with the $Q$s significantly exceeding the radiative limit found from Eq.~(\ref{Qrad}), as can be established by substituting the resonator parameters from e.g.\cite{Vukovic15SHG,Kuo14SHG_GaAs,Mariani14SH} into this equation. Evidently, the model underlying Eq.~(\ref{Qrad}) does not work well for non-spherical shapes, and a more accurate (perhaps numeric) approach is required to establish the radiative loss and find the minimal useful resonator size.

The other two optimization parameters, the ultimate absorption-limited $Q$-factor and nonlinear susceptibility, are both determined by the resonator material. Therefore it is important to explore new optical materials that are strongly nonlinear and at the same time very transparent. These materials also need to be compatible with a stringent surface quality requirements, which includes chemical and mechanical stability, low solubility, etc. Since the second-order susceptibility $\chi^{(2)}$ is much stronger than the third-order susceptibility $\chi^{(3)}$, let us focus on optical crystals with quadratic nonlinearity. A summary of resonator parameters achieved with such crystals  is given in Table~\ref{tbl:crystals} for WGM resonators and in Table~\ref{tbl:crystals1} for ring resonators made from thin crystalline films. Some of these materials have been explored in multiple works. Here we only quote those with the highest demonstrated $Q$.

\begin{table}[htb]
\begin{tabular}{c c c c c c}
		\hline
 Ref.						& Crystal 	& $\lambda$, nm	& $Q$, $10^6$ 	& $d_{ij}$, pm/V  \\
		\hline
\cite{Ilchenko08qtz}		&Quartz		&1550			&5000		&$0.3_{11}$\\
\cite{Lin12_BBO}			&BBO		& 1560			&  740		&$2.3_{22},\,0.16_{31}$\\
\cite{Savchenkov10VCO}		&LiTaO$_3$	& 1550 			&  570		&$16_{33},\,2.7_{31},\,1.6_{22}$ \\
\cite{Ilchenko04SH}			&LiNbO$_3$	& 1310			&  200		& $25_{33},\,4.9_{31},\,2.7_{22}$\\
\cite{Fuerst15SHG}			&LB4		&490			&	200		&$0.55_{33},\,0.073_{31}$\\
\cite{Lin12_BBO}			&BBO		& 370			&  150		&$2.3_{22},\,0.16_{31}$\\
\cite{savchenkov13spie}		&SBN		& 1550			&  75		&  \\
		\hline
\end{tabular}
\caption{Summary of WGM resonators machined from various second-order nonlinear optical crystals. }
\label{tbl:crystals}
\end{table}

\begin{table}[htb]
\begin{tabular}{c c c c c c}
		\hline
 Ref.						& Film 	& $\lambda$, nm	& $Q$, $10^3$ 	& $d_{ij}$, pm/V  \\
		\hline
\cite{Wang15eo}			&LiNbO$_3$		&1550			& 1190		&$25_{33},\,4.9_{31},\,2.7_{22}$\\
\cite{Xiong12AlN}			&AlN		&1555			& 600		&$2.5_{33}$\\
\cite{Lake16SHG} 			&GaP		&1550			& 100		&$68_{36}$	\\
\cite{Vukovic15SHG}			&ZnSe		&1548			&	50		&$33_{36}$\\
\cite{Kuo14SHG_GaAs}		&GaAs		&1985			&	30		&$94_{36}$\\		
\cite{Xiong11GaN_SHG}		&GaN		&1560			& 10		&$3.8_{33},\,2.5_{31},\,2.5_{15}$\\
\cite{Mariani14SH}			&AlGaAs		&1580			&	5		&\\
		\hline	
\end{tabular}
\caption{Summary of on-chip ring resonators fabricated from various second-order nonlinear optical crystalline films. Note that for GaP, ZnSe and GaAs $d_{36}=d_{14}$.}
\label{tbl:crystals1}
\end{table}

In Tables~\ref{tbl:crystals} and \ref{tbl:crystals1} we list the most significant quadratic susceptibility tensor components $d_{ij}=\chi^{(2)}_{ij}/2$.  A combination of these components specific for a given type of phase matching and crystal orientation will determine the $\tilde{\chi}^{(2)}$.

Quality factors reported in Tables~\ref{tbl:crystals} and \ref{tbl:crystals1} are vastly different. Note that in Table~\ref{tbl:crystals} $Q$ is given in millions, and in Table~\ref{tbl:crystals1} in thousands. The inferior quality of the ``large" ($R=40\;\mu$m) ring resonators \cite{Xiong12AlN,Xiong11GaN_SHG} is clearly unaffected by the radiative losses and is most likely due to the poor surface quality. This means that there is a potential for a significant increase of $Q$ by improving the fabrication process. There is also a possibility that the absorption-limited $Q$ in complex crystals such as lithium niobate can be increased by optimizing the crystal composition and reducing the impurities concentration \cite{Leidinger15LN_abs}.

Besides facilitating a strong interaction between two photons, development of strongly nonlinear high-$Q$ resonators is motivated by the possibility to access multipartite and multiphoton  entangled states, such as optical GHZ states \cite{Greenberger90GHZ,Pan2000GHZ}, W-states \cite{Dur2000W,Eibl04W,Jin14W}, cluster \cite{Briegel01cluster} and graph \cite{Hein04graph} states, Smolin states \cite{Smolin01}, and others. Theses states are very important in the context of quantum information applications as well as in fundamental quantum optics \cite{Raussendorf01qcomp,Yeo2006,Agrawal06W}, but due to weak optical nonlinearity, their realization has been so far only possible at low photon rate with very strong pump pulses, a combination usually leading to noisy data. One may expect that nonlienar WGM resonators will advance this field of research towards higher photon numbers (and higher-dimensional Hilbert space) while significantly reducing the requirements to the pump lasers.

While yielding the best $Q$ factors, fabrication of crystalline resonators by diamond turning and polishing unfortunately remains an art. Meanwhile, many research and industrial applications involving classical optical fields could benefit from this technology if a robust and scalable fabrication technique was available. It may be expected that developing such techniques, even at the cost of the quality factor, will be one of the thrust directions in the WGM research in the nearest future.

\subsection{Fabrication challenges and scalability quest}

The complexity of WGM resonators fabrication and integration appears to be the main factor limiting their broad infusion into nonlinear and quantum optics technology. While WGM resonators remain objects of research, it is acceptable to invest considerable time in their individual fabrication, mounting on a mechanically and thermally stable platform, and coupling to the optical input and output. However using a resonator as an instrument component requires repeatable and scalable fabrication process, ideally chip-based. This is partially accomplished with fused silica microdisks made from the naturally formed oxide layer on a silicon substrate. The layer is photolithographically patterned with disk preforms, then the substrate is dry-etched to form pedestals, and the elevated preforms are either reflown into toroidal shapes by laser heating, e.g. \cite{kippenberg04Kerr,Del'Haye07comb,DelHaye08comb,Kippenberg04Raman,Kippenberg04Raman1,Li13Brillouin}, or chemically etched into wedge shapes \cite{Lee12etch,yi15o}. The quality factor of toroids routinely exceeds $10^8$, in wedge-shaped resonators it can almost reach $10^9$ \cite{Lee12etch}.

Such a technologically important material as lithium niobate presents a harder challenge, which has been addressed only most recently. In one approach a lithium niobate wafer was bound on a silicon surface, polished down to one micron thickness and milled with a focused ion beam into a disk shape. The disk was elevated from the silicon substrate by etching the later with XeF$_2$ \cite{Wang14wgmr}. This fabrication technique yielded $R=35\;\mu$m resonators with $Q=4.84\times 10^5$.

An alternative technological approach involving femtosecond laser ablation followed by focused ion beam milling, HF etching and finally high temperature annealing yielded 0.7 $\mu$m thick, $R=41\;\mu$m resonators with $Q=2.5\times 10^5$ \cite{Lin15sh,Lin15fab}.

A combination of chemical (HF) and reactive ion etching of a lithographically patterned lithium niobate - silica - lithium niobate ``sandwich" allowed for fabrication of 0.4 $\mu$m thick, $R=28\;\mu$m resonators with the quality factor reaching $10^5$\cite{Wang14SGH}. This technology was later perfected by a different group, allowing to reach  $Q=1.19\times 10^6$ with $R=39.6\;\mu$m resonators of similar thickness \cite{Wang15eo}.

Diamond is another technologically appealing material with remarkable optical properties. Very small ($R=2.4\;\mu$m) diamond ring-resonators with $Q=5000$ have been fabricated by bonding diamond film onto a silica-on-silicon substrate, followed by a series of lithographic patterning and dry etching steps \cite{Faraon11diamond}. These resonators have been used to enhance NV centers emission at low temperature.

\begin{figure}[t]
\begin{center}
\includegraphics[width=7.5cm]{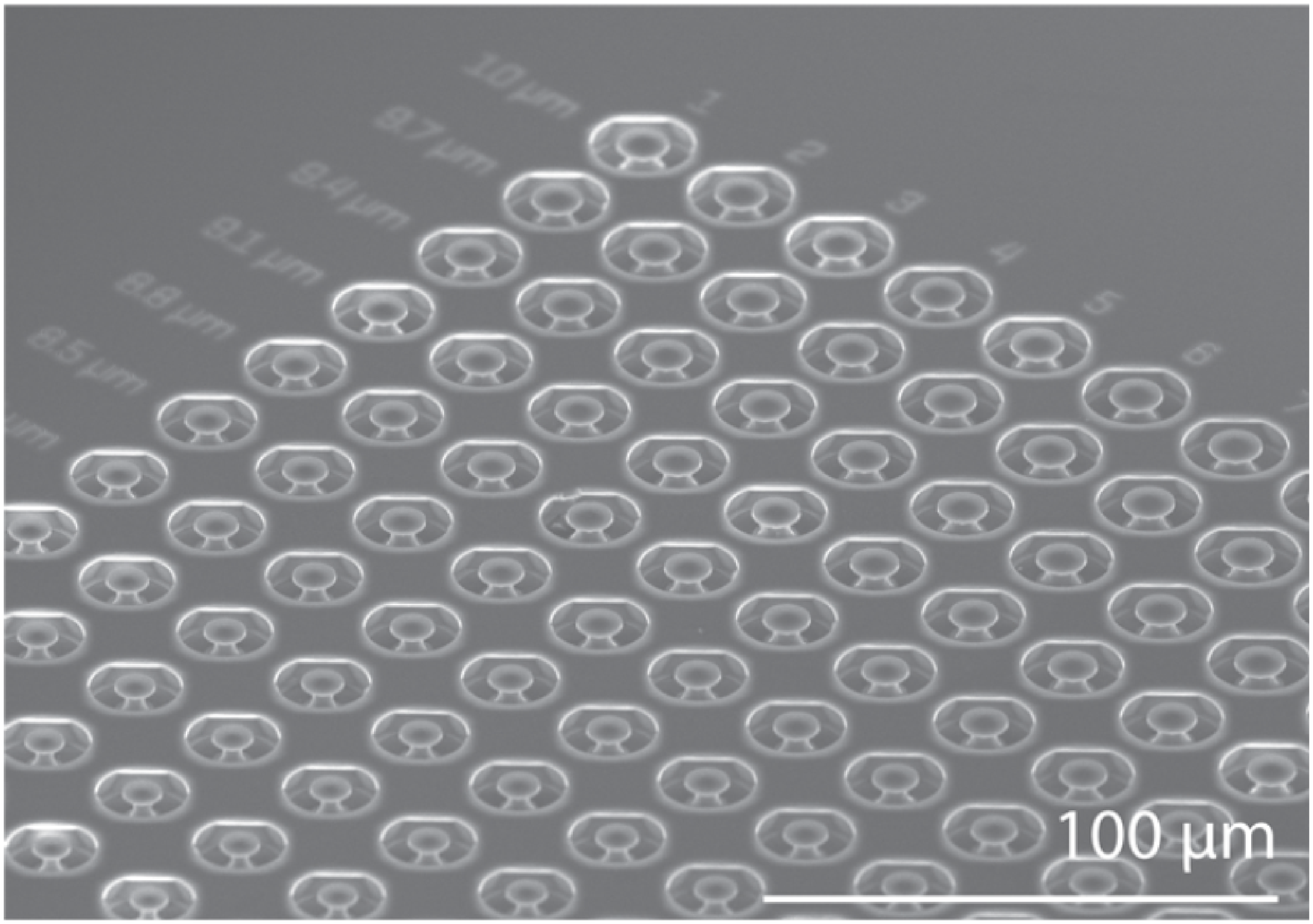}
\includegraphics[width=7.5cm]{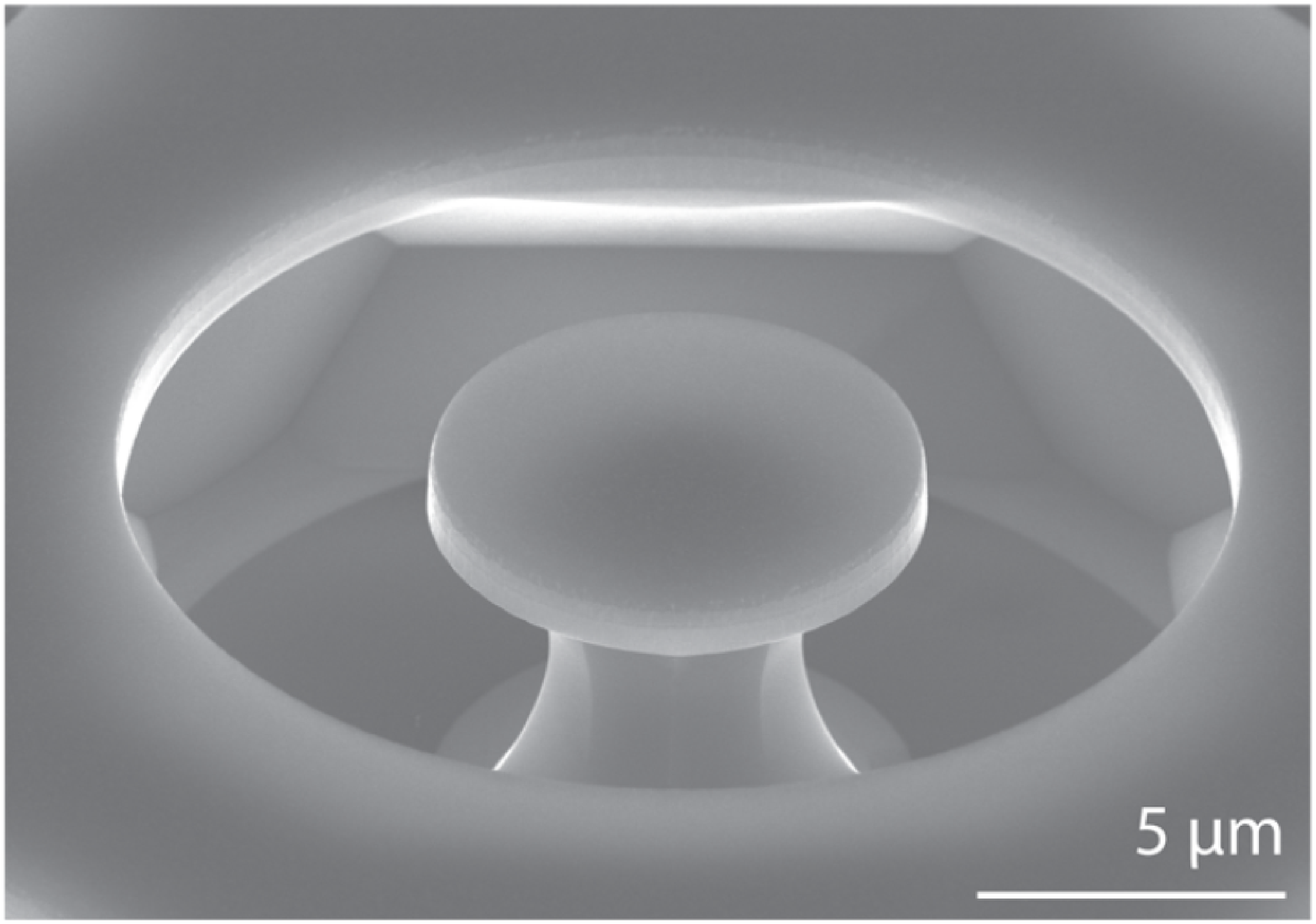}
\end{center}
\caption[]{\label{fig:diamond} Top: an array of microdisk resonators fabricated from a bulk diamond. Bottom: a magnified view of a single resonator. Reprinted from \cite{Khanaliloo15diamond}.}
\end{figure}

A versatile method that may be applicable to a variety of materials has been developed by Burek \textit{et al.} for fabrication of various free-standing nanostructures, including disk, ring, and racetrack resonators, out of single-crystal diamond \cite{Burek12diamond,Burek14diamond}. The method is based on the angle-etching with ions, a technique when the ions trajectories are bent by a Faraday cage, allowing to undercut and eventually to fully separate the lithographically defined structures from the bulk material. With this technique, $Q\approx 1.5\times 10^5$ was achieved for the racetrack resonators \cite{Burek14diamond}.
Etching bulk diamond was also shown to produce arrays of $R=3.95\;\mu$m on-chip disk resonators with $Q\approx 10^5$  \cite{Khanaliloo15diamond}, see Fig.~\ref{fig:diamond}. This process was quite complicated and consisted of seven coating/etching steps.

The above technologies produce well controlled resonators, but the couplers (typically, tapered fibers or waveguides) need to be fabricated and mounted separately, which limits its scalability.
On-chip fabrication of resonator and waveguide coupler arrays, such as shown in Figs.~\ref{fig:MZint} and \ref{fig:SiNswitch}, is possible using reactive ion etching of various resonator material layers grown on a substrate \cite{Xiong11GaN_SHG,Levy10comb,razzari10np,Ferrera08,Levy11SiNmicroring,Pernice12shg,Xiong12AlN,Dutt15on-chip_sqz,Heebner04phase}, sometimes complete with a protective layer. In the case of silicon, the resonator-waveguide structures have been fabricated using the standard silicon-on-insulator microfabrication techniques \cite{Clemmen:2009dn,Azzini:2012rc,Engin:2013qd,Guo:2014db,Grassani:2015zl,Wakabayashi:2015lq,Suo:2015fv,Xu06switch,Xu07switch,Li16conv}. Besides the earlier mentioned four-wave-mixing based applications, such devices have been used as all-optical switches \cite{Xu06switch,Xu07switch}. This application is based on the free carrier induced optical bistability, and leads to extremely low optical pulse switching energy of just several pJ.

These techniques enables production of coupled WGM devices, potentially with multiple resonators and a network of couplers and waveguides on the same chip. One drawback of this approach is the fixed coupling rate: the gaps between the resonators and couplers cannot be changed, although the coupling rate may be adjusted by immersing the device in variable index fluids. Besides, as we already mentioned, at the present technology level the $Q$-factors of etched microring resonators is usually limited, although a very respectable value of $Q\approx 1.2\times 10^6$ was reported in a doped high-index silica glass waveguide-coupled microring at 1544 nm wavelength \cite{razzari10np}. Nearly as high $Q$-factors in the same optical band were measured in single-crystal diamond waveguide-coupled microrings \cite{Hausmann14diamond}, that were also used to generate Kerr combs.  In $\chi^{(2)}$ materials, the quality factors are far more modest. For a lithium niobate microring ($R=100\;\mu$m) resonator $Q\approx 4000$ has been achieved \cite{Guarino07ln}; for a  waveguide-coupled GaAs microdisk ($R=3.7\;\mu$m), $Q\approx 10^4$ \cite{Baker11citical}.

\ack

D.V.S. acknowledges financial support from Alexander von Humboldt Foundation and the DARPA Quiness program, and would like to thank Dr. Maria Chekhova for useful discussions. H.G.L.S. would like to thank Florian Sedlmeir and Alfredo Rueda for useful discussions. The authors appreciate Drs.' Kartik Srinivasan, Serge Rosenblum and Hailin Wang valuable feedback on this paper preprint. The research was partly carried out at the Jet Propulsion Laboratory, California Institute of Technology, under a contract with the National Aeronautics and Space Administration.

\section*{References}

\bibliographystyle{iopart-num}
%\bibliography{./../bibliography}

\end{document}